\documentclass[aps,twocolumn,showpacs,preprintnumbers,floats]{revtex4}\def\@cite#1#2{\textsuperscript{[{#1\if@tempswa , #2\fi}]}}

\usepackage{graphicx}
\usepackage{amsmath}
\usepackage{amsfonts}
\usepackage{amssymb}
\usepackage{color}
\usepackage{subfigure}
\usepackage{epsfig}
\usepackage{morefloats}
\usepackage{multirow}
\usepackage{graphicx,booktabs}
\usepackage{mathrsfs}
\usepackage{txfonts}
\usepackage{indentfirst}
\usepackage{graphicx,booktabs}

\usepackage{longtable,lscape}
\newcommand{\vrho}{\mbox{\boldmath$\rho$\unboldmath}}
\newcommand{\vlab}{\mbox{\boldmath$\lambda$\unboldmath}}

\begin{document}

\title{Strong and radiative decays of the low-lying $D$-wave singly heavy baryons }
\author{
Ya-Xiong Yao$^{1}$, Kai-Lei Wang$^{1}$, Xian-Hui Zhong$^{1,2}$~\footnote {E-mail: zhongxh@hunnu.edu.cn}}

\affiliation{ 1) Department
of Physics, Hunan Normal University, and Key Laboratory of
Low-Dimensional Quantum Structures and Quantum Control of Ministry
of Education, Changsha 410081, China }

\affiliation{ 2) Synergetic Innovation Center for Quantum Effects and Applications (SICQEA),
Hunan Normal University, Changsha 410081, China}

\begin{abstract}

The strong and radiative decays of the low-lying $\lambda$-mode $D$-wave $\Lambda_{c(b)}$,
$\Sigma_{c(b)}$, $\Xi_{c(b)}$, $\Xi_{c(b)}'$, and $\Omega_{c(b)}$ baryons
are studied in a constituent quark model.
Our calculation shows the following: (i) The missing $\lambda$-mode $D$-wave $\Omega_{c(b)}$,
$\Lambda_{b}$, and $\Xi_{b}$ baryons have a relatively narrow decay width of a few MeV or a few tens of MeV and
their dominant strong and radiative decay channels can be ideal for searching
for their signals in future experiments. (ii) The $\lambda$-mode $1D$-wave excitations in the $\Sigma_{c(b)}$
and $\Xi_{c(b)}'$ families appear to have a relatively broad width of $\sim 50-200$ MeV.
Most of the $1D$-wave states have large decay rates into the $1P$-wave heavy baryons
via the pionic or kaonic strong decay processes, which should be taken seriously in future observations.
(iii) Both $\Lambda_c(2860)$ and $\Xi_c(3050)$ seem to favor the $J^P=3/2^+$ excitation
$|^2D_{\lambda\lambda} \frac{3}{2}^+ \rangle$ of $\bar{\mathbf{3}}_F$, while both
$\Lambda_c(2880)$ and $\Xi_c(3080)$ may be assigned as the $J^P=5/2^+$ excitation
$|^2D_{\lambda\lambda} \frac{5}{2}^+ \rangle$ of $\bar{\mathbf{3}}_F$.
The nature of $\Xi_c(3050)$ and $\Xi_c(3080)$ could be tested by the radiative transitions $\Xi_c(3055)^0\to \Xi_c(2790)^0  \gamma$ and $\Xi_c(3080)^0 \to \Xi_c(2815)^0 \gamma$, respectively.
\end{abstract}

\maketitle

\section{Introduction}{\label{introduction}}

The LHC facility provides good opportunities for us
to discover some of the missing heavy baryons. Recently, five extremely narrow $\Omega_c$
states, $\Omega_c(3000)$, $\Omega_c(3050)$, $\Omega_c(3066)$, $\Omega_c(3090)$, and $\Omega_c(3119)$,
were observed in the $\Xi_c^{+}K^-$ channel by the LHCb Collaboration~\cite{Aaij:2017nav}.
Most of them may be interpreted as the $P$-wave excited states of $\Omega_c$~\cite{Wang:2017hej,Cheng:2017ove,
Chen:2017gnu,Padmanath:2017lng,Wang:2017vnc, Karliner:2017kfm,Chen:2017sci,Agaev:2017lip}.
Lately, the LHCb Collaboration observed a new structure $\Xi_b(6227)^-$ in both the $\Lambda_b^0K^-$ and $\Xi_b^0\pi^-$
invariant mass spectra~\cite{Aaij:2018yqz}. The mass of this structure and the observed
decay modes are consistent with expectations of a $P$-wave excited state in the $\Xi_b'$ family~\cite{Wang:2017kfr,Chen:2018orb,Ebert:2007nw,Ebert:2011kk,Roberts:2007ni,Chen:2014nyo,Mao:2015gya,
GarciaRecio:2012db,Karliner:2008sv,Wang:2010it,Valcarce:2008dr,Vijande:2012mk,Thakkar:2016dna}.
Besides the missing $P$-wave heavy baryons, some low-lying $D$-wave singly heavy baryons
should be also observed at the LHC in forthcoming experiments. Furthermore,
the Belle II experiments will also offer
the possibility of studying excited heavy baryons. Thus, the theoretical studies of
the low-lying $P$- and $D$-wave singly heavy baryons will
provide very useful references for searching for them in future experiments.
Considering that the decay properties of heavy baryons should be sensitive to its inner structure,
one may better understand the nature of the heavy baryons by studying their decays.
In our recent work~\cite{Wang:2017kfr}, we systematically studied the
strong and radiative decay properties of the $P$-wave singly heavy baryons.
As a continuation of Ref.~\cite{Wang:2017kfr}, we study the
strong and radiative decays of the low-lying $D$-wave singly heavy baryons in the present work.

In the heavy baryon resonances listed in the Review of Particle Physics (RPP)~\cite{Olive:2016xmw},
there are several good $D$-wave candidates, such as $\Lambda_c(2880)^+$, $\Xi_c(3055)^{0,+}$,
$\Xi_c(3080)^{0,+}$, and $\Xi_c(3123)^+$. Recently, a new $D$-wave candidate in
the $\Lambda_c$ family, i.e., $\Lambda_c(2860)^+$, was observed in the $D^0p$ channel by
the LHCb Collaboration~\cite{Aaij:2017vbw}. However, no candidates of the $D$-wave bottom baryons
have been found in experiments. To look for the missing $D$-wave
singly heavy baryons, and to identify these possible $D$-wave heavy
baryons observed in experiments, many theoretical studies were carried out
with various phenomenological methods. For example, the mass spectra were calculated in various quark models~\cite{Ebert:2011kk,Yoshida:2015tia,Roberts:2007ni,Ebert:2007nw,Ebert:2005xj,
Maltman:1980er,Copley:1979wj,Valcarce:2008dr,Shah:2016mig,
Shah:2016nxi,Karliner:2008sv,Chen:2014nyo,Chen:2016iyi,Capstick:1986bm},
the Faddeev method~\cite{Garcilazo:2007eh}, the lattice QCD~\cite{Bali:2015lka,Padmanath:2013bla}, the QCD sum rules~\cite{Mao:2017wbz,Chen:2016phw,Wang:2017vtv}, and so on.
Furthermore, the strong decay properties of the low-lying $D$-wave charmed baryons were also
studied within some methods, such as the $^3P_0$ model~\cite{Chen:2007xf,Chen:2014nyo,Chen:2016iyi,
Chen:2017aqm,Zhao:2016qmh,Ye:2017dra}, the heavy hadron chiral perturbation theory
~\cite{Cheng:2006dk,Cheng:2015iom,Cheng:2015naa}, the chiral quark model (ChQM)~\cite{Zhong:2007gp,Liu:2012sj,Nagahiro:2016nsx}, and so on.
It should be pointed out that there are few discussions of the radiative decays of
the $D$-wave charmed and bottom baryons and the strong decays of the $D$-wave bottom baryons,
although there are many discussions about the radiative decays~\cite{Cheng:1992xi,Wang:2009ic,Wang:2009cd,Jiang:2015xqa,Zhu:1998ih,Tawfiq:1999cf,Dey:1994qi,
Bernotas:2013eia,Gamermann:2010ga,Aliev:2014bma,Aliev:2009jt,Aliev:2016xvq,Aliev:2011bm,
Chow:1995nw,Bahtiyar:2016dom,Bahtiyar:2015sga,Ivanov:1998wj,Savage:1994wa,Banuls:1999br} and strong decays~\cite{Chen:2017sci,Zhu:2000py,Agaev:2017lip,Agaev:2017nn,Chen:2017gnu,Ye:2017yvl,Cho:1994vg,Pirjol:1997nh,
Chiladze:1997ev,Blechman:2003mq,Huang:1995ke,Tawfiq:1998nk,Tawfiq:1999vz,Ivanov:1999bk,Ivanov:1998qe,
Korner:1994nh,Hussain:1999sp,Albertus:2005zy,Hwang:2006df,Guo:2007qu,Hernandez:2011tx,Limphirat:2010zz,Aliev:2010yx} for the low-lying $S$- and/or $P$-wave singly heavy baryons. More details about the status for the studies of the heavy baryons can be found in Refs.~\cite{Chen:2016spr,Crede:2013sze,
Klempt:2009pi,Korner:1994nh}. A a whole, it is necessary to carry out a systematical study of the strong and radiative decays for the
$D$-wave singly charmed and bottom baryons.

In this work, we apply a nonrelativistic constituent quark model
to study the strong decays with emission of one light
pseudoscalar meson and
the radiative decays with emission of one photon for the low-lying $D$-wave singly heavy
baryons.  By an analysis of the decay properties for the $D$-wave states, we will suggest ideal decay channels to observe
missing states in future experiments. For a simplicity, the harmonic oscillator wave functions of
the heavy baryons are adopted in our calculations.
To deal with the strong decays of a hadron, an effective
chiral Lagrangian at the tree level~\cite{Manohar:1983md} is introduced.
In this interaction, the emitted light pseudoscalar mesons
are treated as Goldstone bosons, which only couple with
the light constituent quarks. Since the quark-meson coupling is invariant under the
chiral transformation, some of the low-energy properties
of QCD are retained~\cite{Manohar:1983md,Li:1997gd,Zhao:2002id}. This method ( i.e., ChQM)
has been successfully applied to study the strong decays of heavy-light mesons and charmed
and strange baryons~\cite{Zhong:2008kd,Zhong:2010vq,Zhong:2009sk,
Zhong:2007gp,Liu:2012sj,Xiao:2013xi,Nagahiro:2016nsx,Wang:2017hej,Xiao:2014ura,Xiao:2017udy}.
The chiral quark model used in this work is different from the often-used $^3P_0$ model~\cite{Micu:1968mk,LeYaouanc:1972vsx,
LeYaouanc:1973ldf}; the differences between them have been pointed out in Ref.~\cite{Wang:2017kfr}.
Meanwhile, to treat the radiative decay of a hadron, we apply
an effective quark-photon electromagnetic (EM) coupling at the tree level.
The higher EM multipole contributions are included by a multipole expansion of the EM interactions.
This approach has been successfully applied to deal with the radiative decays of $c\bar{c}$
and $b\bar{b}$ systems~\cite{Deng:2016stx,Deng:2016ktl}, and recently it has been extended to
study the radiative transitions of heavy baryons~\cite{Wang:2017hej,Lu:2017meb,Xiao:2017udy}.

The paper is organized as follows. Section~\ref{framework} is our framework, in which we give a brief
review of the quark model classification of the singly heavy baryons and the quark model description
of the strong and radiative decays. Then, the numerical results for the heavy baryons belonging to $\bar{\mathbf{3}}_F$ and $\mathbf{6}_F$
are presented and discussed in Secs.~\ref{RD} and ~\ref{RD2}, respectively. Finally, a summary is
given in Sec.~\ref{Summa}.

%p{1.6cm}|p{1.8cm}p{1.8cm}p{1.8cm}p{1.8cm}p{1.8cm}p{1.8cm}p{1.8cm}p{1.8cm}p{1.8cm}
\begin{table*}[htp]
\begin{center}
\caption{\label{sp1}  Mass spectra of the singly heavy baryons of $\bar{\mathbf{3}}_F$ up to $D$ wave from
various quark models~\cite{Ebert:2011kk,Yoshida:2015tia,Chen:2016iyi,Roberts:2007ni} compared
with the data from the Particle Data Group~\cite{Olive:2016xmw}.}
\scalebox{1.0}{
\begin{tabular}{c|ccccccccccccccccccccccccccccccccccccccccccccc}\hline\hline
&\multicolumn{4}{c}{$\underline{~~~~~~~~~~~~~~~~~~~~~~~~~~~~~~~~~~~~~~~~~~~~~~~~~~~~~~~~~\Lambda_c~~~~~~~~~~~~~~~~~~~~~~~~~~~~~~~~~~~~~~~}$}    &\multicolumn{3}{c}{$\underline{~~~~~~~~~~~~~~~~~~~~~~~~~~~~~~~~~~\Lambda_b ~~~~~~~~~~~~~~~~~~~~~~~~~~~}$}\\
State          &~~~~RQM~\cite{Ebert:2011kk}~~~~&~~~~NQM~\cite{Yoshida:2015tia}~~~~    &~~~~NQM~\cite{Chen:2016iyi}~~~~      &~~~~PDG~\cite{Olive:2016xmw}~~~~   &~~~~RQM~\cite{Ebert:2011kk}~~~~ &~~~~NQM~\cite{Yoshida:2015tia}~~~~ &~~~~PDG~\cite{Olive:2016xmw}~~~~\\ \hline
$1^2S \frac{1}{2}^+$                &2286     &2285        &2286      &2286    &5620      &5618      &5620 \\
$1^2P_{\lambda} \frac{1}{2}^-$       &2598     &2628       &2614      &2592     &5930      &5938      &5912\\
$1^2P_{\lambda} \frac{3}{2}^-$       &2627     &2630       &2639      &2628     &5942      &5939      &5920\\
$1^2D_{\lambda\lambda} \frac{3}{2}^+$&2874     &2920       &2843      &2860?         &6190      &6211   & ?\\
$1^2D_{\lambda\lambda} \frac{5}{2}^+$&2880     &2922       &2851      &2880?        &6196      &6212   &?\\ \hline
&\multicolumn{4}{c}{$\underline{~~~~~~~~~~~~~~~~~~~~~~~~~~~~~~~~~~~~~~~~~~~~~~~~~~~~~~~~~\Xi_c~~~~~~~~~~~~~~~~~~~~~~~~~~~~~~~~~~~~~~~}$}    &\multicolumn{3}{c}{$\underline{~~~~~~~~~~~~~~~~~~~~~~~~~~~~~~~~~~\Xi_b ~~~~~~~~~~~~~~~~~~~~~~~~~~~}$}\\
State          &RQM~\cite{Ebert:2011kk}        &NQM~\cite{Roberts:2007ni}      &NQM~\cite{Chen:2016iyi}     &PDG~\cite{Olive:2016xmw}  &RQM~\cite{Ebert:2011kk}          &NQM~\cite{Roberts:2007ni} &PDG~\cite{Olive:2016xmw}\\ \hline
$1^2S \frac{1}{2}^+$                 &2476     &2466       &2470     &$2468$               &5803     &5806         &5795\\
$1^2P_{\lambda} \frac{1}{2}^-$       &2792     &2773       &2793     &2792                 &6120     &6090         &?\\
$1^2P_{\lambda} \frac{3}{2}^-$       &2819     &2783       &2820     &2817                 &6130     &6093         &?\\
$1^2D_{\lambda\lambda} \frac{3}{2}^+$&3059     &3012       &3033     &3055?                   &6366     &6311 & ?\\
$1^2D_{\lambda\lambda} \frac{5}{2}^+$&3076     &3004       &3040     &3080?                  &6373     &6300 &?\\
\hline\hline
\end{tabular}}
\end{center}
\end{table*}

%p{1.6cm}|p{1.8cm}p{1.8cm}p{1.8cm}p{1.8cm}p{1.8cm}p{1.8cm}p{1.8cm}p{1.8cm}p{1.8cm}
\begin{table*}[htp]
\begin{center}
\caption{\label{sp2}  Mass spectra of the singly heavy baryons of $\mathbf{6}_F$ up to $D$ wave from
various quark models~\cite{Ebert:2011kk,Yoshida:2015tia,Chen:2016iyi,Roberts:2007ni} compared
with the data from the Particle Data Group~\cite{Olive:2016xmw}.}
\scalebox{1.0}{
\begin{tabular}{c|ccccccccccccccccccccccccccccccccccccccccccccc}\hline\hline
&\multicolumn{4}{c}{$\underline{ ~~~~~~~~~~~~~~~~~~~~~~~~~~~~~~~~~~\Sigma_c~~~~~~~~~~~~~~~~~~~~~~~~~~~~~~~~~}$}    &\multicolumn{3}{c}{$\underline{~~~~~~~~~~~~~~~~~~~~~~~~~~~~~~~~~~\Sigma_b ~~~~~~~~~~~~~~~~~~~~~~~~~~~}$}\\
State          &RQM~\cite{Ebert:2011kk}       &NQM~\cite{Yoshida:2015tia}   &NQM~\cite{Chen:2016iyi}  &PDG~\cite{Olive:2016xmw} &RQM~\cite{Ebert:2011kk}       &NQM~\cite{Yoshida:2015tia} &PDG~\cite{Olive:2016xmw}\\ \hline
$1^2S \frac{1}{2}^+$                &2443      &2460       &2456     &2455        &5808       &5823         &5811 \\
$1^4S \frac{3}{2}^+$                &2519      &2523       &2515     &2520        &5834       &5845         &5832 \\
$1^2P_{\lambda} \frac{1}{2}^-$      &2713      &2802       &2702     & ?          &6101       &6127         &?\\
$1^2P_{\lambda} \frac{3}{2}^-$      &2798      &2807       &2785     & ?          &6096       &6132 &?\\
$1^4P_{\lambda} \frac{1}{2}^-$      &2799      &2826       &2765     & ?          &6095       &6135 &?\\
$1^4P_{\lambda} \frac{3}{2}^-$      &2773      &2837       &2798     & ?          &6087       &6141 &?\\
$1^4P_{\lambda} \frac{5}{2}^-$      &2789      &2839       &2790     & ?          &6084       &6144 &?\\
$1^2D_{\lambda\lambda} \frac{3}{2}^+$&3043     &3065       &2952     & ?          &6326       &6356 &?\\
$1^2D_{\lambda\lambda} \frac{5}{2}^+$&3038     & 3099      &2942     & ?          &6284       &6397 &?\\
$1^4D_{\lambda\lambda} \frac{1}{2}^+$&3041     &3103       &2949     & ?          &6311       &6395 &?   \\
$1^4D_{\lambda\lambda} \frac{3}{2}^+$&3040     &3094       &2964     & ?          &6285       &6393 &?\\
$1^4D_{\lambda\lambda} \frac{5}{2}^+$&3023     &3114       &2962     & ?          &6270       &6402 &?\\
$1^4D_{\lambda\lambda} \frac{7}{2}^+$&3013     &$\cdot\cdot\cdot$        &2943    & ?         &6260       & $\cdot\cdot\cdot$&?\\ \hline
&\multicolumn{4}{c}{$\underline{~~~~~~~~~~~~~~~~~~~~~~~~~~~~~~~~~~~~~~~\Xi^{\prime}_c~~~~~~~~~~~~~~~~~~~~~~~~~~}$}    &\multicolumn{3}{c}{$\underline{~~~~~~~~~~~~~~~~~~~~~~~~~~~~~~~~~~\Xi^{\prime}_b ~~~~~~~~~~~~~~~~~~~~~~~~~~~}$}\\
State          &RQM~\cite{Ebert:2011kk}       & NQM~\cite{Roberts:2007ni}   &NQM~\cite{Chen:2016iyi}     &PDG~\cite{Olive:2016xmw}   &RQM~\cite{Ebert:2011kk}  & NQM~\cite{Roberts:2007ni}&PDG~\cite{Olive:2016xmw}               \\ \hline
$1^2S \frac{1}{2}^+$                 &2579     & 2592              &2579     &2575  &5936  &5958 &5935                  \\
$1^4S \frac{3}{2}^+$                 &2649     & 2650              &2649     &2645  &5963  &5982 &5955            \\
$1^2P_{\lambda} \frac{1}{2}^-$       &2936     & 2859              &2839     & ?     &6233  &6192 &  ?                     \\
$1^2P_{\lambda} \frac{3}{2}^-$       &2935     & 2871              &2921     & ?     &6234  &6194 & ?                     \\
$1^4P_{\lambda} \frac{1}{2}^-$       &2854     & $\cdot\cdot\cdot$ &2900     & ?     &6227  &$\cdot\cdot\cdot$& ?                     \\
$1^4P_{\lambda} \frac{3}{2}^-$       &2912     &$\cdot\cdot\cdot$  &2932     & ?     &6224  &$\cdot\cdot\cdot$& ?                 \\
$1^4P_{\lambda} \frac{5}{2}^-$       &2929     &2905               &2927     & ?     &6226  &6204 &   ?        \\
$1^2D_{\lambda\lambda} \frac{3}{2}^+$&3167     &$\cdot\cdot\cdot$ &3089     & ?      &6459  &$\cdot\cdot\cdot$ &?                        \\
$1^2D_{\lambda\lambda} \frac{5}{2}^+$&3166     &$\cdot\cdot\cdot$&3091     & ?      &6432   & 6402       &?    \\
$1^4D_{\lambda\lambda} \frac{1}{2}^+$ &3163    &$\cdot\cdot\cdot$&3075     & ?      &6447   &$\cdot\cdot\cdot$ &?                          \\
$1^4D_{\lambda\lambda} \frac{3}{2}^+$&3160     &$\cdot\cdot\cdot$ &3081    & ?      &6431   &$\cdot\cdot\cdot$ &?                      \\
$1^4D_{\lambda\lambda} \frac{5}{2}^+$&3153     &3080       &3077     &     ?       &6420    &$\cdot\cdot\cdot$ &?            \\
$1^4D_{\lambda\lambda} \frac{7}{2}^+$&3147     &3094       &3078     &    ?       &6414     & 6405           &?                          \\
\hline
&\multicolumn{4}{c}{$\underline{~~~~~~~~~~~~~~~~~~~~~~~~~~~~~~~~~~~~~~~~~\Omega_c~~~~~~~~~~~~~~~~~~~~~~~~~~~~~}$}    &\multicolumn{3}{c}{$\underline{ ~~~~~~~~~~~~~~~~~~~~~~~~~~~~~~~\Omega_b ~~~~~~~~~~~~~~~~~~~~~~~~~~~}$}\\
State          &RQM~\cite{Ebert:2011kk}  &NQM~\cite{Yoshida:2015tia} &NQM~\cite{Roberts:2007ni}     &PDG~\cite{Olive:2016xmw}    &RQM~\cite{Ebert:2011kk}     &NQM~\cite{Yoshida:2015tia} &PDG~\cite{Olive:2016xmw}\\ \hline
$1^2S \frac{1}{2}^+$                 &2698     &2731   &2718             &2695       &6064       &6076        &6046            \\
$1^4S \frac{3}{2}^+$                 &2768     &2779   &2776             &2770       &6088       &6094       &   ?             \\
$1^2P_{\lambda} \frac{1}{2}^-$       &2966     &3030   &2977             &  ?        &6330       &6333      &   ? \\
$1^2P_{\lambda} \frac{3}{2}^-$       &3029     &3033   &2986             &  ?        &6331       &6336     &   ? \\
$1^4P_{\lambda} \frac{1}{2}^-$       &3055     &3048   &2990             &   ?       &6339       &6340   &   ?  \\
$1^4P_{\lambda} \frac{3}{2}^-$       &3054     &3056   &2994             &   ?       &6340       &6344   &   ? \\
$1^4P_{\lambda} \frac{5}{2}^-$       &3051     &3057   &3014             &   ?       &6334       &6345   &   ? \\
$1^2D_{\lambda\lambda} \frac{3}{2}^+$&3282     &3257   &3262             &   ?       &6530       &6528  &   ?          \\
$1^2D_{\lambda\lambda} \frac{5}{2}^+$&3286     &3288   &3273             &  ?        &6520       &6561  &   ?  \\
$1^4D_{\lambda\lambda} \frac{1}{2}^+$&3287     &3292   &3275             &  ?        &6540       &6561  &   ?   \\
$1^4D_{\lambda\lambda} \frac{3}{2}^+$&3298     &3285  &3280              &   ?       &6549      &6559   &   ?          \\
$1^4D_{\lambda\lambda} \frac{5}{2}^+$&3297     &3299  &$\cdot\cdot\cdot$  &  ?       &6529       &6566  &   ?  \\
$1^4D_{\lambda\lambda} \frac{7}{2}^+$&3283     &$\cdot\cdot\cdot$&3327    & ?        &6517       &$\cdot\cdot\cdot$ &   ?\\
\hline\hline
\end{tabular}}
\end{center}
\end{table*}

\section{framework}\label{framework}

\subsection{Spectra}

The heavy baryon containing a heavy quark violates
the SU(4) symmetry. However, the SU(3) symmetry between the other two
light quarks ($u$, $d$, or $s$) is approximately kept.
The heavy baryons containing a single heavy quark
belong to two different SU(3) flavor representations: the symmetric sextet $\mathbf{6}_F$ and
antisymmetric antitriplet $\bar{\mathbf{3}}_F$~\cite{Wang:2017kfr}. In the singly charmed (bottom) baryons, there are two families,
$\Lambda_{c}$ and $\Xi_{c}$ ($\Lambda_{b}$ and $\Xi_{b}$) belonging to $\bar{\mathbf{3}}_F$,
while there are three families, $\Sigma_{c}$, $\Xi_{c}'$, and $\Omega_{c}$
($\Sigma_{b}$, $\Xi_{b}'$, and $\Omega_{b}$),
belonging to $\mathbf{6}_F$~\cite{Wang:2017kfr}.

The spatial wave function of a heavy baryon is adopted the
harmonic oscillator form in the constituent quark model~\cite{Zhong:2007gp}.
For a $q_1q_2Q$ basis state, it contains
two light quarks $q_1$ and $q_2$ with
equal mass $m$ and a heavy quark $Q$ with mass $m'$. The
basis states are generated by the oscillator Hamiltonian
\begin{eqnarray} \label{hm2}
\mathcal{H}=\frac{P^2_{cm}}{2 M}+\frac{1}{2m_\rho}\mathbf{p}^2_\rho+\frac{1}{2m_\lambda}\mathbf{p}^2_\lambda+
\frac{3}{2}K(\rho^2+\lambda^2).
\end{eqnarray}
The constituent quarks are confined in an oscillator potential with the potential parameter $K$ independent
of the flavor quantum number. The Jacobi coordinates $\vrho$ and $\vlab$ and c.m. coordinate
$\mathbf{R}_{c.m.}$ can be related to the coordinate $\textbf{r}_{j}$ of the $j$th
quark. The momenta $\mathbf{p}_\rho$, $\mathbf{p}_\lambda $, and $\mathbf{P}_{c.m.}$ are defined by
$\mathbf{p}_\rho=m_\rho\dot{\vrho}$,
$\mathbf{p}_\lambda=m_\lambda\dot{\vlab}$,
$\mathbf{P}_{c.m.}=M \mathbf{\dot{R}}_{c.m.}$, with
$M=2m+m'$, $m_\rho=m$, and $m_\lambda=\frac{3m m'}{2m+m'}$.
The wave function of an oscillator is give by
\begin{eqnarray}
\psi^{n_\sigma}_{l_\sigma m}(\sigma)=R_{n_\sigma
l_\sigma}(\sigma)Y_{l_\sigma m}(\sigma),
\end{eqnarray}
where $\sigma\equiv\rho,\lambda$.
In the wave functions, there are two oscillator parameters, i.e., the
potential strengths $\alpha_\rho$ and $\alpha_\lambda$.
The parameters $\alpha_\rho$ and $\alpha_\lambda$
satisfy the following relation~\cite{Zhong:2007gp}:
\begin{eqnarray} \label{rhol}
\alpha^2_\lambda=\sqrt{\frac{3m'}{2m+m'}}\alpha^2_\rho.
\end{eqnarray}

The details of the classifications of the heavy baryons in the constituent quark model
can be found in Refs.~\cite{Zhong:2007gp}.
Since  the bottom and charm quark masses are much larger than the light quark mass ($m_Q > m_q$),  the $\lambda$-mode excitations of singly heavy baryons should be
easily formed than the $\rho$-mode excitations~\cite{Yoshida:2015tia}.
Thus, in the present work, we only study the $\lambda$-mode excitations.
The mass spectra of the single
heavy baryons up to the $1D$-wave excitations predicted within various quark models are summarized in Tables~\ref{sp1} and~\ref{sp2}.

\begin{table*}[htp]
\begin{center}
\caption{\label{a1} Quark model parameters adopted in present work. }
\scalebox{1.0}{
\begin{tabular}{cccccccccccccccccccccccccc}\hline\hline
Parameter&~~$m_{u(d)}$~~&~~$m_{s}$~~&~~$m_{c}$~~&~~$m_{b}$~~&~~$\alpha_\rho$ (for $\Lambda_{c(b)}$ and $\Sigma_{c(b)}$)~~& ~~$\alpha_\rho$ (for $\Xi^{(\prime)}_{c(b)}$)~~&~~$\alpha_\rho$ (for $\Omega_{c(b)}$)~~&~~$f_\pi$~~&~~$f_K$\\
Value (MeV)&330&450&1480&5000&400&420&440& 132&160\\
\hline\hline
\end{tabular}}
\end{center}
\end{table*}

\subsection{Decays}

In this work,  strong decays of the $D$-wave singly heavy baryons with emission of one light pseudoscalar meson
are studied within ChQM~\cite{Manohar:1983md}.
This model has been successfully applied to study the strong decays of heavy-light mesons and charmed
and strange baryons~\cite{Zhong:2008kd,Zhong:2010vq,Zhong:2009sk,
Zhong:2007gp,Liu:2012sj,Xiao:2013xi,Nagahiro:2016nsx,Wang:2017hej,Xiao:2014ura,Xiao:2017udy}.
In this model, the light pseudoscalar mesons, i.e., $\pi$, $K$, and $\eta$,
are treated as fundamental states, which only couple with
the light constituent quarks of a hadron via the simple chiral Lagrangian~\cite{Manohar:1983md}
\begin{equation}\label{coup}
H_{m}=\sum_j
\frac{1}{f_m}\bar{\psi}_j\gamma^{j}_{\mu}\gamma^{j}_{5}\psi_j\partial^{\mu}\phi_m,
\end{equation}
where $\psi_j$ represents the $j$th quark field in the hadron,
$\phi_m$ is the pseudoscalar meson field, and $f_m$ is the
pseudoscalar meson decay constant.

Meanwhile, to treat the radiative decay of a hadron, we apply
the constituent quark model, which has been successfully applied to study
the radiative decays of $c\bar{c}$ and $b\bar{b}$ systems~\cite{Deng:2016stx,Deng:2016ktl}.
In this model, the quark-photon EM coupling at the tree level
is adopted as
\begin{eqnarray}\label{he}
H_e=-\sum_j
e_{j}\bar{\psi}_j\gamma^{j}_{\mu}A^{\mu}(\mathbf{k},\mathbf{r}_j)\psi_j,
\end{eqnarray}
where $A^{\mu}$ represents the photon field with 3-momenta $\mathbf{k}$. $e_j$ and $\mathbf{r}_j$
stand for the charge and coordinate of the constituent quark $\psi_j$, respectively.

To match the nonrelativistic harmonic
oscillator wave functions, in the calculations, one should adopt the nonrelativistic forms
for the quark-pseudoscalar and quark-photon EM couplings listed in Eqs.~(\ref{coup}) and (\ref{he}),
which have been given in the previous works~\cite{Lu:2017meb,Xiao:2017udy,Zhong:2008kd,Zhong:2010vq,Zhong:2009sk,
Zhong:2007gp,Liu:2012sj,Xiao:2013xi,Wang:2017hej,Xiao:2014ura,Deng:2016stx,Deng:2016ktl,Brodsky:1968ea,Li:1997gd,Zhao:1998fn,Zhao:2002id,Xiao:2015gra,Zhong:2011ti,Zhong:2011ht,Li:1994cy}.

For a strong decay process, the partial decay width can be calculated with~\cite{Zhong:2007gp}
\begin{equation}\label{dww}
\Gamma_m=\left(\frac{\delta}{f_m}\right)^2\frac{(E_f+M_f)|\mathbf{q}|}{4\pi
M_i(2J_i+1)} \sum_{J_{fz},J_{iz}}|\mathcal{M}_{J_{fz},J_{iz}}|^2 ,
\end{equation}
while for a radiative decay process, the partial decay width can be calculated with~\cite{Deng:2016stx,Deng:2016ktl}
\begin{equation}\label{dww}
\Gamma_\gamma=\frac{|\mathbf{k}|^2}{\pi}\frac{2}{2J_i+1}\frac{M_{f}}{M_{i}}\sum_{J_{fz},J_{iz}}|\mathcal{A}_{J_{fz},J_{iz}}|^2,
\end{equation}
where $\mathcal{M}_{J_{fz},J_{iz}}$ and $\mathcal{A}_{J_{fz},J_{iz}}$ correspond to
the strong and radiative transition amplitudes, respectively.
The quantum numbers $J_{iz}$ and $J_{fz}$ stand for the third components of the total
angular momenta of the initial and final heavy baryons,
respectively. $E_f$ and $M_f$ are the energy and mass of the final heavy baryon,
and $M_i$ is the mass of the initial heavy baryon.
$\delta$ as a global parameter accounts for the
strength of the quark-meson couplings. It has been determined in our previous study of the strong
decays of the charmed baryons and heavy-light mesons
\cite{Zhong:2007gp,Zhong:2008kd}. Here, we fix its value the same as
that in Refs.~\cite{Zhong:2008kd,Zhong:2007gp}, i.e., $\delta=0.557$.

In the calculation, we adopt the same quark model parameter set as that in Ref.~\cite{Wang:2017kfr},
which has been collected in Table~\ref{a1}.
The masses of the well-established hadrons used in the calculations are
adopted from the RPP~\cite{Olive:2016xmw}.

\begin{table}[htb]
\begin{center}
\caption{ \label{Lambdacb} Partial widths of strong decays for the $\lambda$-mode $D$-wave $\Lambda_c$ and  $\Lambda_b$ baryons.
The masses of the $D$-wave $\Lambda_c$ ($\Lambda_b$) states $|^2D_{\lambda\lambda} \frac{3}{2}^+ \rangle$ and
$|^2D_{\lambda\lambda} \frac{5}{2}^+ \rangle$ are taken as 2856 and 2881 (6190 and 6196) MeV, respectively.
The superscript (subscript) stands for the uncertainty of a prediction with a $+ 10\%$ ($-10\%$) uncertainty of the oscillator parameter $\alpha_{\rho}$.}
%\footnotesize
\begin{tabular}{c|ccccc}
\hline\hline
\multirow{2}{*}{~~~Decay mode~~~}       & $\underline{|\Lambda_c~^2D_{\lambda\lambda} \frac{3}{2}^+ \rangle(2856)}$ & $\underline{|\Lambda_c~^2D_{\lambda\lambda} \frac{5}{2}^+ \rangle(2881)}$\\
              & $\Gamma_i$ (MeV)      & $\Gamma_i$ (MeV)     \\ \hline
 $\Sigma_c\pi$                             &$4.57_{-1.20}^{+1.09}$  & $1.33_{+0.50}^{-0.35}$  \\
 $\Sigma_c^{\ast}\pi$                      &$0.95_{+0.09}^{-0.03}$  & $4.38_{-0.74}^{+0.67}$  \\
Sum                                        &$5.52_{-1.11}^{+1.06}$  &$5.71_{-0.24}^{+0.32}$  \\ \hline\hline
 \multirow{2}{*}{Decay mode}        & $\underline{|\Lambda_b~^2D_{\lambda\lambda} \frac{3}{2}^+ \rangle(6190)}$ & $\underline{|\Lambda_b~^2D_{\lambda\lambda} \frac{5}{2}^+ \rangle(6196)}$\\
             & $\Gamma_i$ (MeV)       & $\Gamma_i$ (MeV)     \\ \hline
 $\Sigma_b\pi$                                       &$6.32_{-1.91}^{+1.78}$  &$1.83_{+0.67}^{-0.47}$   \\
 $\Sigma_b^*\pi$                                     &$2.76_{+0.41}^{-0.22}$  &$7.42_{-1.43}^{+1.38}$   \\
Sum                                                  &$9.08_{-1.50}^{+1.56}$    &$9.25_{-0.76}^{+0.91}$  \\
\hline
\hline
\end{tabular}
\end{center}
\end{table}

\begin{figure}[htbp]
\begin{center}
\centering  \epsfxsize=6.8 cm \epsfbox{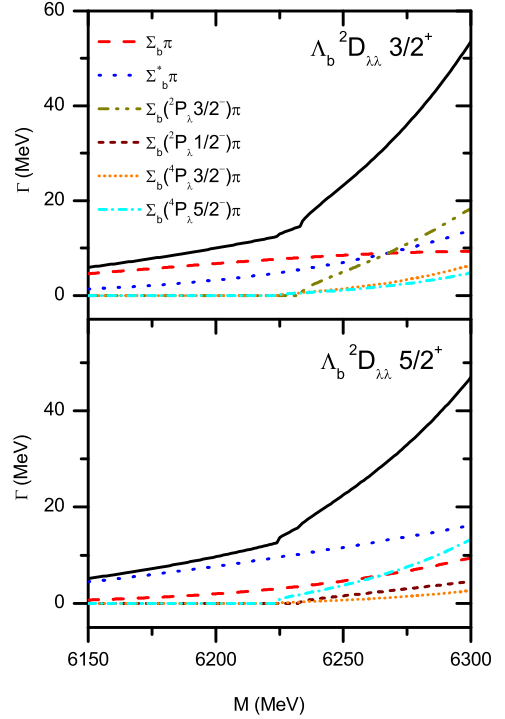}
\vspace{-0.8 cm} \caption{ Strong decay partial widths of the main decay modes for the $\lambda$-mode $1D$-wave excited $\Lambda_b$ states $|\Lambda_b~D_{\lambda\lambda}3/2^+\rangle$ and $|\Lambda_b~D_{\lambda\lambda}5/2^+\rangle$ as functions of their mass. The solid curves stand for the total widths. The masses of the $P$-wave heavy baryons in final states are adopted from the quark model predictions in Ref.~\cite{Ebert:2011kk} (see Table~\ref{sp2}).   } \label{DLB}
\end{center}
\end{figure}

\begin{table}[htp]
\begin{center}
\caption{\label{RadL}  Partial widths of radiative decays for the $\lambda$-mode $D$-wave $\Lambda_c$ and $\Lambda_b$ baryons.
The masses of the $D$-wave $\Lambda_c$ ($\Lambda_b$) states $|^2D_{\lambda\lambda} \frac{3}{2}^+ \rangle$ and $|^2D_{\lambda\lambda} \frac{5}{2}^+ \rangle$ are taken as 2856 and 2881 (6190 and 6196) MeV, respectively. $M_f$ stands for the masses of $P$-wave heavy baryons (MeV) in the final states, which are adopted from the RPP~\cite{Olive:2016xmw} and Ref.~\cite{Ebert:2011kk}. The superscript (subscript) stands for the uncertainty of a prediction with a $+ 10\%$ ($-10\%$) uncertainty of the oscillator parameter $\alpha_{\rho}$. }
\begin{tabular}{cc|cccccccccc}\hline\hline
%                      &       &&         &     &  \underline{$1P$ $\Sigma_c$ states }      &   &&      &      \\
\multirow{2}{*}{~~~Decay mode~~~} &\multirow{2}{*}{$M_f$}&$\underline{|\Lambda_c~^2D_{\lambda\lambda} \frac{3}{2}^+ \rangle(2856)}$
&$\underline{|\Lambda_c~^2D_{\lambda\lambda} \frac{5}{2}^+ \rangle(2881)}$
 \\
&                               &$\Gamma_i$ (keV)         &$\Gamma_i$ (keV)
                                                 \\ \hline
$|\Lambda_{c}^{+}~ ^2P_{\lambda}\frac{1}{2}^- \rangle\gamma$    &2592         &0.01                       &0.13$^{-0.03}_{+0.03}$\\
$|\Lambda_{c}^{+}~ ^2P_{\lambda}\frac{3}{2}^- \rangle\gamma$    &2628         &0.07                       &0.26$^{-0.02}_{+0.03}$\\
$|\Sigma_{c}^{+}~ ^2P_{\lambda}\frac{1}{2}^- \rangle\gamma$    &2713          &0.23$^{-0.04}_{+0.05}$     &0.80$^{-0.12}_{+0.15}$\\
$|\Sigma_{c}^{+}~ ^2P_{\lambda}\frac{3}{2}^- \rangle\gamma$    &2798          &0.01                       &0.05\\
$|\Sigma_{c}^{+}~ ^4P_{\lambda}\frac{1}{2}^- \rangle\gamma$    &2799          &$<0.01$                    &$<0.01$\\
$|\Sigma_{c}^{+}~ ^4P_{\lambda}\frac{3}{2}^- \rangle\gamma$    &2773          &0.08                       &0.13$^{-0.02}_{+0.03}$\\
$|\Sigma_{c}^{+}~ ^4P_{\lambda}\frac{5}{2}^- \rangle\gamma$    &2789          &$<0.01$                    &0.19$^{-0.03}_{+0.04}$\\

\hline\hline
%                      &  &&              &     &  \underline{$1P$ $\Sigma_b$ states }      &   &&      &       \\
\multirow{2}{*}{~~~Decay mode~~~} &\multirow{2}{*}{$M_f$}&$\underline{|\Lambda_b~^2D_{\lambda\lambda} \frac{3}{2}^+ \rangle(6190)}$
&$\underline{|\Lambda_b~^2D_{\lambda\lambda} \frac{5}{2}^+ \rangle(6196)}$
 \\
&                                &$\Gamma_i$ (keV)        &$\Gamma_i$ (keV)
                                                 \\ \hline
$|\Lambda_{b}^{0}~ ^2P_{\lambda}\frac{1}{2}^- \rangle\gamma$    &5912         &19.7$^{-3.0}_{+3.7}$       &1.67$^{-0.34}_{+0.49}$\\
$|\Lambda_{b}^{0}~ ^2P_{\lambda}\frac{3}{2}^- \rangle\gamma$    &5920         &6.26$^{-1.08}_{+1.45}$     &24.1$^{-3.7}_{+4.7}$\\
$|\Sigma_{b}^{0}~ ^2P_{\lambda}\frac{1}{2}^- \rangle\gamma$    &6101          &0.04                       &0.09\\
$|\Sigma_{b}^{0}~ ^2P_{\lambda}\frac{3}{2}^- \rangle\gamma$    &6096          &0.17$^{-0.03}_{+0.04}$     &0.18$^{-0.03}_{+0.03}$\\
$|\Sigma_{b}^{0}~ ^4P_{\lambda}\frac{1}{2}^- \rangle\gamma$    &6095          &0.08                       &0.01\\
$|\Sigma_{b}^{0}~ ^4P_{\lambda}\frac{3}{2}^- \rangle\gamma$    &6087          &0.34$^{-0.05}_{+0.07}$     &0.21$^{-0.04}_{+0.04}$\\
$|\Sigma_{b}^{0}~ ^4P_{\lambda}\frac{5}{2}^- \rangle\gamma$    &6084          &0.11$^{-0.02}_{+0.02}$     &0.75$^{-0.12}_{+0.16}$\\
\hline\hline
\end{tabular}
\end{center}
\end{table}

\section{Results for singly heavy baryons of $\bar{3}_F$}\label{RD}

\subsection{$\Lambda_c$ states }

In the $\Lambda_c$ family, there are two $\lambda$-mode $1D$-wave
excitations $|\Lambda_c~^2D_{\lambda\lambda} \frac{3}{2}^+ \rangle$
and $|\Lambda_c~^2D_{\lambda\lambda} \frac{5}{2}^+ \rangle$ according to the quark model classification.
The masses for the $\lambda$-mode $1D$-wave $\Lambda_c$
excitations are predicted to be $\sim 2.9$ GeV in various models (see Table~\ref{sp1}).
The resonances $\Lambda_c(2860)$ with $J^P=3/2^+$ and $\Lambda_c(2880)$ with $J^P=5/2^+$
listed in RPP~\cite{Olive:2016xmw} most likely correspond to the
two $\lambda$-mode $1D$-wave $\Lambda_c$ excitations $|\Lambda_c~^2D_{\lambda\lambda} \frac{3}{2}^+ \rangle$
and $|\Lambda_c~^2D_{\lambda\lambda} \frac{5}{2}^+ \rangle$, respectively.

\subsubsection{$J^P=5/2^+$ state and $\Lambda_c(2880)$ }

The $\Lambda_c(2880)$ state was first observed in $\Lambda^+_c \pi^+\pi^-$ by CLEO
\cite{Artuso:2000xy}. It was confirmed in $\Sigma_c\pi$ and $\Sigma_c(2520)\pi$ channels by
Belle~\cite{Abe:2006rz} and in the $D^0p$ channel by $BABAR$
\cite{Aubert:2006sp} and LHCb~\cite{Aaij:2017vbw}.
It has a narrow decay width of $\Gamma\simeq 5.6$ MeV~\cite{Olive:2016xmw}.
The spin-parity numbers were determined to be $J^P=5/2^+$ by Belle~\cite{Abe:2006rz} and were
confirmed by LHCb ~\cite{Aaij:2017vbw} recently.

The $\Lambda_c(2880)$ state may be classified as the $1D$-wave charmed baryons
~\cite{Cheng:2006dk,Chen:2007xf,Capstick:1986bm,Ebert:2007nw,Garcilazo:2007eh}.
If $\Lambda_c(2880)$ is a conventional $\lambda$-mode $1D$-wave excitation, it should
be assigned to $|\Lambda_c~^2D_{\lambda\lambda} \frac{5}{2}^+ \rangle$.
With this assignment, the width of $\Lambda_c(2880)$ can be
reasonably understood by ChQM~\cite{Zhong:2007gp}.
It is found that the main decay channel of $\Lambda_c(2880)$ should be $\Sigma_c(2520)\pi$ (see Table~\ref{Lambdacb}).
The partial width ratio, $\mathcal{R}=\frac{\Gamma[\Sigma_c(2520)\pi]}{\Gamma[\Sigma_c(2455)\pi]}\simeq 3.3$, predicted by us is too large to compare with the measured value $\mathcal{R}\simeq 0.225$ at Belle~\cite{Abe:2006rz}.
The recent $^3P_0$ analysis of the strong decays of $\Lambda_c(2880)$ in Ref.~\cite{Chen:2017aqm}
is consistent with our predictions.

It should be mentioned that the measured ratio $\mathcal{R}=\frac{\Gamma[\Sigma_c(2520)\pi]}{\Gamma[\Sigma_c(2455)\pi]}\simeq 0.225$
of $\Lambda_c(2880)$ may be strongly affected by its nearby state $\Lambda_c(2860)\frac{3}{2}^+$
newly observed in the $D^0p$ channel at LHCb~\cite{Aaij:2017vbw}.
Thus, the measured ratio from Belle~\cite{Abe:2006rz} should not be
a genuine ratio for $\Lambda_c(2880)$. This situation is very similar to that of $D_{sJ}(2860)$
before two largely overlapping states $D_{s1}(2860)$ and $D_{s3}(2860)$ were found by LHCb~\cite{Aaij:2014xza}.
Considering $D_{sJ}(2860)$ as the $J^P=3^-$ state $1^3D_3$,
the measured partial width ratio $\mathcal{R}=\frac{\Gamma[D^*K]}{\Gamma[DK]}\simeq 1.1$ cannot be explained
by ChQM~\cite{Zhong:2008kd,Zhong:2009sk} and many other approaches~\cite{Zhang:2006yj,Colangelo:2006rq,DeFazio:2008nr}.
Then, people proposed an alternative solution that there might exist two largely overlapping
resonances at about 2.86 GeV~\cite{vanBeveren:2009jq,Zhong:2009sk}, which was confirmed by LHCb recently~\cite{Aaij:2014xza}.
%Meanwhile, some people also proposed that $\Lambda_c(2880)$ may not be a conventional three-quark state, for example,
%in Ref.~\cite{Lutz:2003jw} it was suggested to be a dynamically generated state.

To better understand the properties of $\Lambda_c(2880)$, considering it as the $1D$-wave
state $|\Lambda_c~^2D_{\lambda\lambda} \frac{5}{2}^+ \rangle$, we further study
its radiative decays into the $1P$-wave charmed baryon states. Our results are listed in Table~\ref{RadL}.
It is found that most of the partial radiative widths of $\Lambda_c(2880)$ into the $1P$-wave states
are $\mathcal{O}(100)$ eV. Combining these partial widths with the total decay width of $\Lambda_c(2880)$,
we find the branching fractions for the main radiative decay channels are $\mathcal{O}(10^{-5})$.
The small decay rates indicate that the radiative decays of $\Lambda_c(2880)$ into the $1P$-wave states might
be hard to observe in experiments.

\subsubsection{$J^P=3/2^+$ state and $\Lambda_c(2860)$ }

Recently, besides the confirmation of $\Lambda_c(2880)$ in the $D^0p$ channel, the LHCb Collaboration observed a new charmed baryon state,
$\Lambda_c(2860)$, with a broad width of $\Gamma\simeq 67.6^{+10.1}_{-8.1}$ MeV
in the same channel~\cite{Aaij:2017vbw}.
The determined spin-parity quantum numbers are $J^P=3/2^+$~\cite{Aaij:2017vbw}.
Both the mass and decay modes of $\Lambda_c(2860)$ indicate that
it might be assigned to the $\lambda$-mode excited
$1D$-wave charmed baryon state $|\Lambda_c~^2D_{\lambda\lambda} \frac{3}{2}^+ \rangle$~\cite{Ebert:2011kk,Chen:2016iyi,Chen:2017aqm}.
Considering $\Lambda_c(2860)$ as the $|\Lambda_c~^2D_{\lambda\lambda} \frac{3}{2}^+ \rangle$ state,
we predict its partial widths into the $\Sigma_c(2455)\pi$ and $\Sigma_c(2520)\pi$ channels,
\begin{eqnarray}\label{1DL}
\Gamma[\Sigma_c(2455)\pi]\simeq 4.6 \ \mathrm{MeV}, \ \ \
\Gamma[\Sigma_c(2520)\pi]\simeq 1.0 \ \mathrm{MeV},
\end{eqnarray}
which roughly agree with the predictions in Ref.~\cite{Chen:2017aqm}.
Combining these predicted partial widths with the measured width of $\Lambda_c(2860)$, we further estimate
that the branching fractions of the $\Sigma_c(2455)\pi$ and
$\Sigma_c(2520)\pi$ channels can reach up to $7\%$ and $2\%$, respectively.
The relatively large branching fractions indicate that $\Lambda_c(2860)$
might be observed in the $\Sigma_c(2455)\pi$ and $\Sigma_c(2520)\pi$
channels as well.

Considering $\Lambda_c(2860)$ as the $1D$-wave
state $|\Lambda_c~^2D_{\lambda\lambda} \frac{3}{2}^+ \rangle$, we also study
its radiative decays into the $1P$-wave states. Our results are listed in Table~\ref{RadL} as well.
It is found that the radiative decay rates into the $1P$-wave states are small.
Their partial decay widths are $\mathcal{O}(10)$ eV. Combining these partial widths with the total decay width of $\Lambda_c(2860)$,
we find the branching fractions, $\mathcal{B}[\Lambda_c(2860)\to 1P \gamma]$, are $\mathcal{O}(10^{-6})$,
which indicates the radiative decays of $\Lambda_c(2860)$ into the $1P$-wave states might
be hard to observe in experiments.

\subsection{$\Lambda_b$ states }

In the $\Lambda_b$ family, there are two $\lambda$-mode $1D$-wave
excitations $|\Lambda_b~^2D_{\lambda\lambda} \frac{3}{2}^+ \rangle$
and $|\Lambda_b~^2D_{\lambda\lambda} \frac{5}{2}^+ \rangle$ according to the quark model classification.
The masses for the $\lambda$-mode
$1D$-wave $\Lambda_b$ excitations are predicted to be $\sim 6.2 $ GeV in various models (see Table~\ref{sp1}).
In the possible mass region of the $1D$-wave $\Lambda_b$ excitations, we study their strong decay properties,
which have been shown in Fig.~\ref{DLB}. To be more specific, taking the masses of the $1D$-wave states
as predicted in the relativistic quark-diquark picture~\cite{Ebert:2011kk} we further present
the results in Table~\ref{Lambdacb}.

\subsubsection{$J^P=3/2^+$ state}

From Fig.~\ref{DLB}, it is found that if the mass of $|\Lambda_b~^2D_{\lambda\lambda} \frac{3}{2}^+ \rangle$ is
$\sim 6200 $ MeV as predicted in theory~\cite{Ebert:2011kk,Yoshida:2015tia}
it should be a narrow state with a width of $\Gamma_{\mathrm{total}}\simeq 10$ MeV.
The decays may be saturated by the $\Sigma_b\pi$ and $\Sigma_b^*\pi$ channels,
and the partial width ratio between them is predicted to be
\begin{eqnarray}\label{aa}
\frac{\Gamma(\Sigma_b\pi)}{\Gamma(\Sigma_b^*\pi)}\simeq 2,
\end{eqnarray}
which is less sensitive to the mass of $|\Lambda_b~^2D_{\lambda\lambda} \frac{3}{2}^+ \rangle$.
On the other hand, if the mass of $|\Lambda_b~^2D_{\lambda\lambda} \frac{3}{2}^+ \rangle$
is larger than 6240 MeV, more strong decay channels may open. The
$|\Lambda_b~^2D_{\lambda\lambda} \frac{3}{2}^+ \rangle$ state
may have a large decay rate into the $| \Sigma_b~^2P_\lambda \frac{3}{2}^- \rangle \pi$ channel
as well (see Fig.~\ref{DLB}).
To establish the missing $D$-wave state $|\Lambda_b~^2D_{\lambda\lambda} \frac{3}{2}^+ \rangle$,
the decay channel $\Sigma_b\pi$ might be the ideal channel to be observed in future experiments.

We further estimate the radiative decays of $|\Lambda_b~^2D_{\lambda\lambda} \frac{3}{2}^+ \rangle$ into the $1P$-wave states. Our results are listed in Table~\ref{RadL}.
It is found that $|\Lambda_b~^2D_{\lambda\lambda} \frac{3}{2}^+ \rangle$ has a relatively large decay rate into
$\Lambda_b(5912)\frac{1}{2}^- \gamma$, and the partial width of $\Gamma[|\Lambda_b~^2D_{\lambda\lambda} \frac{3}{2}^+ \rangle\to \Lambda_b(5912)\frac{1}{2}^- \gamma]$ can reach up to $\sim 20$ keV. Combining it with our predicted total width, we find the branching fraction of
$\mathcal{B}[|\Lambda_b~^2D_{\lambda\lambda} \frac{3}{2}^+ \rangle\to \Lambda_b(5912) \gamma]$ is $\mathcal{O}(10^{-3})$, which indicates that
$|\Lambda_b~^2D_{\lambda\lambda} \frac{3}{2}^+ \rangle$ has the possibility of being observed in the $\Lambda_b(5912)\frac{1}{2}^- \gamma$ channel.

\subsubsection{$J^P=5/2^+$ state}

If the mass of $|\Lambda_b~^2D_{\lambda\lambda} \frac{5}{2}^+ \rangle$ is less
than 6200 MeV as predicted in various quark models~\cite{Ebert:2011kk,Yoshida:2015tia},
the decays of $|\Lambda_b~^2D_{\lambda\lambda} \frac{5}{2}^+ \rangle$ may be saturated by
the $\Sigma_b\pi$ and $\Sigma_b^*\pi$ channels. The $|\Lambda_b~^2D_{\lambda\lambda} \frac{5}{2}^+ \rangle$ state
has a narrow width of $\Gamma_{\mathrm{total}}\simeq 10$ MeV, which is comparable with that
of $|\Lambda_b~^2D_{\lambda\lambda} \frac{3}{2}^+ \rangle$ (see Fig.~\ref{DLB}).
However, the strong decays of $|\Lambda_b~^2D_{\lambda\lambda} \frac{5}{2}^+ \rangle$
are governed by the $\Sigma_b^*\pi$ channel.
The partial width ratio between $\Sigma_b\pi$ and $\Sigma_b^*\pi$
is predicted to be
\begin{eqnarray}\label{cc}
\frac{\Gamma(\Sigma_b\pi)}{\Gamma(\Sigma_b^*\pi)}\simeq 0.25,
\end{eqnarray}
which shows few sensibilities to the mass of $|\Lambda_b~^2D_{\lambda\lambda} \frac{5}{2}^+ \rangle$.
On the other hand, if the mass of $|\Lambda_b~^2D_{\lambda\lambda} \frac{5}{2}^+ \rangle$ is larger
than 6240 MeV, more strong decay channels may open.
The $| \Sigma_b~^4P_\lambda \frac{5}{2}^- \rangle \pi$ decay mode may
play an important role in the decays as well. To establish the missing $D$-wave state $|\Lambda_b~^2D_{\lambda\lambda} \frac{5}{2}^+ \rangle$, the decay channel $\Sigma_b^*\pi$ should be the ideal channel to be observed in future experiments.

To know more properties of $|\Lambda_b~^2D_{\lambda\lambda} \frac{5}{2}^+ \rangle$, we further estimate its
radiative decays into the $1P$-wave states. Our results are listed in Table~\ref{RadL}.
It is found that the radiative process $|\Lambda_b~^2D_{\lambda\lambda} \frac{5}{2}^+ \rangle\to \Lambda_b(5920)\frac{3}{2}^- \gamma$
has a relatively large partial width $\sim 24$ keV. Combining it with our predicted total width, we find the branching fraction of
$\mathcal{B}[|\Lambda_b~^2D_{\lambda\lambda} \frac{5}{2}^+ \rangle\to \Lambda_b(5920)\frac{3}{2}^- \gamma]$ is $\mathcal{O}(10^{-3})$, which indicates that $|\Lambda_b~^2D_{\lambda\lambda} \frac{5}{2}^+ \rangle$ has the possibility of being observed in the $\Lambda_b(5920)\frac{3}{2}^- \gamma$ channel.

\begin{figure}[htbp]
\begin{center}
\centering  \epsfxsize=7.2 cm \epsfbox{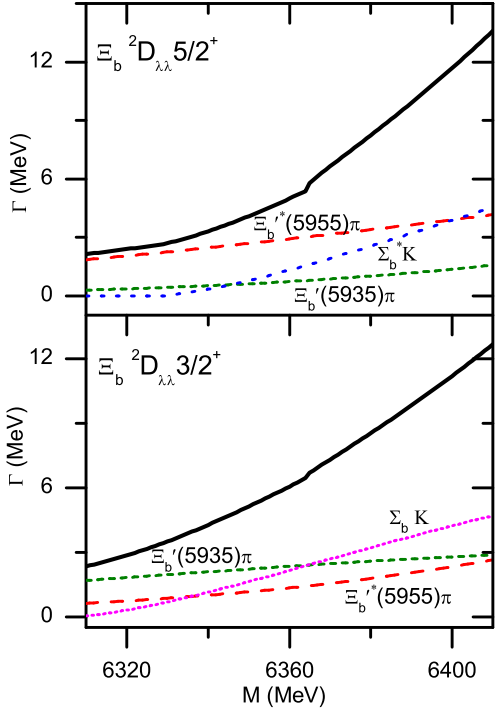}
\vspace{-0.8 cm} \caption{  Strong decay partial widths of the main decay modes for the $\lambda$-mode $1D$-wave excited $\Xi_b$ states $|\Xi_b~D_{\lambda\lambda}3/2^+\rangle$ and $|\Xi_b~D_{\lambda\lambda}5/2^+\rangle$ as functions of their mass. The solid curves stand for the sum of the strong decay partial widths.  } \label{Xib}
\end{center}
\end{figure}

\begin{table}[htb]
\begin{center}
\caption{ \label{Xicb} Partial widths of strong decays for the $\lambda$-mode $D$-wave $\Xi_c$ and  $\Xi_b$ baryons.
The masses of the $D$-wave $\Xi_c$ ($\Xi_b$) states $|^2D_{\lambda\lambda} \frac{3}{2}^+ \rangle$ and
$|^2D_{\lambda\lambda} \frac{5}{2}^+ \rangle$ are taken as 3055 and 3080 (6373 and 6366) MeV, respectively.
$M_f$ stands for the masses of $P$-wave heavy baryons (MeV) in the final states, which are adopted from
the RPP~\cite{Olive:2016xmw} and Ref.~\cite{Ebert:2011kk}. The superscript (subscript) stands for the uncertainty of a prediction with a $+ 10\%$
($-10\%$) uncertainty of the oscillator parameter $\alpha_{\rho}$.}
%\footnotesize
\begin{tabular}{cc|cccc}
\hline\hline
\multirow{3}{*}{~~~Decay mode~~~}       &\multirow{3}{*}{$M_f$}& $\underline{|\Xi_c~^2D_{\lambda\lambda} \frac{3}{2}^+ \rangle(3055)}$ & $\underline{|\Xi_c~^2D_{\lambda\lambda} \frac{5}{2}^+ \rangle(3080)}$\\
                &  & $\Gamma_i$ (MeV)      & $\Gamma_i$ (MeV)     \\ \hline
 $\Xi'_c\pi$                                         & 2575  &1.93$^{+0.57}_{-0.61}$    &0.75$^{-0.18}_{+0.27}$ \\
 $\Xi^{\prime*}_c\pi$                                         & 2645  &0.60$^{-0.04}_{+0.07}$    &2.08$^{+0.38}_{-0.40}$  \\
 $\Sigma_cK$                                         & 2455  &2.49$^{+0.38}_{-0.29}$    &0.22$^{<-0.01}_{+0.10}$  \\
 $\Sigma_c^{\ast}K$                                  & 2520  &0.14$^{<-0.01}_{<+0.01}$  &1.68$^{+0.11}_{-0.14}$  \\
 $|\Xi'_c~ 1^4P_{\lambda} \frac{1}{2}^-\rangle \pi$  & 2854  &$<$0.01                   &1.42$^{+0.39}_{-0.34}$  \\
 $|\Xi'_c~ 1^4P_{\lambda} \frac{3}{2}^-\rangle \pi$  & 2912  &0.04                      &0.20$^{+0.04}_{-0.03}$  \\
 $|\Xi'_c~ 1^4P_{\lambda} \frac{5}{2}^-\rangle \pi$  & 2929  & ...                      &0.54$^{+0.10}_{-0.10}$  \\
 Sum                                                 &       &5.20$^{+0.91}_{-0.83}$    &6.90$^{+0.84}_{-0.64}$  \\ \hline\hline
 \multirow{2}{*}{Decay mode}       &\multirow{2}{*}{$M_f$}& $\underline{|\Xi_b~^2D_{\lambda\lambda} \frac{3}{2}^+ \rangle(6366)}$ & $\underline{|\Xi_b~^2D_{\lambda\lambda} \frac{5}{2}^+ \rangle(6373)}$\\
             &  & $\Gamma_i$ (MeV)      & $\Gamma_i$ (MeV)     \\ \hline
 $\Xi'_b \pi$                                        &5935  &2.41$^{+0.79}_{-0.81}$  &0.90$^{-0.22}_{+0.30}$   \\
 $\Xi'^{*}_b\pi$                                     &5955  &1.45$^{-0.13}_{+0.25}$  &3.22$^{+0.68}_{-0.68}$   \\
 $\Sigma_bK$                                         &5811  &2.47$^{+0.20}_{-0.27}$  &0.07   \\
 $\Sigma_b^*K$                                       &5832  &0.30$^{+0.04}_{-0.01}$  &2.10$^{+0.12}_{-0.16}$   \\
 $|\Xi'_b~ 1^4P_{\lambda} \frac{1}{2}^-\rangle \pi$  &6227  &$<$0.01                 &0.28$^{+0.06}_{-0.06}$   \\
  $|\Xi'_b~ 1^4P_{\lambda} \frac{3}{2}^-\rangle \pi$ &6224  &0.11$^{+0.02}_{-0.02}$  &0.10$^{+0.02}_{-0.02}$   \\
 $|\Xi'_b~ 1^4P_{\lambda} \frac{5}{2}^-\rangle \pi$  &6226  &0.06                    &0.45$^{+0.09}_{-0.09}$   \\
 Sum                                                 &      &6.8$^{+0.92}_{-0.02}$   &7.11$^{+0.75}_{-0.71}$  \\
\hline
\hline
\end{tabular}
\end{center}
\end{table}

\begin{table}[htp]
\begin{center}
\caption{\label{RadX} Partial widths of radiative decays for the $\lambda$-mode $D$-wave $\Xi_c$ and $\Xi_b$ baryons.
The masses of the $D$-wave $\Xi_c$ ($\Xi_b$) states $|^2D_{\lambda\lambda} \frac{3}{2}^+ \rangle$ and $|^2D_{\lambda\lambda} \frac{5}{2}^+ \rangle$ are taken as 3055 and 3080 (6366 and 6373) MeV, respectively. $M_f$ stands for the masses of $P$-wave heavy baryons (MeV) in the final states, which are adopted from the RPP~\cite{Olive:2016xmw} and Ref.~\cite{Ebert:2011kk}. The superscript (subscript) stands for the uncertainty of a prediction with a $+ 10\%$ ($-10\%$) uncertainty of the oscillator parameter $\alpha_{\rho}$. }
\begin{tabular}{cc|cccccccccc}\hline\hline
%                      &       &&         &     &  \underline{$1P$ $\Sigma_c$ states }      &   &&      &      \\
\multirow{2}{*}{Decay mode}       &\multirow{2}{*}{$M_f$} &$\underline{|\Xi_c~^2D_{\lambda\lambda} \frac{3}{2}^+ \rangle(3055)}$
&$\underline{|\Xi_c~^2D_{\lambda\lambda} \frac{5}{2}^+ \rangle(3080)}$
 \\
&                                  &$\Gamma_i$ (keV)    &$\Gamma_i$ (keV)
                                                 \\ \hline
$|\Xi_{c}^{+}~ ^2P_{\lambda}\frac{1}{2}^- \rangle\gamma$    &2792           &1.09$^{-0.22}_{+0.33}$      &0.36$^{-0.06}_{+0.08}$\\
$|\Xi_{c}^{0}~ ^2P_{\lambda}\frac{1}{2}^- \rangle\gamma$    &               &79.0$^{-12.8}_{+16.7}$      &7.62$^{-1.41}_{+1.97}$\\

$|\Xi_{c}^{+}~ ^2P_{\lambda}\frac{3}{2}^- \rangle\gamma$    &2815           &0.57$^{-0.11}_{+0.15}$      &0.28$^{-0.07}_{+0.11}$ \\
$|\Xi_{c}^{0}~ ^2P_{\lambda}\frac{3}{2}^- \rangle\gamma$    &               &21.1$^{-3.6}_{+4.9}$        &85.1$^{-13.9}_{+18.2}$\\

$|\Xi_{c}^{'+}~ ^2P_{\lambda}\frac{1}{2}^- \rangle\gamma$    &2936          &0.06                        &0.26$^{-0.04}_{+0.05}$\\
$|\Xi_{c}^{'0}~ ^2P_{\lambda}\frac{1}{2}^- \rangle\gamma$    &              &0.0                         &0.0\\
$|\Xi_{c}^{'+}~ ^2P_{\lambda}\frac{3}{2}^- \rangle\gamma$    &2935          &0.22$^{-0.04}_{+0.05}$      &0.42$^{-0.07}_{+0.08}$\\
$|\Xi_{c}^{'0}~ ^2P_{\lambda}\frac{3}{2}^- \rangle\gamma$    &              &0.0                         &0.0\\

$|\Xi_{c}^{'+}~ ^4P_{\lambda}\frac{1}{2}^- \rangle\gamma$    &2854          &0.97$^{-0.13}_{+0.16}$      &0.21$^{-0.02}_{+0.03}$\\
$|\Xi_{c}^{'0}~ ^4P_{\lambda}\frac{1}{2}^- \rangle\gamma$    &              &0.0                         &0.0\\

$|\Xi_{c}^{'+}~ ^4P_{\lambda}\frac{3}{2}^- \rangle\gamma$    &2912          &0.66$^{-0.11}_{+0.13}$      &0.63$^{-0.10}_{+0.12}$\\
$|\Xi_{c}^{'0}~ ^4P_{\lambda}\frac{3}{2}^- \rangle\gamma$    &              &0.0                         &0.0\\
$|\Xi_{c}^{'+}~ ^4P_{\lambda}\frac{5}{2}^- \rangle\gamma$    &2929          &0.09                        &1.24$^{-0.20}_{+0.27}$\\
$|\Xi_{c}^{'0}~ ^4P_{\lambda}\frac{5}{2}^- \rangle\gamma$    &              &0.0                         &0.0\\
\hline\hline
%                      &  &&              &     &  \underline{$1P$ $\Sigma_b$ states }      &   &&      &       \\
\multirow{2}{*}{Decay mode}       &\multirow{2}{*}{$M_f$}&$\underline{|\Xi_b~^2D_{\lambda\lambda} \frac{3}{2}^+ \rangle(6366)}$
&$\underline{|\Xi_b~^2D_{\lambda\lambda} \frac{5}{2}^+ \rangle(6373)}$
 \\
&                                  &$\Gamma_i$ (keV)   &$\Gamma_i$ (keV)
                                                 \\ \hline
$|\Xi_{b}^{0}~ ^2P_{\lambda}\frac{1}{2}^- \rangle\gamma$    &6120         &3.62$^{-0.51}_{+0.64}$        &0.33$^{-0.07}_{+0.10}$\\
$|\Xi_{b}^{-}~ ^2P_{\lambda}\frac{1}{2}^- \rangle\gamma$    &             &32.0$^{-4.9}_{+6.2}$          &2.58$^{-0.53}_{+0.76}$\\

$|\Xi_{b}^{0}~ ^2P_{\lambda}\frac{3}{2}^- \rangle\gamma$    &6130         &1.09$^{-0.19}_{+0.26}$        &4.78$^{-0.71}_{+0.88}$\\
$|\Xi_{b}^{-}~ ^2P_{\lambda}\frac{3}{2}^- \rangle\gamma$    &             &9.40$^{-1.63}_{+2.19}$        &39.5$^{-6.2}_{+7.9}$\\

$|\Xi_{b}^{'0}~ ^2P_{\lambda}\frac{1}{2}^- \rangle\gamma$    &6233          &0.17$^{-0.02}_{+0.04}$      &0.37$^{-0.05}_{+0.07}$\\
$|\Xi_{b}^{'-}~ ^2P_{\lambda}\frac{1}{2}^- \rangle\gamma$    &              &0.0       &0.0\\
$|\Xi_{b}^{'0}~ ^2P_{\lambda}\frac{3}{2}^- \rangle\gamma$    &6234          &0.57$^{-0.09}_{+0.13}$      &0.56$^{-0.09}_{+0.12}$\\
$|\Xi_{b}^{'-}~ ^2P_{\lambda}\frac{3}{2}^- \rangle\gamma$    &              &0.0       &0.0\\

$|\Xi_{b}^{'0}~ ^4P_{\lambda}\frac{1}{2}^- \rangle\gamma$    &6227          &0.31$^{-0.05}_{+0.05}$      &0.05\\
$|\Xi_{b}^{'-}~ ^4P_{\lambda}\frac{1}{2}^- \rangle\gamma$    &              &0.0       &0.0\\

$|\Xi_{b}^{'0}~ ^4P_{\lambda}\frac{3}{2}^- \rangle\gamma$    &6224          &1.04$^{-0.16}_{+0.22}$      &0.59$^{-0.08}_{+0.11}$\\
$|\Xi_{b}^{'-}~ ^4P_{\lambda}\frac{3}{2}^- \rangle\gamma$    &              &0.0       &0.0\\
$|\Xi_{b}^{'0}~ ^4P_{\lambda}\frac{5}{2}^- \rangle\gamma$    &6226          &0.27$^{-0.04}_{+0.06}$      &1.82$^{-0.29}_{+0.38}$\\
$|\Xi_{b}^{'-}~ ^4P_{\lambda}\frac{5}{2}^- \rangle\gamma$    &              &0.0                         &0.0\\
\hline\hline
\end{tabular}
\end{center}
\end{table}

\subsection{$\Xi_c$ states}

In the $\Xi_c$ family, there are two
$\lambda$-mode $1D$-wave excitations $|\Xi_c~^2D_{\lambda\lambda} \frac{3}{2}^+ \rangle$
and $|\Xi_c~^2D_{\lambda\lambda} \frac{5}{2}^+ \rangle$.
The typical masses of the $\lambda$-mode
$1D$-wave $\Xi_c$ excitations are $\sim 3.05$ GeV
within various quark model predictions (see Table~\ref{sp1}). From the point of view of mass,
the charmed-strange baryons $\Xi_c(3055)^+$ and $\Xi_c(3080)^+$ observed
in the $\Lambda_c\bar{K}\pi$ final state by the Belle~\cite{Chistov:2006zj}
and $BABAR$~\cite{Aubert:2007dt} Collaborations are good candidates of the $\lambda$-mode $1D$-wave states.
Recently, a new decay mode $D^+\Lambda$ for both $\Xi_c(3055)^+$ and $\Xi_c(3080)^+$
was observed by the Belle Collaboration~\cite{Kato:2016hca}. They first reported the following
partial width ratios:
\begin{eqnarray}\label{3055}
\frac{\Gamma[\Xi_c(3055)^+\rightarrow \Lambda D^+]}{\Gamma[\Xi_c(3055)^+\rightarrow \Sigma_c(2455)^{++} K^-]}=5.09\pm1.01\pm0.76,
\end{eqnarray}
\begin{eqnarray}\label{11}
\frac{\Gamma[\Xi_c(3080)^+\rightarrow \Lambda D^+]}{\Gamma[\Xi_c(3080)^+\rightarrow \Sigma_c(2455)^{++} K^-]}=1.29\pm0.30\pm0.15,
\end{eqnarray}
and
\begin{eqnarray}\label{22}
\frac{\Gamma[\Xi_c(3080)^+\rightarrow \Sigma_c(2520)^{++} K^-]}{\Gamma[\Xi_c(3080)^+\rightarrow \Sigma_c(2455)^{++} K^-]}=1.07\pm0.27\pm0.01.
\end{eqnarray}
Furthermore, more accurate widths for
both $\Xi_c(3055)^+$ and $\Xi_c(3080)^+$ were obtained by the Belle Collaboration, i.e.,
$\Gamma_{\Xi_c(3055)^+}=7.8\pm 1.1\pm 1.5$ and $\Gamma_{\Xi_c(3080)^+}=3.0\pm 0.7\pm 0.4$ MeV~\cite{Kato:2016hca}.

\subsubsection{$J^P=3/2^+$ state and $\Xi_c(3055)$}

In Ref.~\cite{Liu:2012sj}, the strong decay properties
of the $1D$-wave states were studied within ChQM. It is found that
$\Xi_c(3055)$ seems to favor the $J^P=3/2^+$ state
$|\Xi_c~^2D_{\lambda\lambda} \frac{3}{2}^+ \rangle$,
which is consistent with the predictions in Refs.~\cite{Ebert:2011kk,Chen:2017aqm}.
Based on the SU(4) symmetry we estimated the partial width of
$\Gamma[\Xi_c(3055)^+\rightarrow \Lambda D^+]$, which is too small
to compare with the observation at Belle~\cite{Kato:2016hca}.
The serious SU(4) symmetry breaking might lead to our failed
description of the decays into the $D^+\Lambda$ channel.
Assigning $\Xi_c(3055)$ as $|\Xi_c~^2D_{\lambda\lambda} \frac{3}{2}^+ \rangle$,
it should have relatively large decay rates into $\Xi_c'\pi$ and $\Sigma_c(2455)K$ channels (see Table~\ref{Xicb}).
The predicted partial width ratio between these two channels is
\begin{eqnarray}
\frac{\Gamma[\Xi_c'^0\pi^+]}{\Gamma[\Sigma_c(2455)^{++} K^-]}\simeq 0.78.
\end{eqnarray}
Combining the predicted partial width of $\Gamma[\Xi_c(3055)^+\rightarrow \Sigma_c(2455)^{++} K^-]\simeq 2.4$
MeV with Eq.~(\ref{3055}), we estimate the partial width into the $\Lambda D$ channel:
$\Gamma[\Xi_c(3055)^+\rightarrow \Lambda D^+]\simeq 12.2\pm 4.2$ MeV. Finally, the total width of $\Xi_c(3055)$ is estimated to be $\Gamma\simeq 17.5\pm 4.2$ MeV,
which is close to the upper limit of the observation.
Other interpretations of $\Xi_c(3055)$ also can be found in the literature~\cite{Ye:2017dra,Zhao:2016qmh,Cheng:2015naa}.
To further confirm the nature of $\Xi_c(3055)$, the ratio of $\Gamma[ \Xi_c'^0\pi^+]/\Gamma[\Sigma_c(2455)^{++} K^-]$ is worth observing in future experiments.

Furthermore, the nature of $\Xi_c(3055)$ can be tested by its radiative decays. Assigning $\Xi_c(3055)$
as the $J^P=3/2^+$ state $|\Xi_c~^2D_{\lambda\lambda} \frac{3}{2}^+ \rangle$, we study
its radiative decays into the $1P$-wave charmed baryon states. Our results are listed in Table~\ref{RadX}.
It is found that the $\Xi_c(3055)^0\to \Xi_c(2790)^0 \gamma$
process has a relatively large partial decay width, $\sim 80 $ keV.
Combining it with the measured width of $\Xi_c(3055)$, we predict the branching fraction
$\mathcal{B}[\Xi_c(3055)^0\to \Xi_c(2790)^0 \gamma]\simeq 1.0\% $. Thus,
the neutral state $\Xi_c(3055)^0$ is most likely to be observed in the $\Xi_c(2790)^0 \gamma$ channel
if it corresponds to the $J^P=3/2^+$ state $|\Xi_c~^2D_{\lambda\lambda} \frac{3}{2}^+ \rangle$ indeed.

\subsubsection{$J^P=5/2^+$ state and $\Xi_c(3080)$}

The $\Xi_c(3080)$ resonance is suggested to be the $\rho$-mode $2S$-wave state with $J^P=1/2^-$
in Ref.~\cite{Liu:2012sj}. The observation of $\Xi_c(3080)^+$ in the $D^+\Lambda$ channel
excludes this assignment because the  $D^+\Lambda$ decay mode should be forbidden~\cite{Liu:2012sj}.
The mass and decay modes observed in experiments indicate that
$\Xi_c(3080)$ is most likely to be the $\lambda$-mode $1D$ excitation
of $\Xi_c$ with $J^P=5/2^+$ (i.e., $|\Xi_c~^2D_{\lambda\lambda} \frac{5}{2}^+ \rangle$)
~\cite{Ebert:2011kk,Cheng:2006dk,Chen:2017aqm}. Considering $\Xi_c(3080)^+$ as
the $|\Xi_c~^2D_{\lambda\lambda} \frac{5}{2}^+ \rangle$ state, we find that
it has relatively large decay rates into the $\Sigma_c^*(2520)K$ and $\Xi_c'^*(2645)\pi$ (see Table~\ref{Xicb}).
The partial width ratio between these two main channels is predicted to be
\begin{eqnarray}
\frac{\Gamma[\Xi_c^{*0}(2645)\pi^+]}{\Gamma[\Sigma_c(2520)^{++} K^-]}\simeq 1.2.
\end{eqnarray}
Combining it with the predicted partial width of $\Gamma[\Xi_c(3080)^+\rightarrow \Sigma_c(2455)^{++} K^-]\sim 0.22$ MeV,
we estimate that $\Gamma[\Xi_c(3080)^+\rightarrow \Lambda D^+]\simeq 0.2$ MeV.
Finally, the total width of $\Xi_c(3080)$ is estimated to be $\Gamma\simeq 6.9$ MeV,
which is close to the upper limit of the observation from the Belle Collaboration~\cite{Kato:2016hca}. However,
our predicted partial width ratio between the $\Sigma_cK$ and $\Sigma_c^*K$ channels
\begin{eqnarray}
\frac{\Gamma[ \Sigma_c^{++}(2520)K^-]}
{\Gamma[\Sigma_c(2455)^{++} K^-]}\simeq 7.6,
\end{eqnarray}
is about an order of magnitude larger than the observed ratio listed
in Eq.~(\ref{22}), and a similar phenomenon is found by Chen \emph{et al.} within their
$^3P_0$ analysis~\cite{Chen:2017aqm}. It should be mentioned that the measured ratio $\mathcal{R}=\frac{\Gamma[ \Sigma_c^{++}(2520)K^-]}{\Gamma[\Sigma_c(2455)^{++} K^-]}\simeq 1.07\pm 0.27$
of $\Xi_c(3080)$ may be strongly affected by its nearby states, such as $\Xi_c(3055)$.
Thus, the measured ratio from Belle~\cite{Kato:2016hca} may not be
a genuine ratio for $\Xi_c(3080)$.

Furthermore, assigning $\Xi_c(3080)$ as the $J^P=5/2^+$ state $|\Xi_c~^2D_{\lambda\lambda} \frac{5}{2}^+ \rangle$, we study
its radiative decays. Our results are listed in Table~\ref{RadX}.
It is found that the $\Xi_c(3080)^0$ should have a relatively large decay rate into $\Xi_c(2815)^0\frac{3}{2}^- \gamma$.
The partial decay width is predicted to be $\Gamma[\Xi_c(3080)^0 \to \Xi_c(2815)^0 \gamma]\simeq 85 $ keV.
Combining it with the measured width of $\Xi_c(3080)$, we predict the branching fraction
$\mathcal{B}[\Xi_c(3055)^0\to \Xi_c(2815)^0 \gamma]\simeq 3\% $.
The neutral state $\Xi_c(3080)^0$ is most likely to be observed in the $\Xi_c(2815)^0 \gamma$ channel
if it corresponds to the $J^P=5/2^+$ state $|\Xi_c~^2D_{\lambda\lambda} \frac{5}{2}^+ \rangle$ indeed.

\subsection{$\Xi_b$ states}

In the $\Xi_b$ family, there are two
$\lambda$-mode $1D$-wave excitations $|\Xi_b~^2D_{\lambda\lambda} \frac{3}{2}^+ \rangle$
and $|\Xi_b~^2D_{\lambda\lambda} \frac{5}{2}^+ \rangle$.
The typical masses of the $\lambda$-mode
$1D$-wave $\Xi_b$ excitations are $6.3-6.4$ GeV
within various quark model predictions (see Table~\ref{sp1}).
In the possible mass regions, the strong decays of these $1D$-wave states
are studied with ChQM. Our results have been shown in
Fig.~\ref{Xib}. To be more specific, taking the masses of the $1D$-wave states
obtained in the relativistic quark-diquark picture~\cite{Ebert:2011kk}, we give the predicted
widths in Table~\ref{Xicb}.

\subsubsection{$J^P=3/2^+$ state}

The $J^P=3/2^+$ state $|\Xi_b~^2D_{\lambda\lambda} \frac{3}{2}^+\rangle$ might be a narrow state
with a width of a few MeV. It mainly decays into $\Xi_b'\pi$, $\Xi_b'^*\pi$ and $\Sigma_b K$ channels.
The partial widths of $\Xi_b'\pi$, $\Xi_b'^*\pi$ are less sensitive to the mass of
$|\Xi_b~^2D_{\lambda\lambda} \frac{3}{2}^+\rangle$; however,  the partial width for the $\Sigma_b K$ channel
shows a significant linear dependence on the mass (see Fig.~\ref{Xib}). If the mass of $|\Xi_b~^2D_{\lambda\lambda} \frac{3}{2}^+\rangle$
takes the predicted value $\sim6.37$ GeV in Ref.~\cite{Ebert:2011kk}, the branching fractions for the main channels are predicted to be
\begin{eqnarray}
\frac{\Gamma[\Xi_b'\pi,\Xi_b'^*\pi,\Sigma_b K]}{\Gamma_{\mathrm{total}}}=35\%,21\%,36\%.
\end{eqnarray}
The $\Xi_b'\pi$ and $\Sigma_b K$ decay channels may be ideal channels for our search for
this missing $1D$-wave $\Xi_b$ baryon in future experiments.

Furthermore, we study the radiative decays of $|\Xi_b~^2D_{\lambda\lambda} \frac{3}{2}^+\rangle$ into the
$1P$-wave bottom baryon states.  Our results are listed in Table~\ref{RadX}. It is found that
the charged state $|\Xi_b^-~^2D_{\lambda\lambda} \frac{3}{2}^+\rangle$ might have
a relatively large decay rate into $\Xi_b^-(\frac{1}{2}^-)\gamma$.
The partial decay width can reach up to $\sim 30$ keV if the mass
for $|\Xi_b~^2D_{\lambda\lambda} \frac{3}{2}^+\rangle$ is taken to be $\sim6366$ MeV as predicted in the relativistic quark model~\cite{Ebert:2011kk}. Combining the predicted total width of $|\Xi_b~^2D_{\lambda\lambda} \frac{3}{2}^+\rangle$,
we estimate the branching fraction $\mathcal{B}[|\Xi_b^-~^2D_{\lambda\lambda} \frac{3}{2}^+\rangle\to \Xi_b^-(\frac{1}{2}^-)\gamma]\simeq \mathcal{O}( 10^{-3})$.

\subsubsection{$J^P=5/2^+$ state}

The $J^P=5/2^+$ state $|\Xi_b~^2D_{\lambda\lambda} \frac{5}{2}^+\rangle$ may be a narrow state
with a width comparable  to the $J^P=3/2^+$ state $|\Xi_b~^2D_{\lambda\lambda} \frac{3}{2}^+\rangle$ (i.e., a few MeV).
The decays of $|\Xi_b~^2D_{\lambda\lambda} \frac{5}{2}^+\rangle$ are governed by $\Xi_b'^*\pi$,
which is less sensitive to its mass. If the mass of $|\Xi_b~^2D_{\lambda\lambda} \frac{5}{2}^+\rangle$ is taken to be $\sim6.37$ GeV
as the prediction in Ref.~\cite{Ebert:2011kk}, the decay channel $\Sigma_b^* K$ becomes important as well (see Table~\ref{Xicb}).
In this case, the branching fractions for the $\Xi_b'\pi$, $\Xi_b'^*\pi$ and $\Sigma_b K$ channels are predicted to be
\begin{eqnarray}
\frac{\Gamma[\Xi_b'\pi,\Xi_b'^*\pi,\Sigma_b K]}{\Gamma_{\mathrm{total}}}=12\%,45\%,29\%.
\end{eqnarray}
To establish this missing $1D$-wave $\Xi_b$ baryon with $J^P=5/2^+$, its dominant decay
modes $\Xi_b'^*\pi$ and $\Sigma_b K$ are worth observing in future experiments.

We also study the radiative decays of $|\Xi_b~^2D_{\lambda\lambda} \frac{5}{2}^+\rangle$ into the
$1P$-wave bottom baryon states.  Our results are listed in Table~\ref{RadX} as well. It is found that
the charged state $|\Xi_b^-~^2D_{\lambda\lambda} \frac{5}{2}^+\rangle$ might have
a relatively large decay rate into $\Xi_b^-(\frac{3}{2}^-)\gamma$.
The partial decay width can reach up to $\sim 40$ keV. If the mass
for $|\Xi_b~^2D_{\lambda\lambda} \frac{3}{2}^+\rangle$ is taken to be $\sim6373$ MeV as predicted in the relativistic quark model~\cite{Ebert:2011kk}. Combining the predicted total width of $|\Xi_b~^2D_{\lambda\lambda} \frac{5}{2}^+\rangle$,
we estimate the branching fraction $\mathcal{B}[|\Xi_b^-~^2D_{\lambda\lambda} \frac{5}{2}^+\rangle\to \Xi_b^-(\frac{3}{2}^-)\gamma]\simeq \mathcal{O}( 10^{-3})$.

\begin{figure*}[htbp]
\begin{center}
\centering  \epsfxsize=16.6 cm \epsfbox{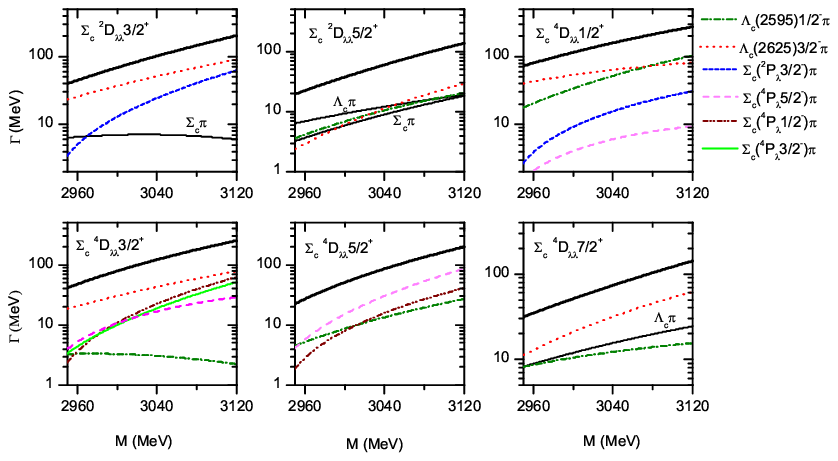}
\vspace{-0.0 cm} \caption{ Strong decay partial widths of the main decay modes for the $\lambda$-mode $1D$-wave excited $\Sigma_c$ states as functions of their mass. The bold solid curves stand for the sum of the partial widths. The masses of the $P$-wave heavy baryons in final states are adopted from the quark model predictions in Ref.~\cite{Ebert:2011kk} (see Table~\ref{sp2}). } \label{sc}
\end{center}
\end{figure*}

\begin{figure*}[htbp]
\begin{center}
\centering  \epsfxsize=16.6 cm \epsfbox{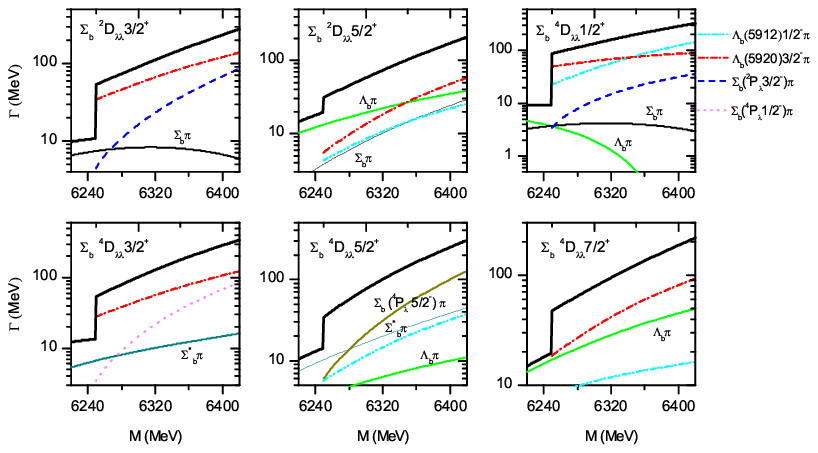}
\vspace{-0.0 cm} \caption{ Strong decay partial widths of the main decay modes for the $\lambda$-mode $1D$-wave excited $\Sigma_b$ states as functions of their mass. The bold solid curves stand for the sum of the partial widths. The masses of the $P$-wave heavy baryons in final states are adopted from the quark model predictions in Ref.~\cite{Ebert:2011kk}, if there are no observations (see Table~\ref{sp2}). } \label{sb}
\end{center}
\end{figure*}

\begin{table*}[htp]
\begin{center}
\caption{ \label{scb}  Strong decay partial widths of the main decay modes for the $\lambda$-mode $D$-wave $\Sigma_c$ and $\Sigma_b$
baryons, the masses (MeV) of which are taken from the quark model predictions of Ref.~\cite{Ebert:2011kk}. $M_f$ stands for the masses of $P$-wave
heavy baryons (MeV) in the final states, which are adopted from the RPP~\cite{Olive:2016xmw} and Ref.~\cite{Ebert:2011kk}. The superscript (subscript) stands for the uncertainty of a prediction with a $+ 10\%$
($-10\%$) uncertainty of the oscillator parameter $\alpha_{\rho}$.}
%\footnotesize
\begin{tabular}{cc|cccccccccccccccccccccccc}
\hline\hline
\multirow{2}{*}{Decay mode}       &\multirow{2}{*}{$M_f$}  &{$\underline{{|\Sigma_c~^2D_{\lambda\lambda} \frac{3}{2}^+ \rangle}(3043)}$}
               &{$\underline{{|\Sigma_c~^2D_{\lambda\lambda} \frac{5}{2}^+ \rangle}(3038)}$}
               &{$\underline{{|\Sigma_c~^4D_{\lambda\lambda}\frac{1}{2}^+ \rangle}(3041)}$}
               &{$\underline{{|\Sigma_c~^4D_{\lambda\lambda}\frac{3}{2}^+ \rangle}(3040)}$}
               &{$\underline{{|\Sigma_c~^4D_{\lambda\lambda}\frac{5}{2}^+ \rangle}(3023)}$}
               &{$\underline{{|\Sigma_c~^4D_{\lambda\lambda}\frac{7}{2}^+ \rangle}(3013)}$}\\
&   & $\Gamma_i$ (MeV)    & $\Gamma_i$ (MeV)    & $\Gamma_i$ (MeV)    & $\Gamma_i$ (MeV)     & $\Gamma_i$ (MeV)    & $\Gamma_i$ (MeV)  \\ \hline
$ \Lambda_{c}\pi$  &2286      &1.29$^{+2.54}_{-1.22}$      &11.9$^{-1.65}_{+1.51}$    &2.62$^{+5.09}_{-2.46}$       &1.15$^{+2.32}_{-1.09}$        &3.1$^{-0.46}_{+0.42}$      &13.1$^{-2.06}_{+1.83}$      \\
$\Sigma_{c}\pi$    &2455      &7.06$^{+4.66}_{-3.95}$      &8.78$^{-1.86}_{+2.25}$    &3.53$^{+2.31}_{-1.96}$       &1.77$^{+1.14}_{-0.98}$        &0.54$^{-0.11}_{+0.14}$     &2.18$^{-0.48}_{+0.60}$  \\
$\Sigma^{\ast}_{c}\pi$ &2520  &2.44$^{-0.33}_{+0.48}$      &2.91$^{+0.72}_{-0.61}$    &1.71$^{+0.79}_{-0.76}$      &7.47$^{+2.04}_{-1.77}$         &11.6$^{+0.80}_{-0.34}$     &1.45$^{+0.13}_{-0.02}$  \\
$ \Xi_{c}K$            &2470  &1.17$^{+0.12}_{-0.14}$      &0.03                      &2.27$^{+0.22}_{-0.27}$       &1.12$^{+0.10}_{-0.13}$        &$<$0.01        &0.01    \\
$ |\Lambda_{c}~^2P_{\lambda}\frac{1}{2}^- \rangle\pi$   &2592   &4.93$^{+0.25}_{-0.19}$     &10.3$^{-0.83}_{+0.68}$     &52.6$^{-3.2}_{+2.7}$       &3.06$^{+2.44}_{-1.70}$         &11.2$^{-0.8}_{+0.9}$       &11.1$^{+3.5}_{-3.0}$    \\
$ |\Lambda_{c}~^2P_{\lambda}\frac{3}{2}^- \rangle\pi$   &2628   &52.8$^{+6.76}_{-5.77}$     &10.9$^{+2.1}_{-2.4}$       &64.0$^{+22.2}_{-18.8}$       &43.1$^{+4.4}_{-3.7}$        &1.92$^{+0.02}_{<+0.01}$       &24.2$^{+0.05}_{-0.04}$    \\
$ |\Sigma_{c}~^2P_{\lambda}\frac{1}{2}^- \rangle\pi$    &2713   &4.09$^{+0.35}_{-0.29}$     &5.85$^{-0.75}_{+0.86}$     &7.56$^{-0.77}_{+0.89}$       &2.10$^{+0.99}_{-0.82}$      &1.13$^{-0.16}_{+0.18}$       &2.83$^{+0.76}_{-0.68}$    \\
$ |\Sigma_{c}~^2P_{\lambda}\frac{3}{2}^- \rangle\pi$    &2798   &25.6$^{+4.82}_{-4.29}$     &0.76$^{-0.05}_{+0.18}$     &15.8$^{+4.0}_{-3.6}$       &3.58$^{+0.39}_{-0.34}$                         &0.01       &0.95$^{+0.06}_{-0.04}$    \\
$ |\Sigma_{c}~^4P_{\lambda}\frac{1}{2}^- \rangle\pi$    &2799   &0.09                       &1.30$^{+0.38}_{-0.33}$     &0.46$^{-0.07}_{+0.10}$       &22.4$^{+3.4}_{-3.0}$         &12.1$^{+2.1}_{-1.9}$       &$<$0.01     \\
$|\Sigma_{c}~^4P_{\lambda}\frac{3}{2}^- \rangle\pi$     &2773   &3.42$^{+0.22}_{-0.16}$     &1.26$^{+0.11}_{-0.09}$     &4.18$^{+0.05}_{+0.02}$       &18.9$^{+2.4}_{-2.0}$        &4.54$^{+2.16}_{-1.80}$       &1.03$^{-0.03}_{+0.04}$   \\
$ |\Sigma_{c}~^4P_{\lambda}\frac{5}{2}^- \rangle\pi$    &2789   &2.20$^{+0.06}_{<+0.01}$    &4.51$^{+0.57}_{-0.45}$     &6.1$^{+1.96}_{-1.72}$      &16.8$^{+4.8}_{-4.3}$           &23.0$^{+3.3}_{-2.9}$       &1.26$^{-0.05}_{+0.09}$    \\
Sum                  &   &105$^{+19.5}_{-15.5}$      &58.5$^{-1.3}_{+1.6}$      &160.8$^{+32.6}_{-25.9}$       &121.5$^{+24.4}_{-19.8}$          &69.1$^{+6.9}_{-5.3}$       &58.1$^{+1.9}_{-1.2}$   \\
\hline\hline
\multirow{3}{*}{Decay mode}    &\multirow{3}{*}{$M_f$} &{$\underline{{|\Sigma_b~^2D_{\lambda\lambda} \frac{3}{2}^+ \rangle}(6326)}$}
               &{$\underline{{|\Sigma_b~^2D_{\lambda\lambda} \frac{5}{2}^+ \rangle}(6284)}$}
               &{$\underline{{|\Sigma_b~^4D_{\lambda\lambda}\frac{1}{2}^+ \rangle}(6311)}$}
               &{$\underline{{|\Sigma_b~^4D_{\lambda\lambda}\frac{3}{2}^+ \rangle}(6285)}$}
               &{$\underline{{|\Sigma_b~^4D_{\lambda\lambda}\frac{5}{2}^+ \rangle}(6270)}$}
               &{$\underline{{|\Sigma_b~^4D_{\lambda\lambda}\frac{7}{2}^+ \rangle}(6260)}$}\\
&    & $\Gamma_i$ (MeV)    & $\Gamma_i$ (MeV)    & $\Gamma_i$ (MeV)    & $\Gamma_i$ (MeV)     & $\Gamma_i$ (MeV)    & $\Gamma_i$ (MeV)  \\ \hline
$ \Lambda_{b}\pi$      &5620   &0.56$^{+2.76}_{-0.46}$    &17.3$^{-1.1}_{+1.6}$     &1.56$^{+5.81}_{-1.51}$      &1.21$^{+3.07}_{-1.20}$     &4.43$^{-0.58}_{+0.45}$       &18.5$^{-2.6}_{+2.1}$      \\
$ \Sigma_{b}\pi$       &5811   &8.28$^{+5.54}_{-4.66}$    &6.32$^{-1.44}_{+1.82}$   &4.16$^{+2.51}_{-2.20}$      &2.03$^{+1.03}_{-0.96}$     &0.37$^{-0.09}_{+0.11}$       &1.43$^{-0.33}_{+0.46}$  \\
$\Sigma^{\ast}_{b}\pi$ &5832   &4.21$^{-0.58}_{+0.80}$    &3.26$^{+0.81}_{-0.70}$   &2.05$^{+1.08}_{-1.00}$      &8.36$^{+2.26}_{-2.01}$      &12.6$^{+0.9}_{-0.5}$       &1.54$^{+0.15}_{-0.05}$   \\
$ \Xi_{b}K$            &5794   &0.89$^{+0.06}_{-0.06}$    &$\cdot\cdot\cdot$        &0.88$^{+0.03}_{-0.04}$     &  $\cdot\cdot\cdot$ & $\cdot\cdot\cdot$ &$\cdot\cdot\cdot$  \\
$|\Lambda_{b}~^2P_{\lambda}\frac{1}{2}^- \rangle\pi$ & 5912   &5.70$^{+0.22}_{-0.13}$   &6.96$^{-0.61}_{+0.54}$    &51.6$^{-2.9}_{+2.2}$     &3.05$^{+2.13}_{-1.55}$     &7.68$^{-0.65}_{+0.66}$       &8.89$^{+2.77}_{-2.34}$   \\
$| \Lambda_{b}~^2P_{\lambda}\frac{3}{2}^- \rangle\pi$& 5920   &70.0$^{+8.3}_{-7.5}$     &10.4$^{-2.0}_{+2.3}$    &66.0$^{+24.1}_{-20.1}$     &40.5$^{+4.1}_{-3.4}$       &1.90$^{-0.001}_{+0.02}$       &21.1$^{+0.06}_{-0.1}$ \\
$| \Sigma_{b}~^2P_{\lambda}\frac{1}{2}^- \rangle\pi$ & 6101   &1.18$^{+0.15}_{-0.13}$   &0.13$^{-0.02}_{+0.03}$    &0.71$^{-0.07}_{+0.09}$     &0.81$^{+0.22}_{-0.20}$     &0.01                          &0.43$^{+0.09}_{-0.08}$   \\
$| \Sigma_{b}~^2P_{\lambda}\frac{3}{2}^- \rangle\pi$ & 6096   &25.6$^{+4.8}_{-4.2}$     &0.28$^{+0.09}_{-0.05}$    &13.1$^{+3.2}_{-2.9}$     &1.49$^{+0.20}_{-0.18}$     &$<$0.01                       &0.24$^{+0.03}_{-0.03}$   \\
$| \Sigma_{b}~^4P_{\lambda}\frac{1}{2}^- \rangle\pi$ & 6095   &0.10$^{-0.04}_{+0.07}$    &0.79$^{+0.21}_{-0.18}$    &0.33$^{-0.06}_{+0.07}$     &10.5$^{+1.81}_{-1.55}$     &5.30$^{+1.0}_{-0.90}$         &$<$0.01    \\
$|\Sigma_{b}~^4P_{\lambda}\frac{3}{2}^- \rangle\pi$  & 6087   &2.53$^{+0.16}_{-0.12}$    &0.41$^{+0.05}_{-0.05}$    &2.24$^{+0.04}_{-0.01}$     &6.44$^{+1.01}_{-0.88}$     &3.52$^{+1.09}_{-0.97}$       &0.15$^{+0.01}_{<-0.01}$   \\
$| \Sigma_{b}~^4P_{\lambda}\frac{5}{2}^- \rangle\pi$ & 6084   &2.39$^{+0.02}_{+0.03}$    &2.09$^{+0.32}_{-0.27}$    &5.46$^{+1.72}_{-1.52}$     &10.7$^{+2.8}_{-2.5}$       &10.8$^{+1.8}_{-1.6}$      &0.34$^{+0.02}_{-0.01}$ \\
Sum                &  &121.4$^{+21.4}_{-16.4}$   &47.9$^{-3.7}_{+5.0}$    &148.0$^{+35.5}_{-26.9}$      &85.1$^{+18.6}_{-14.4}$     &48.5$^{+3.5}_{-2.7}$      &52.6$^{+0.2}_{-0.05}$    \\
\hline\hline
\end{tabular}
\end{center}
\end{table*}

\begin{table*}[htp]
\begin{center}
\caption{\label{RadSDW} Partial widths of radiative decays for the $\lambda$-mode $D$-wave $\Sigma_c$ and $\Sigma_b$
baryons, the masses (MeV) of which are taken from the quark model predictions of Ref.~\cite{Ebert:2011kk}. $M_f$ stands for the masses of $P$-wave
heavy baryons (MeV) in the final states, which are adopted from the RPP~\cite{Olive:2016xmw} and Ref.~\cite{Ebert:2011kk}. The superscript (subscript) stands for the uncertainty of a prediction with a $+ 10\%$ ($-10\%$) uncertainty of the oscillator parameter $\alpha_{\rho}$.}
\begin{tabular}{cc|cccccccccccccccccccccccccccccc|}\hline\hline
\multirow{2}{*}{~~~Decay mode~~~}       &\multirow{2}{*}{$M_f$} &\multicolumn{1}{c}{$\underline{|\Sigma_c~^2D_{\lambda\lambda} \frac{3}{2}^+ \rangle(3043)}$}
               &\multicolumn{1}{c}{$\underline{|\Sigma_c~^2D_{\lambda\lambda} \frac{5}{2}^+ \rangle(3038)}$}
               &\multicolumn{1}{c}{$\underline{|\Sigma_c~^4D_{\lambda\lambda}\frac{1}{2}^+ \rangle(3041)}$}
               &\multicolumn{1}{c}{$\underline{|\Sigma_c~^4D_{\lambda\lambda}\frac{3}{2}^+ \rangle(3040)}$}
               &\multicolumn{1}{c}{$\underline{|\Sigma_c~^4D_{\lambda\lambda}\frac{5}{2}^+ \rangle(3023)}$}
               &\multicolumn{1}{c}{$\underline{|\Sigma_c~^4D_{\lambda\lambda}\frac{7}{2}^+ \rangle(3013)}$}\\
&&$\Gamma_i$ (keV)    &$\Gamma_i$ (keV)    &$\Gamma_i$ (keV) &$\Gamma_i$ (keV) &     $\Gamma_i$ (keV) &$\Gamma_i$ (keV)  \\
\hline
$ |\Sigma_{c}^{++}~ ^2P_{\lambda}\frac{1}{2}^- \rangle\gamma$ &2713 &231.9$^{-27.7}_{+30.6}$     &58.4$^{-7.1}_{+8.1}$     &15.01$^{-1.9}_{+2.3}$    &23.6$^{-2.7}_{+2.9}$   &7.40$^{-0.92}_{+1.04}$    &0.017       \\
$ |\Sigma_c^{+}~ ^2P_{\lambda} \frac{1}{2}^- \rangle\gamma$   &     &1.50$^{-0.01}_{-0.08}$      &2.27$^{-0.21}_{+0.19}$   &2.56$^{-0.37}_{+0.46}$   &2.58$^{-0.33}_{+0.37}$    &0.80$^{-0.11}_{+0.13}$    &0.001   \\
$ |\Sigma_c^{0}~ ^2P_{\lambda} \frac{1}{2}^- \rangle\gamma$   &     &164.2$^{-23.2}_{+28.2}$     &22.6$^{-3.2}_{+4.0}$     &0.45$^{-0.02}_{+0.02}$   &2.71$^{-0.24}_{+0.22}$    &0.87$^{-0.09}_{+0.09}$    &0.004   \\
$ |\Sigma_{c}^{++}~ ^2P_{\lambda}\frac{3}{2}^- \rangle\gamma$ &2798 &78.5$^{+12.0}_{-15.1}$      &34.9$^{-5.5}_{+7.1}$     &1.93$^{-0.29}_{+0.36}$   &0.59$^{-0.10}_{+0.13}$    &4.90$^{-0.69}_{+0.85}$    &3.94$^{-0.55}_{+0.67}$     \\
$ |\Sigma_{c}^{+}~ ^2P_{\lambda}\frac{3}{2}^- \rangle\gamma$  &     &1.82$^{-0.23}_{+0.28}$      &0.81$^{-0.14}_{+0.18}$   &0.20$^{-0.03}_{+0.04}$   &0.06                      &0.52$^{-0.08}_{+0.09}$    &0.42$^{-0.07}_{+0.07}$     \\
$ |\Sigma_{c}^{0}~ ^2P_{\lambda}\frac{3}{2}^- \rangle\gamma$  &     &38.6$^{-6.1}_{+8.0}$        &47.52$^{-8.0}_{+10.6}$   &0.24$^{-0.03}_{+0.04}$   &0.08                      &0.60$^{-0.08}_{+0.09}$    &0.49$^{-0.07}_{+0.07}$     \\
$ |\Sigma_{c}^{++}~ ^4P_{\lambda}\frac{1}{2}^- \rangle\gamma$ &2799 &5.09$^{-0.63}_{+0.70}$      &0.29$^{-0.03}_{+0.02}$   &136.5$^{-19.3}_{+23.5}$  &52.7$^{-7.6}_{+9.4}$      &9.79$^{-1.48}_{+1.88}$    &1.30$^{-0.28}_{+0.41}$     \\
$ |\Sigma_{c}^{+}~ ^4P_{\lambda}\frac{1}{2}^- \rangle\gamma$  &     &0.55$^{-0.08}_{+0.08}$      &0.03                     &0.97$^{-0.07}_{+0.06}$   &0.45$^{-0.05}_{+0.05}$    &0.14$^{-0.01}_{+0.02}$    &0.005     \\
$ |\Sigma_{c}^{0}~ ^4P_{\lambda}\frac{1}{2}^- \rangle\gamma$  &     &0.60$^{-0.06}_{+0.06}$      &0.03                     &94.3$^{-14.5}_{+18.4}$   &38.61$^{-6.0}_{+7.8}$     &5.64$^{-0.90}_{+1.17}$    &1.00$^{-0.20}_{+0.28}$      \\
$ |\Sigma_{c}^{++}~ ^4P_{\lambda}\frac{3}{2}^- \rangle\gamma$ &2773 &11.9$^{-1.6}_{+1.8}$        &4.73$^{-0.54}_{+0.59}$   &94.6$^{-13.6}_{+16.5}$   &121.8$^{-17.0}_{+20.5}$   &48.5$^{-7.2}_{+8.9}$   &13.5$^{-2.0}_{+2.5}$    \\
$ |\Sigma_{c}^{+}~ ^4P_{\lambda}\frac{3}{2}^- \rangle\gamma$  &     &1.27$^{-0.17}_{+0.21}$       &0.51$^{-0.06}_{+0.08}$  &1.26$^{-0.13}_{+0.14}$   &0.90$^{-0.06}_{+0.04}$    &0.47$^{-0.06}_{+0.08}$    &0.17$^{-0.01}_{+0.02}$     \\
$ |\Sigma_{c}^{0}~ ^4P_{\lambda}\frac{3}{2}^- \rangle\gamma$  &     &1.42$^{-0.16}_{+0.17}$       &0.55$^{-0.05}_{+0.04}$  &55.9$^{-8.6}_{+10.9}$    &84.22$^{-12.8}_{+16.3}$   &44.2$^{-7.1}_{+9.2}$   &8.22$^{-1.3}_{+1.7}$     \\
$ |\Sigma_{c}^{++}~ ^4P_{\lambda}\frac{5}{2}^- \rangle\gamma$ &2789 &2.66$^{-0.39}_{+0.48}$       &12.93$^{-1.8}_{+2.3}$   &10.1$^{-2.0}_{+2.7}$     &32.3$^{-5.1}_{+6.5}$      &49.4$^{-7.5}_{+9.5}$   &23.3$^{-3.8}_{+4.9}$     \\
$ |\Sigma_{c}^{+}~ ^4P_{\lambda}\frac{5}{2}^- \rangle\gamma$  &     &0.28$^{-0.04}_{+0.05}$       &1.36$^{-0.20}_{+0.25}$  &0.10$^{-0.02}_{+0.02}$   &0.42$^{-0.06}_{+0.06}$    &0.35$^{-0.03}_{+0.04}$    &0.55$^{-0.08}_{+0.11}$      \\
$ |\Sigma_{c}^{0}~ ^4P_{\lambda}\frac{5}{2}^- \rangle\gamma$  &     &0.33$^{-0.04}_{+0.05}$       &1.59$^{-0.20}_{+0.25}$  &6.72$^{-1.28}_{+.80}$    &19.6$^{-3.3}_{+4.2}$      &35.0$^{-5.7}_{+7.3}$    &30.5$^{-5.2}_{+6.9}$     \\
$|\Lambda_c^{+}~ ^2P_{\lambda}\frac{1}{2}^- \rangle\gamma$    &2592 &17.53$^{-0.02}_{+0.84}$      &25.36$^{-0.62}_{-2.2}$    &24.7$^{-1.7}_{+1.2}$   &88.8$^{-4.6}_{+2.3}$      &31.3$^{-2.2}_{+1.7}$   &0.35$^{-0.14}_{+0.24}$      \\
$|\Lambda_c^{+}~ ^2P_{\lambda}\frac{3}{2}^- \rangle\gamma$    &2628 &64.6$^{-5.6}_{+5.0}$         &42.95$^{-3.0}_{+2.1}$     &21.2$^{-2.2}_{+2.2}$   &7.71$^{-1.07}_{+1.27}$    &53.9$^{-3.5}_{+2.5}$   &45.4$^{-2.4}_{+1.2}$     \\
\hline\hline
\multirow{2}{*}{~~~Decay mode~~~}       &\multirow{2}{*}{$M_f$}     &\multicolumn{1}{c}{$\underline{|\Sigma_b~^2D_{\lambda\lambda} \frac{3}{2}^+ \rangle(6326)}$}
               &\multicolumn{1}{c}{$\underline{|\Sigma_b~^2D_{\lambda\lambda} \frac{5}{2}^+ \rangle(6284)}$}
               &\multicolumn{1}{c}{$\underline{|\Sigma_b~^4D_{\lambda\lambda}\frac{1}{2}^+ \rangle(6311)}$}
               &\multicolumn{1}{c}{$\underline{|\Sigma_b~^4D_{\lambda\lambda}\frac{3}{2}^+ \rangle(6285)}$}
               &\multicolumn{1}{c}{$\underline{|\Sigma_b~^4D_{\lambda\lambda}\frac{5}{2}^+ \rangle(6270)}$}
               &\multicolumn{1}{c}{$\underline{|\Sigma_b~^4D_{\lambda\lambda}\frac{7}{2}^+ \rangle(6260)}$}\\
& &$\Gamma_i$ (keV)    &$\Gamma_i$ (keV)  &$\Gamma_i$ (keV)  &$\Gamma_i$ (keV) &     $\Gamma_i$ (keV) &$\Gamma_i$ (keV) \\
\hline
$|\Sigma_b^{+}~ ^2P_{\lambda}\frac{1}{2}^- \rangle\gamma$  &6101  &216.0$^{-32.4}_{+40.65}$   &15.90$^{-2.59}_{+3.42}$   &2.22$^{-0.31}_{+0.39}$     &2.79$^{-0.41}_{+0.50}$    &0.72$^{-0.11}_{+0.14}$  &$<0.01$\\
$|\Sigma_b^{0}~ ^2P_{\lambda}\frac{1}{2}^- \rangle\gamma$  &      &15.9$^{-2.4}_{+3.0}$       &1.09$^{-0.18}_{+0.24}$    &0.11$^{-0.01}_{+0.02}$     &0.17$^{-0.03}_{+0.03}$    &0.04            &$<0.01$ \\
$|\Sigma_b^{-}~ ^2P_{\lambda}\frac{1}{2}^- \rangle\gamma$  &      &45.3$^{-6.7}_{+8.4}$       &3.61$^{-0.58}_{+0.77}$    &0.67$^{-0.10}_{+0.12}$     &0.72$^{-0.11}_{+0.13}$    &0.19$^{-0.03}_{+0.03}$  &$<0.01$ \\
$|\Sigma_b^{+}~ ^2P_{\lambda}\frac{3}{2}^- \rangle\gamma$  &6096  &141.0$^{-21.9}_{+28.2}$    &74.4$^{-12.5}_{+46.5}$   &1.50$^{-0.22}_{+0.28}$      &0.26$^{-0.04}_{+0.06}$    &2.04$^{-0.30}_{+0.36}$  &1.53$^{-0.22}_{+0.27}$       \\
$|\Sigma_b^{0}~ ^2P_{\lambda}\frac{3}{2}^- \rangle\gamma$  &      &9.84$^{-1.55}_{+1.98}$     &5.86$^{-0.99}_{+1.31}$    &0.09                       &0.02                      &0.12$^{-0.01}_{+0.03}$   &0.09        \\
$|\Sigma_b^{-}~ ^2P_{\lambda}\frac{3}{2}^- \rangle\gamma$  &      &31.4$^{-4.9}_{+6.2}$       &14.3$^{-2.36}_{+3.18}$   &0.39$^{-0.06}_{+0.07}$      &0.06                      &0.52$^{-0.07}_{+0.10}$  &0.39$^{-0.05}_{+0.07}$         \\
$|\Sigma_b^{+}~ ^4P_{\lambda}\frac{1}{2}^- \rangle\gamma$  &6095  &4.92$^{-0.50}_{+0.49}$     &0.13$^{-0.01}_{+0.02}$    &243.8$^{-35.9}_{+44.7}$    &67.54$^{-10.5}_{+13.6}$   &9.69$^{-1.58}_{+2.05}$  &1.64$^{-0.34}_{+0.48}$        \\
$|\Sigma_b^{0}~ ^4P_{\lambda}\frac{1}{2}^- \rangle\gamma$  &      &0.30$^{-0.03}_{+0.03}$     &0.008                     &17.6$^{-2.6}_{+3.3}$       &4.94$^{-0.78}_{+1.00}$    &0.68$^{-0.11}_{+0.15}$  &0.12$^{-0.02}_{+0.04}$        \\
$|\Sigma_b^{-}~ ^4P_{\lambda}\frac{1}{2}^- \rangle\gamma$  &      &1.27$^{-0.13}_{+0.13}$     &0.034                     &52.1$^{-7.5}_{+9.4}$       &14.2$^{-2.2}_{+2.9}$      &2.14$^{-0.35}_{+0.45}$  &0.34$^{-0.07}_{+0.10}$         \\
$|\Sigma_b^{+}~ ^4P_{\lambda}\frac{3}{2}^- \rangle\gamma$  &6087  &9.18$^{-1.11}_{+1.25}$     &1.74$^{-0.21}_{+0.25}$    &124.8$^{-19.0}_{+24.0}$    &124.9$^{-19.4}_{+24.9}$   &62.7$^{-10.2}_{+13.3}$ &11.0$^{-1.8}_{+2.3}$      \\
$|\Sigma_b^{0}~ ^4P_{\lambda}\frac{3}{2}^- \rangle\gamma$  &      &0.56$^{-0.07}_{+0.07}$     &0.11$^{-0.02}_{+0.01}$    &8.78$^{-1.34}_{+1.71}$     &9.05$^{-1.42}_{+1.82}$    &4.75$^{-0.77}_{+1.02}$  &0.78$^{-0.13}_{+0.16}$       \\
$|\Sigma_b^{-}~ ^4P_{\lambda}\frac{3}{2}^- \rangle\gamma$  &      &2.37$^{-0.29}_{+0.32}$     &0.45$^{-0.06}_{+0.06}$    &27.5$^{-4.1}_{+5.2}$       &26.6$^{-4.1}_{+5.3}$      &12.7$^{-2.1}_{+2.7}$ &2.39$^{-0.39}_{+0.51}$        \\
$|\Sigma_b^{+}~ ^4P_{\lambda}\frac{5}{2}^- \rangle\gamma$  &6084  &2.88$^{-0.39}_{+0.48}$     &6.53$^{-0.94}_{+1.15}$    &19.9$^{-4.0}_{+5.8}$       &37.4$^{-6.3}_{+8.3}$      &62.8$^{-10.2}_{+13.3}$ &50.4$^{-8.6}_{+11.4}$      \\
$|\Sigma_b^{0}~ ^4P_{\lambda}\frac{5}{2}^- \rangle\gamma$  &      &0.17$^{-0.02}_{+0.03}$     &0.40$^{-0.07}_{+0.06}$    &1.43$^{-0.29}_{+0.40}$     &2.65$^{-0.45}_{+0.59}$    &4.57$^{-0.75}_{+0.97}$  &4.01$^{-0.68}_{+0.91}$        \\
$|\Sigma_b^{-}~ ^4P_{\lambda}\frac{5}{2}^- \rangle\gamma$  &      &0.74$^{-0.10}_{+0.13}$     &1.68$^{-0.24}_{+0.30}$    &4.31$^{-0.87}_{+1.25}$     &8.18$^{-1.36}_{+1.80}$    &13.3$^{-2.1}_{+2.8}$ &9.59$^{-1.62}_{+2.15}$        \\
$|\Lambda_b^{0}~ ^2P_{\lambda}\frac{1}{2}^- \rangle\gamma$ &5912  &19.0$^{+0.9}_{-2.1}$    &21.6$^{+0.7}_{-2.2}$   &27.6$^{-1.6}_{+0.8}$    &79.0$^{-4.3}_{+2.3}$   &27.4$^{-2.0}_{+1.6}$ &0.35$^{-0.14}_{+0.25}$         \\
$|\Lambda_b^{0}~ ^2P_{\lambda}\frac{3}{2}^- \rangle\gamma$ &5920  &99.0$^{-5.9}_{+3.7}$    &45.8$^{-2.6}_{+1.6}$   &30.2$^{-2.7}_{+2.5}$    &9.06 $^{-1.28}_{+1.56}$   &56.1$^{-3.0}_{+1.5}$ &45.4$^{-1.7}_{+0.03}$        \\
\hline\hline
\end{tabular}
\end{center}
\end{table*}

\begin{figure*}[htbp]
\begin{center}
\centering  \epsfxsize=16.6 cm \epsfbox{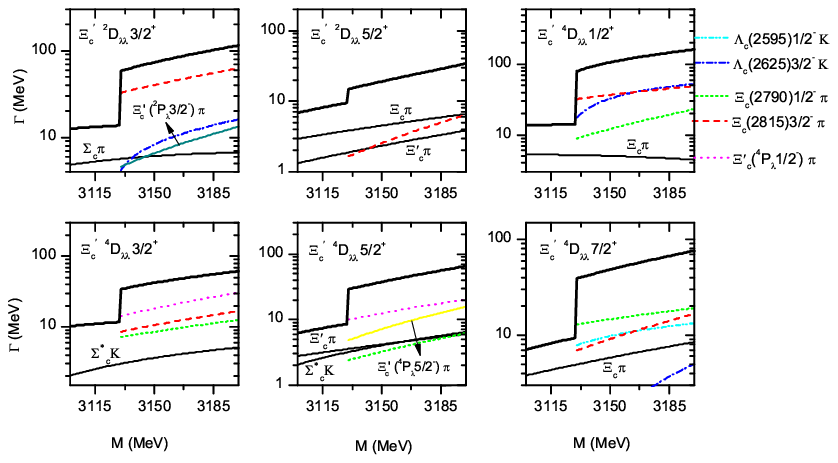}
\vspace{-0.0 cm} \caption{ Strong decay partial widths of the main decay modes for the $\lambda$-mode $1D$-wave excited $\Xi_c'$ states as functions of their mass. The bold solid curves stand for the sum of the partial widths. The masses of the $P$-wave heavy baryons in final states are adopted from the quark model predictions in Ref.~\cite{Ebert:2011kk}, if there are no observations (see Table~\ref{sp2}). } \label{Xipc}
\end{center}
\end{figure*}

\begin{figure*}[htbp]
\begin{center}
\centering  \epsfxsize=16.6 cm \epsfbox{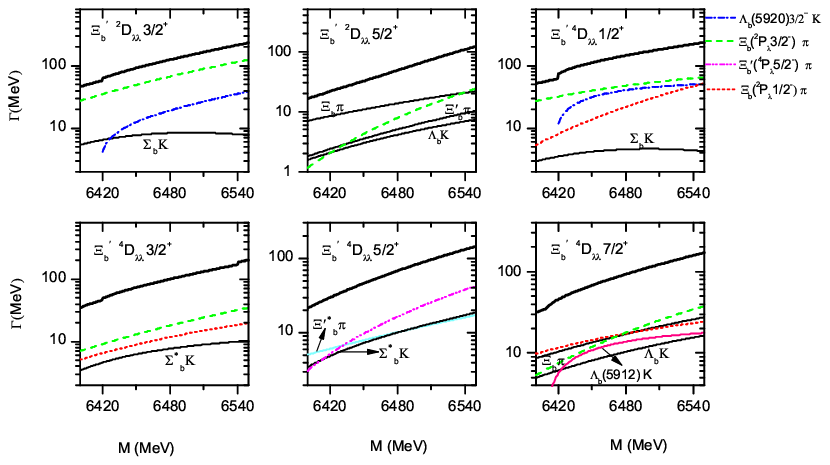}
\vspace{-0.0 cm} \caption{ Strong decay partial widths of the main decay modes for the $\lambda$-mode $1D$-wave excited $\Xi_b'$ states as functions of their mass. The bold solid curves stand for the sum of the partial widths. The masses of the $P$-wave heavy baryons in final states are adopted from the quark model predictions in Ref.~\cite{Ebert:2011kk}, if there are no observations (see Table~\ref{sp2}).  } \label{Xipb}
\end{center}
\end{figure*}

\begin{figure*}[htbp]
\begin{center}
\centering  \epsfxsize=8.4 cm \epsfbox{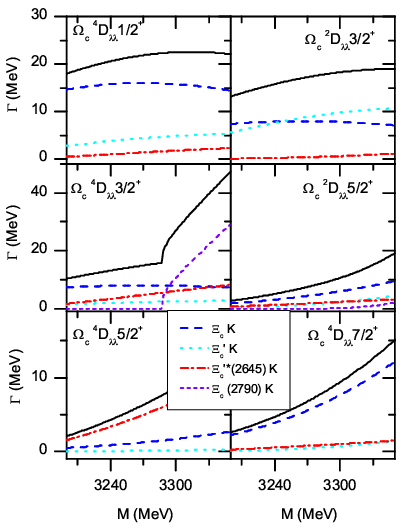} \epsfxsize=8.4 cm \epsfbox{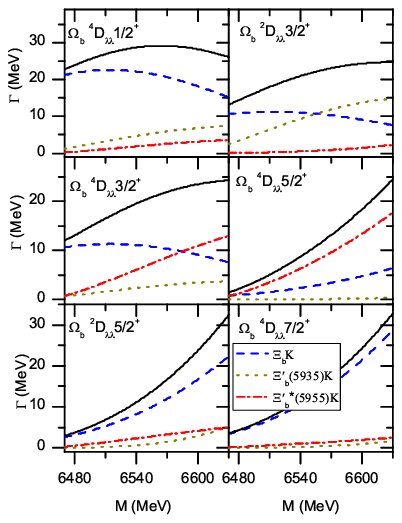}
\vspace{-0.4 cm} \caption{ Strong decay partial widths of the main decay modes for the $\lambda$-mode $1D$-wave excited $\Omega_c$ and $\Omega_b$ states as functions of their mass. The bold solid curves stand for the total widths.   } \label{omcb}
\end{center}
\end{figure*}

\section{Results for singly heavy baryons of $6_F$}\label{RD2}

\subsection{$\Sigma_c$}

In the $\Sigma_c$ family, according to the quark model classification,
there are six $\lambda$-mode $1D$-wave excitations: $|\Sigma_c~^4D_{\lambda\lambda} \frac{1}{2}^+ \rangle$,
$|\Sigma_c~^4D_{\lambda\lambda} \frac{3}{2}^+ \rangle$, $|\Sigma_c~^2D_{\lambda\lambda} \frac{3}{2}^+ \rangle$,
$|\Sigma_c~^2D_{\lambda\lambda} \frac{5}{2}^+ \rangle$, $|\Sigma_c~^4D_{\lambda\lambda} \frac{5}{2}^+ \rangle$,
and $|\Sigma_c~^4D_{\lambda\lambda} \frac{7}{2}^+ \rangle$.
However, no $D$-wave states have been established. The typical masses of the $\lambda$-mode
$1D$-wave $\Sigma_c$ excitations are predicted to be $\sim 3.0$ GeV within various quark models (see Table~\ref{sp2}).
In the possible mass range, we study their strong decay transitions within ChQM.
Our results are shown in Fig.~\ref{sc}.
To be more specific, taking the masses of the $1D$-wave states
as predicted in the relativistic quark-diquark picture~\cite{Ebert:2011kk}, we further present
the results in Table~\ref{scb}.

\subsubsection{$J^P=1/2^+$ state}

The $J^P=1/2^+$ state $|\Sigma_c~^4D_{\lambda\lambda} \frac{1}{2}^+ \rangle$ might be a broad state.
If its mass is taken as the prediction 3041 MeV in Ref.~\cite{Ebert:2011kk}, the sum of the partial widths for the pionic and kaonic decays
can reach up to $\Gamma_{\mathrm{Sum}}\sim 160$ MeV (see Table~\ref{scb}). This state has large decay rates into $\Lambda_c(2595) \pi$
and $\Lambda_c(2625) \pi$ final states.
The ratios between the partial decay widths for the $\Lambda_c(2595)\pi$ and $\Lambda_c(2625)\pi$ channels and $\Gamma_{\mathrm{Sum}}$ are predicted to be
\begin{eqnarray}
\frac{\Gamma[\Lambda_c(2595) \pi]}{\Gamma_{\mathrm{Sum}}}\simeq 33\%,\ \frac{\Gamma[\Lambda_c(2625)\pi]}{\Gamma_{\mathrm{Sum}}}\simeq 40\%.
\end{eqnarray}
Both $\Lambda_c(2595) \pi$ and $\Lambda_c(2625) \pi$ may be ideal channels for our search for $|\Sigma_c~^4D_{\lambda\lambda} \frac{1}{2}^+ \rangle$ in future experiments.

We also estimate its radiative transitions into the $1P$-wave charmed baryon states. Our results are listed in Table~\ref{RadSDW}.
It is found that $|\Sigma_c^{++(0)}~^4D_{\lambda\lambda} \frac{1}{2}^+ \rangle$ might have
relatively large decay rates into $ |\Sigma_{c}^{++(0)}~ ^4P_{\lambda}\frac{1}{2}^- \rangle\gamma$ and
$ |\Sigma_{c}^{++(0)}~ ^4P_{\lambda}\frac{3}{2}^- \rangle\gamma$, and their partial radiative decay widths
are estimated to be $\mathcal{O}(10)-\mathcal{O}(100)$ keV. The branching fractions for these main radiative
decay processes may reach up to $\mathcal{O}(10^{-4})-\mathcal{O}(10^{-3})$.

\subsubsection{$J^P=3/2^+$ states}

The $J^P=3/2^+$ state $|\Sigma_c~^2D_{\lambda\lambda} \frac{3}{2}^+ \rangle$ dominantly decays into
the $P$-wave states through the poinic decay modes $\Lambda_c(2625)\pi$ and $ |\Sigma_{c}~^2P_{\lambda}\frac{3}{2}^- \rangle \pi$, while its decay rate into $\Sigma_c\pi$ is sizable. It has a width of $\mathcal{O}(10)-\mathcal{O}(100)$ MeV, which significantly depends on its mass. If the mass is taken as the prediction 3043 MeV in Ref.~\cite{Ebert:2011kk}, the sum of the partial widths of the pionic decays is $\Gamma_{\mathrm{Sum}}\sim 100$ MeV (see Table~\ref{scb}), and the ratios between the partial decay widths for the $\Lambda_c(2625)\pi$ and $\Sigma_c\pi$ channels and $\Gamma_{\mathrm{Sum}}$ are predicted to be
\begin{eqnarray}
\frac{\Gamma[\Lambda_c(2625) \pi]}{\Gamma_{\mathrm{Sum}}}\simeq 50\%,\ \frac{\Gamma[\Sigma_c\pi]}{\Gamma_{\mathrm{Sum}}}\simeq 7\%.
\end{eqnarray}
The $\Sigma_c\pi$ and $\Lambda_c(2625) \pi$ decay channels may be ideal channels for our search for $|\Sigma_c~^2D_{\lambda\lambda} \frac{3}{2}^+ \rangle$ in future experiments.

For the other $J^P=3/2^+$ state $|\Sigma_c~^4D_{\lambda\lambda} \frac{3}{2}^+ \rangle$, one finds that it has large decay rates into the $P$-wave states through the pionic decay modes $\Lambda_c(2625)\pi$, $ |\Sigma_{c}~^4P_{\lambda}\frac{1}{2}^- \rangle \pi$, $ |\Sigma_{c}~^4P_{\lambda}\frac{3}{2}^- \rangle \pi$, and $ |\Sigma_{c}~^4P_{\lambda}\frac{5}{2}^- \rangle \pi$.
Its width should be broader than that of $|\Sigma_c~^2D_{\lambda\lambda} \frac{3}{2}^+ \rangle$. Furthermore, the decay rate into $\Sigma_c(2520)\pi$ is sizable as well. If its mass is taken as the prediction 3040 MeV in Ref.~\cite{Ebert:2011kk}, the sum of the partial widths of the pionic decays can reach up to $\Gamma_{\mathrm{Sum}}\sim 120$ MeV (see Table~\ref{scb}), while the ratios between the partial widths of $\Lambda_c(2625)\pi$ and $\Sigma_c(2520)\pi$ and $\Gamma_{\mathrm{Sum}}$ are predicted to be
\begin{eqnarray}
\frac{\Gamma[\Lambda_c(2625) \pi]}{\Gamma_{\mathrm{Sum}}}\simeq 35\%, \ \ \frac{\Gamma[\Sigma_c(2520)\pi]}{\Gamma_{\mathrm{Sum}}}\simeq 6\%.
\end{eqnarray}
The $\Lambda_c(2625) \pi$ and $\Sigma_c(2520)\pi$ may be ideal channels for our search for $|\Sigma_c~^4D_{\lambda\lambda} \frac{3}{2}^+ \rangle$ in future experiments.

We also estimate the radiative transitions of these $J^P=3/2^+$ states into the $1P$-wave states.
Our results are listed in Table~\ref{RadSDW}. It is found that $|\Sigma_c^{++(0)}~^4D_{\lambda\lambda} \frac{3}{2}^+ \rangle$ might have
relatively large decay rates into $ |\Sigma_{c}^{++(0)}~ ^4P_{\lambda}\frac{3}{2}^- \rangle\gamma$,
while $|\Sigma_c^{+}~^4D_{\lambda\lambda} \frac{3}{2}^+ \rangle$ might have
a relatively large decay rates into $ \Lambda_c(2595)^+\gamma $, and their partial radiative decay widths
are estimated to be $\mathcal{O}(10)-\mathcal{O}(100)$ keV. The branching fractions for these main radiative
decay processes may reach up to $\mathcal{O}(10^{-4})-\mathcal{O}(10^{-3})$.
The $|\Sigma_c^{+}~^4D_{\lambda\lambda} \frac{3}{2}^+ \rangle$ may have the possibility of being observed
in the $ \Lambda_c(2595)^+\gamma $ channel.

\subsubsection{$J^P=5/2^+$ states}

The $J^P=5/2^+$ state $|\Sigma_c~^2D_{\lambda\lambda} \frac{5}{2}^+ \rangle$ might be a narrow state with a width of $\mathcal{O}(10)$ MeV (see Fig.~\ref{sc}). It has large decay rates into $\Lambda_c\pi$, $\Sigma_c\pi$, $\Lambda_c(2595) \pi$, and $\Lambda_c(2625) \pi$
with comparable partial decay widths. If its mass is taken as the prediction 3038 MeV in Ref.~\cite{Ebert:2011kk}, the sum of the partial widths of the pionic decays is about $\Gamma_{\mathrm{Sum}}\sim 60$ MeV (see Table~\ref{scb}),
and the ratios between the partial widths for the main decay modes, $\Lambda_c\pi$, $\Sigma_c\pi$, $\Lambda_c(2595) \pi$, and $\Lambda_c(2625) \pi$, and $\Gamma_{\mathrm{Sum}}$ are predicted to be
\begin{eqnarray}
&& \frac{\Gamma[\Lambda_c\pi,\Sigma_c\pi,\Lambda_c(2595) \pi,\Lambda_c(2625) \pi]}{\Gamma_{\mathrm{Sum}}}\nonumber\\
&& \simeq 20\%,15\%,18\%,19\%.
\end{eqnarray}
The $\Lambda_c\pi$, $\Sigma_c\pi$, $\Lambda_c(2595) \pi$, and $\Lambda_c(2625) \pi$ decay channels may be
ideal channels for our search for the missing $J^P=5/2^+$ state $|\Sigma_c~^2D_{\lambda\lambda} \frac{5}{2}^+ \rangle$.

The other $J^P=5/2^+$ state $|\Sigma_c~^4D_{\lambda\lambda} \frac{5}{2}^+ \rangle$ might also be a narrow state with a width of $\mathcal{O}(10)$ MeV. This state has relatively large decay rates into $\Sigma_c(2520)\pi$, $\Lambda_c(2595) \pi$, $ |\Sigma_{c}~^4P_{\lambda}\frac{1}{2}^- \rangle \pi$, and $ |\Sigma_{c}~^4P_{\lambda}\frac{5}{2}^- \rangle \pi$.
If its mass is taken as the prediction 3023 MeV in Ref.~\cite{Ebert:2011kk}, the sum of the partial widths of the pionic decays is about $\Gamma_{\mathrm{Sum}}\sim 70$ MeV (see Table~\ref{scb}),
while the ratios between the partial widths for the main decay modes, $\Sigma_c(2520)\pi$ and $\Lambda_c(2595) \pi$, and $\Gamma_{\mathrm{Sum}}$ are predicted to be
\begin{eqnarray}
\frac{\Gamma[\Sigma_c(2520)\pi,\Lambda_c(2595) \pi]}{\Gamma_{\mathrm{Sum}}}
\simeq 17\%,16\%.
\end{eqnarray}
The $\Sigma_c(2520)\pi$ and $\Lambda_c(2595) \pi$ may be ideal channels
for a search for $|\Sigma_c~^4D_{\lambda\lambda} \frac{5}{2}^+ \rangle$ in future experiments.

The radiative transitions of these $J^P=5/2^+$ states into the $1P$-wave charmed baryon states are estimated as well. Our results are listed in Table~\ref{RadSDW}. It is found that for the $|\Sigma_c~^2D_{\lambda\lambda} \frac{5}{2}^+ \rangle$ state the main
radiative decay processes are $|\Sigma_c^{++(0)}~^2D_{\lambda\lambda} \frac{5}{2}^+ \rangle\to |\Sigma_{c}^{++(0)}~ ^2P_{\lambda}\frac{1}{2}^- \rangle\gamma$, $|\Sigma_{c}^{++(0)}~ ^2P_{\lambda}\frac{3}{2}^- \rangle\gamma$, and $|\Sigma_c^{+}~^2D_{\lambda\lambda} \frac{5}{2}^+ \rangle\to \Lambda_c(2595,2625)^+\gamma$, while for the other $J^P=5/2^+$ state $|\Sigma_c~^4D_{\lambda\lambda} \frac{5}{2}^+ \rangle$, the main radiative decay processes are $|\Sigma_c^{++(0)}~^4D_{\lambda\lambda} \frac{5}{2}^+ \rangle\to |\Sigma_{c}^{++(0)}~ ^4P_{\lambda}\frac{3}{2}^- \rangle\gamma$, $|\Sigma_{c}^{++(0)}~ ^4P_{\lambda}\frac{5}{2}^- \rangle\gamma$, and $|\Sigma_c^{+}~^4D_{\lambda\lambda} \frac{5}{2}^+ \rangle\to \Lambda_c(2595,2625)^+\gamma$. The partial radiative decay widths for these processes are estimated to be $\mathcal{O}(10)$ keV, while the branching fractions may be $\mathcal{O}(10^{-4})$. These radiative transitions may be hard to observe in experiments.

\subsubsection{$J^P=7/2^+$ state}

The $J^P=7/2^+$ state $|\Sigma_c~^4D_{\lambda\lambda} \frac{7}{2}^+ \rangle$ might have a relatively narrow width of $\mathcal{O}(10)$ MeV.
It mainly decays into $\Lambda_c\pi$, $\Lambda_c(2595) \pi$,
and $\Lambda_c(2625) \pi$ channels. If one adopts the predicted mass 3013 MeV in Ref.~\cite{Ebert:2011kk}, the sum of the partial widths of the pionic decays
is estimated to be $\Gamma_{\mathrm{Sum}}\sim 60$ MeV (see Table~\ref{scb}), and the ratios between the partial decay widths for these main channels, $\Lambda_c\pi$, $\Lambda_c(2595) \pi$, and $\Lambda_c(2625) \pi$, and $\Gamma_{\mathrm{Sum}}$ are predicted to be
\begin{eqnarray}
\frac{\Gamma[\Lambda_c\pi,\Lambda_c(2595) \pi,\Lambda_c(2625) \pi]}{\Gamma_{\mathrm{Sum}}}
\simeq 22\%,19\%,41\%.
\end{eqnarray}
The $\Lambda_c\pi$, $\Lambda_c(2595) \pi$, and $\Lambda_c(2625) \pi$ may be ideal channels for a search for $|\Sigma_c~^4D_{\lambda\lambda} \frac{7}{2}^+ \rangle$ in future experiments.

The radiative transitions into the $1P$-wave states are estimated as well. Our results are listed in Table~\ref{RadSDW}.
It is found that the main radiative decay processes are $|\Sigma_c^{++(0)}~^4D_{\lambda\lambda} \frac{7}{2}^+ \rangle\to |\Sigma_{c}^{++(0)}~ ^4P_{\lambda}\frac{5}{2}^- \rangle\gamma$
and $|\Sigma_c^{+}~^4D_{\lambda\lambda} \frac{7}{2}^+ \rangle \to \Lambda_c(2595)^+\gamma $.
Their partial radiative decay widths are estimated to be $\mathcal{O}(10)$ keV,
while the branching fractions are $\mathcal{O}(10^{-4})$.
These radiative transitions may be hard to observe in experiments.

\subsection{$\Sigma_b$}

In the $\Sigma_b$ family, according to the quark model classification,
there are six $\lambda$-mode $1D$-wave excitations: $|\Sigma_b~^4D_{\lambda\lambda} \frac{1}{2}^+ \rangle$,
$|\Sigma_b~^4D_{\lambda\lambda} \frac{3}{2}^+ \rangle$, $|\Sigma_b~^2D_{\lambda\lambda} \frac{3}{2}^+ \rangle$,
$|\Sigma_b~^2D_{\lambda\lambda} \frac{5}{2}^+ \rangle$, $|\Sigma_b~^4D_{\lambda\lambda} \frac{5}{2}^+ \rangle$,
and $|\Sigma_b~^4D_{\lambda\lambda} \frac{7}{2}^+ \rangle$.
However, no $1D$-wave states have been established. The typical masses of the $\lambda$-mode
$1D$-wave $\Sigma_b$ excitations are predicted to be $\sim 6.3$ within various quark models (see Table~\ref{sp2}).
In the possible mass ranges, we study their strong decay transitions within ChQM.
Our results are shown in Fig.~\ref{sb}.
To be more specific, taking the masses of the $1D$-wave states
obtained in the relativistic quark-diquark picture~\cite{Ebert:2011kk}, we present
the results in Table~\ref{scb}.

\subsubsection{$J^P=1/2^+$ state}

The $J^P=1/2^+$ state $|\Sigma_b~^4D_{\lambda\lambda} \frac{1}{2}^+ \rangle$ might be a broad state with a width of $\mathcal{O}(100)$ MeV.
If its mass is taken as the prediction in Ref.~\cite{Ebert:2011kk}, the sum of the partial widths for the pionic decays
can reach up to $\Gamma_{\mathrm{Sum}}\sim 150$ MeV (see Table~\ref{scb}). This state might mainly decay into the $P$-wave $\Lambda_b$
states $\Lambda_b(5912)\frac{1}{2}^-$ and $\Lambda_b(5920)\frac{3}{2}^-$ via pionic decay modes $\Lambda_b(5912) \pi$
and $\Lambda_b(5920) \pi$.
The ratios between the partial decay widths for the $\Lambda_b(5912) \pi$ and $\Lambda_b(5920) \pi$ channels and $\Gamma_{\mathrm{Sum}}$ are predicted to be
\begin{eqnarray}
\frac{\Gamma[\Lambda_b(5912) \pi]}{\Gamma_{\mathrm{Sum}}}\simeq 35\%,\ \frac{\Gamma[\Lambda_b(5920)\pi]}{\Gamma_{\mathrm{Sum}}}\simeq 44\%.
\end{eqnarray}
Both $\Lambda_b(5912) \pi$ and $\Lambda_b(5920) \pi$ may be ideal channels for our search for $|\Sigma_b~^4D_{\lambda\lambda} \frac{1}{2}^+ \rangle$ in future experiments.

We also estimate its radiative transitions. Our results are listed in Table~\ref{RadSDW}.
It is found that $|\Sigma_b^{+}~^4D_{\lambda\lambda} \frac{1}{2}^+ \rangle$ might have
relatively large decay rates into $ |\Sigma_{b}^{+}~ ^4P_{\lambda}\frac{1}{2}^- \rangle\gamma$ and
$ |\Sigma_{b}^{+}~ ^4P_{\lambda}\frac{3}{2}^- \rangle\gamma$, and their partial radiative decay widths
are estimated to be $\mathcal{O}(100)$ keV. The branching fractions for these main radiative
decay processes may reach up to $\mathcal{O}(10^{-3})$.

\subsubsection{$J^P=3/2^+$ states}

The $J^P=3/2^+$ state $|\Sigma_b~^2D_{\lambda\lambda} \frac{3}{2}^+ \rangle$ has large decay rates into
the $P$-wave states through the pionic decay modes $\Lambda_b(5920)\frac{3}{2}^- \pi$ and $ |\Sigma_{b}~^2P_{\lambda}\frac{3}{2}^- \rangle \pi$. It has a width of $\mathcal{O}(10)-\mathcal{O}(100)$ MeV, which obviously depends on its mass (see Fig.~\ref{sb}). If the mass is taken as the prediction 6326 MeV in Ref.~\cite{Ebert:2011kk}, the sum of the partial widths of the pionic decays can reach up to $\Gamma_{\mathrm{Sum}}\sim 120$ MeV (see Table~\ref{scb}), and the ratio between the partial decay width for the $\Lambda_b(5920)\pi$ channel and $\Gamma_{\mathrm{Sum}}$ is predicted to be
\begin{eqnarray}
\frac{\Gamma[\Lambda_b(5920) \pi]}{\Gamma_{\mathrm{Sum}}}\simeq 58\%.
\end{eqnarray}
The $\Lambda_b(5920)\pi$ decay channel may be an ideal channel for our search for
$|\Sigma_b~^2D_{\lambda\lambda} \frac{3}{2}^+ \rangle$ in future experiments.

For the other $J^P=3/2^+$ state $|\Sigma_b~^4D_{\lambda\lambda} \frac{3}{2}^+ \rangle$,
one finds that this state mainly decays into the $P$-wave states through the pionic decay modes $\Lambda_b(5920)\pi$, $ |\Sigma_{b}~^4P_{\lambda}\frac{1}{2}^- \rangle \pi$, and $ |\Sigma_{b}~^4P_{\lambda}\frac{5}{2}^- \rangle \pi$. Furthermore, the decay rate into $\Sigma_b^*(5832)\pi$ is sizable as well.
Its width should be narrower than that of $|\Sigma_b~^2D_{\lambda\lambda} \frac{3}{2}^+ \rangle$. If its mass is taken as the predictions in Ref.~\cite{Ebert:2011kk}, the sum of the partial widths of the pionic decays can reach up to $\Gamma_{\mathrm{Sum}}\sim 85$ MeV (see Table~\ref{scb}), while the ratios between the partial widths of $\Gamma[\Sigma_b^*(5832)\pi]$ and $\Gamma[\Lambda_b(5920)\pi]$ and $\Gamma_{\mathrm{Sum}}$ are predicted to be
\begin{eqnarray}
\frac{\Gamma[\Sigma_b^*(5832)\pi]}{\Gamma_{\mathrm{Sum}}}\simeq 10\%,\ \ \frac{\Gamma[\Lambda_b(5920)\pi]}{\Gamma_{\mathrm{Sum}}}\simeq 46\%.
\end{eqnarray}
Both $\Sigma_b^*(5832)\pi$ and $\Lambda_b(5920)\pi$ may be ideal channels for a search for $|\Sigma_b~^4D_{\lambda\lambda} \frac{3}{2}^+ \rangle$ in future experiments.

The radiative decays of the $J^P=3/2^+$ states into the $1P$-wave bottom baryon states are estimated as well. Our results are listed in Table~\ref{RadSDW}. It is found that $|\Sigma_b^{+}~^2D_{\lambda\lambda} \frac{3}{2}^+ \rangle$ might have
relatively large decay rates into $ |\Sigma_{b}^{+}~ ^2P_{\lambda}\frac{1}{2}^- \rangle\gamma$ and
$ |\Sigma_{b}^{+}~ ^2P_{\lambda}\frac{3}{2}^- \rangle\gamma$, while $|\Sigma_b^{+}~^4D_{\lambda\lambda} \frac{3}{2}^+ \rangle$ might have
a relatively large decay rate into $ |\Sigma_{b}^{+}~ ^4P_{\lambda}\frac{3}{2}^- \rangle\gamma$. Their partial radiative decay widths
are estimated to be $\mathcal{O}(100)$ keV, while the branching fractions may reach up to $\mathcal{O}(10^{-3})$.

\subsubsection{$J^P=5/2^+$ states}

The $|\Sigma_b~^2D_{\lambda\lambda} \frac{5}{2}^+ \rangle$ might be a relatively narrow state with a width of $\mathcal{O}(10)$ MeV. It
has large decay rates into $\Lambda_b\pi$, $\Sigma_b\pi$, $\Lambda_b(5912) \pi$, and $\Lambda_b(5920) \pi$. If its mass is taken as the prediction 6284 MeV in Ref.~\cite{Ebert:2011kk}, the sum of the partial widths of the pionic decays is about $\Gamma_{\mathrm{Sum}}\sim 50$ MeV (see Table~\ref{scb}),
and the ratios between the partial decay widths for these main channels, $\Lambda_b\pi$, $\Sigma_b\pi$, $\Lambda_b(5912) \pi$, and $\Lambda_b(5920) \pi$, and $\Gamma_{\mathrm{Sum}}$ are predicted to be
\begin{eqnarray}
\frac{\Gamma[\Lambda_b\pi,\Sigma_b\pi,\Lambda_b(5912) \pi,\Lambda_b(5920) \pi]}{\Gamma_{\mathrm{Sum}}}\nonumber\\
\ \ \ \ \ \ \ \ \ \ \simeq 35\%,13\%,14\%,21\%.
\end{eqnarray}
The $\Lambda_b\pi$, $\Sigma_b\pi$, $\Lambda_b(5912) \pi$, and $\Lambda_b(5920) \pi$ decay channels may be ideal channels for our search for $|\Sigma_b~^2D_{\lambda\lambda} \frac{5}{2}^+ \rangle$ in future experiments.

For the other $J^P=5/2^+$ state $|\Sigma_b~^4D_{\lambda\lambda} \frac{5}{2}^+ \rangle$, it has a width of $\mathcal{O}(10)-\mathcal{O}(100)$ MeV, which significantly depends on the mass.
This state has large decay rates into $\Lambda_b\pi$, $\Sigma_b^*\pi$, $\Lambda_b(5912) \pi$, $ |\Sigma_{b}~^4P_{\lambda}\frac{1}{2}^- \rangle \pi$, and $ |\Sigma_{b}~^4P_{\lambda}\frac{5}{2}^- \rangle \pi$.
If its mass is taken as the prediction 6270 MeV in Ref.~\cite{Ebert:2011kk}, the sum of the partial widths of the pionic decays is about $\Gamma_{\mathrm{Sum}}\sim 50$ MeV (see Table~\ref{scb}), and the ratios between the partial decay widths for the $\Lambda_b\pi$, $\Sigma_b^*\pi$, and $\Lambda_b(5912) \pi$ final states and $\Gamma_{\mathrm{Sum}}$ are predicted to be
\begin{eqnarray}
\frac{\Gamma[\Lambda_b\pi,\Sigma_b^*\pi,\Lambda_b(5912) \pi]}{\Gamma_{\mathrm{Sum}}}\simeq 9\%,26\%,16\%.
\end{eqnarray}
The $\Lambda_b\pi$, $\Sigma_b^*\pi$, and $\Lambda_b(5912) \pi$ decay channels might be ideal channels for our search for $|\Sigma_b~^4D_{\lambda\lambda} \frac{5}{2}^+ \rangle$ in future experiments.

The radiative decays of these $J^P=5/2^+$ states into the $1P$-wave bottom baryon states are estimated as well. Our results are listed in Table~\ref{RadSDW}. It is found that for the $|\Sigma_b~^2D_{\lambda\lambda} \frac{5}{2}^+ \rangle$ state the main
radiative decay processes are $|\Sigma_b^{+}~^2D_{\lambda\lambda} \frac{5}{2}^+ \rangle\to |\Sigma_{b}^{+}~ ^2P_{\lambda}\frac{1}{2}^- \rangle\gamma$, $|\Sigma_{b}^{+}~ ^2P_{\lambda}\frac{1}{2}^- \rangle\gamma$, and $|\Sigma_b^{0}~^2D_{\lambda\lambda} \frac{5}{2}^+ \rangle\to \Lambda_b(5912,5920)^+\gamma$, while for the other $J^P=5/2^+$ state $|\Sigma_b~^4D_{\lambda\lambda} \frac{5}{2}^+ \rangle$, the main radiative decay processes are $|\Sigma_b^{+}~^4D_{\lambda\lambda} \frac{5}{2}^+ \rangle\to |\Sigma_{b}^{+}~ ^4P_{\lambda}\frac{3}{2}^- \rangle\gamma$, $|\Sigma_{b}^{+}~ ^4P_{\lambda}\frac{5}{2}^- \rangle\gamma$, and $|\Sigma_b^{0}~^4D_{\lambda\lambda} \frac{5}{2}^+ \rangle\to \Lambda_b(5912,5920)^+\gamma$. The partial radiative decay widths for these processes are estimated to be $\mathcal{O}(10)$ keV, while the branching fractions may be $\mathcal{O}(10^{-4})-\mathcal{O}(10^{-3})$.

\subsubsection{$J^P=7/2^+$ state}

The $J^P=7/2^+$ state $|\Sigma_b~^4D_{\lambda\lambda} \frac{7}{2}^+ \rangle$ might have a width of $\mathcal{O}(10)-\mathcal{O}(100)$ MeV, which strongly depends on its mass (see Fig.~\ref{sb}). This state has large decay rates into $\Lambda_b\pi$, $\Lambda_b(5912) \pi$,
and $\Lambda_b(5920) \pi$ channels. If one adopts the predicted mass 6260 MeV in Ref.~\cite{Ebert:2011kk}, the sum of the partial widths of the pionic decays
is estimated to be $\Gamma_{\mathrm{Sum}}\sim 50$ MeV (see Table~\ref{scb}), and the ratios between the partial decay widths for these main channels, $\Lambda_b\pi$, $\Lambda_b(5912) \pi$, and $\Lambda_b(5920) \pi$, and $\Gamma_{\mathrm{Sum}}$ are predicted to be
\begin{eqnarray}
\frac{\Gamma[\Lambda_b\pi,\Lambda_b(5912) \pi,\Lambda_b(5920) \pi]}{\Gamma_{\mathrm{Sum}}}\simeq 35\%,16\%,40\%.
\end{eqnarray}
The $\Lambda_b\pi$, $\Lambda_b(5912) \pi$, and $\Lambda_b(5920) \pi$ decay channels may be ideal channels
for our search for $|\Sigma_b~^4D_{\lambda\lambda} \frac{7}{2}^+ \rangle$ in future experiments.

We also estimate its radiative decays into the $1P$-wave bottom baryon states. Our results are listed in Table~\ref{RadSDW}.
It is found that the main radiative decay processes are $|\Sigma_b^{+}~^4D_{\lambda\lambda} \frac{7}{2}^+ \rangle\to |\Sigma_{b}^{+}~ ^4P_{\lambda}\frac{5}{2}^- \rangle\gamma$
and $|\Sigma_b^{0}~^4D_{\lambda\lambda} \frac{7}{2}^+ \rangle \to \Lambda_b(5920)^0\gamma $.
Their partial radiative decay widths are estimated to be $\mathcal{O}(10)$ keV,
while the branching fractions may be $\mathcal{O}(10^{-4})-\mathcal{O}(10^{-3})$.
The neutral $J^P=7/2^+$ state $|\Sigma_b^{0}~^4D_{\lambda\lambda} \frac{7}{2}^+ \rangle$ may have the possibility of being observed
in the $ \Lambda_b(5920)^0\gamma$ channel.

\begin{table*}[htb]
\begin{center}
\caption{ \label{Xicbp} Partial widths of strong decays for the $\lambda$-mode $D$-wave $\Xi'_c$ and  $\Xi'_b$
baryons, the masses (MeV) of which are taken from the quark model predictions of Ref.~\cite{Ebert:2011kk}. $M_f$ stands for the masses of $P$-wave
heavy baryons (MeV) in the final states, which are adopted from the RPP~\cite{Olive:2016xmw} and Ref.~\cite{Ebert:2011kk}. The superscript (subscript) stands for the uncertainty of a prediction with
a $+ 10\%$ ($-10\%$) uncertainty of the oscillator parameter $\alpha_{\rho}$.}
%\footnotesize
\begin{tabular}{cc|cccccccccccccccc}
\hline\hline
\multirow{2}{*}{Decay mode}       &\multirow{2}{*}{$M_f$}   &{$\underline{{|\Xi'_c~^2D_{\lambda\lambda} \frac{3}{2}^+ \rangle}(3167)}$}
               &{$\underline{{|\Xi'_c~^2D_{\lambda\lambda} \frac{5}{2}^+ \rangle}(3166)}$}
               &{$\underline{{|\Xi'_c~^4D_{\lambda\lambda}\frac{1}{2}^+ \rangle}(3163)}$}
               &{$\underline{{|\Xi'_c~^4D_{\lambda\lambda}\frac{3}{2}^+ \rangle}(3160)}$}
               &{$\underline{{|\Xi'_c~^4D_{\lambda\lambda}\frac{5}{2}^+ \rangle}(3153)}$}
               &{$\underline{{|\Xi'_c~^4D_{\lambda\lambda}\frac{7}{2}^+ \rangle}(3147)}$}\\
  &   & $\Gamma_i$ (MeV)    & $\Gamma_i$ (MeV)    & $\Gamma_i$ (MeV)    & $\Gamma_i$ (MeV)     & $\Gamma_i$ (MeV)    & $\Gamma_i$ (MeV)  \\ \hline
\hline
$ \Xi_{c}\pi$    &2470    &2.43$^{+2.13}_{-1.59}$     &5.13$^{-0.96}_{+1.05}$     &4.92$^{+4.18}_{-3.19}$        &2.48$^{+2.06}_{-1.60}$         &1.32$^{-0.25}_{+0.29}$      &5.68$^{-1.10}_{+1.24}$ &     \\
$ \Xi'_{c}\pi$   &2578    &3.51$^{+1.79}_{-1.66}$     &2.77$^{-0.62}_{+0.77}$     &1.75$^{+0.88}_{-0.81}$        &0.87$^{+0.44}_{-0.40}$         &0.17$^{-0.03}_{+0.05}$      &0.73$^{-0.17}_{+0.21}$ & \\
$\Xi'^*_{c}\pi$  &2645    &0.74$^{-0.09}_{+0.13}$     &1.18$^{+0.27}_{-0.26}$     &0.76$^{+0.27}_{-0.28}$        &2.94$^{+0.72}_{-0.69}$         &4.28$^{+0.34}_{-0.24}$      &0.56$^{+0.06}_{-0.04}$ & \\
$ \Lambda_{c}K$  &2286    &1.23$^{+1.13}_{-0.83}$     &1.25$^{-0.43}_{+0.53}$     &2.51$^{+2.24}_{-1.67}$        &1.27$^{+1.12}_{-0.84}$         &0.72$^{-0.13}_{+0.15}$      &3.09$^{-0.58}_{+0.63}$  &  \\
$ \Sigma_{c}K$   &2455    &6.37$^{+1.97}_{-2.07}$     &1.72$^{-0.43}_{+0.58}$     &3.16$^{+0.96}_{-1.0}$        &1.57$^{+0.46}_{-0.49}$         &0.11$^{-0.03}_{+0.04}$      &0.40$^{-0.10}_{+0.14}$  &  \\
$\Sigma^{\ast}_{c}K$  &2520 &0.47$^{-0.01}_{+0.04}$   &1.57$^{+0.25}_{-0.28}$     &1.22$^{+0.23}_{-0.26}$       &3.97$^{+0.62}_{-0.70}$      &4.18$^{+0.37}_{-0.40}$       &0.59$^{+0.08}_{-0.08}$  &  \\
$| \Lambda_{c}~^2P_{\lambda}\frac{1}{2}^- \rangle K$  &2592    &0.84$^{+0.11}_{-0.10}$    &0.55$^{+0.08}_{-0.06}$       &3.66$^{-0.30}_{+0.38}$      &3.55$^{+1.22}_{-1.06}$         &0.65$^{-0.05}_{+0.07}$      &10.1$^{+2.5}_{-2.3}$  &  \\
$| \Lambda_{c}~^2P_{\lambda}\frac{3}{2}^- \rangle K$  &2628    &10.9$^{+2.4}_{-2.1}$      &0.24$^{+0.10}_{-0.05}$       &39.1$^{+9.5}_{-8.6}$        &7.25$^{+1.53}_{-1.36}$         &0.15$^{+0.02}_{-0.02}$      &1.15$^{+0.15}_{-0.13}$ &   \\
$| \Xi_{c}~^2P_{\lambda}\frac{1}{2}^- \rangle\pi$    &2792     &1.85$^{+0.15}_{-0.13}$    &1.83$^{+0.15}_{-0.13}$       &14.5$^{-1.28}_{+2.11}$      &9.28$^{+1.53}_{-1.43}$         &3.37$^{-0.31}_{+0.36}$      &14.4$^{+4.0}_{-3.5}$  &   \\
$|\Xi_{c}~^2P_{\lambda}\frac{3}{2}^- \rangle\pi$     &2815     &47.3$^{+7.7}_{-7.3}$      &3.54$^{-0.82}_{+1.13}$       &40.2$^{+11.9}_{-10.5}$      &11.8$^{+1.7}_{-1.4}$         &0.67$^{+0.03}_{-0.02}$      &8.81$^{+0.04}_{-0.04}$&  \\
$| \Xi'_{c}~^2P_{\lambda}\frac{1}{2}^- \rangle\pi$   &2936     &0.39$^{+0.05}_{-0.05}$    &0.21$^{-0.03}_{+0.04}$       &0.26$^{-0.03}_{<+0.01}$     &0.51$^{+0.15}_{-0.14}$                        &0.03                        &0.92$^{+0.21}_{-0.19}$  &  \\
$| \Xi'_{c}~^2P_{\lambda}\frac{3}{2}^- \rangle\pi$   &2935     &8.53$^{+1.70}_{-1.50}$    &0.22$^{-0.03}_{+0.04}$       &5.50$^{+1.33}_{-1.21}$      &1.0$^{+0.13}_{-0.11}$                      &$<$0.01                     &0.33$^{+0.02}_{-0.02}$  &  \\
$| \Xi'_{c}~^4P_{\lambda}\frac{1}{2}^- \rangle\pi$   &2854     &0.23$^{-0.07}_{+0.10}$    &1.00$^{+0.32}_{-0.30}$       &0.75$^{-0.11}_{+0.14}$      &20.5$^{+2.9}_{-2.6}$          &13.0$^{+2.1}_{-2.0}$      &$<$0.01      &  \\
$| \Xi'_{c}~^4P_{\lambda}\frac{3}{2}^- \rangle\pi$   &2912     &1.01$^{+0.10}_{-0.07}$    &0.41$^{+0.04}_{-0.04}$       &1.10$^{+0.03}_{-0.02}$      &5.62$^{+0.82}_{-0.72}$         &2.85$^{+1.08}_{-0.94}$      &0.34$^{<-0.01}_{<+0.01}$ &  \\
$| \Xi'_{c}~^4P_{\lambda}\frac{5}{2}^- \rangle\pi$   &2929     &0.59$^{+0.04}_{-0.02}$    &1.46$^{+0.22}_{-0.18}$       &2.23$^{+0.64}_{-0.57}$      &5.75$^{+1.51}_{-1.35}$         &7.55$^{+1.17}_{-1.05}$      &0.37$^{<-0.01}_{+0.01}$  &  \\
Sum           &   &86.4$^{+19.1}_{-17.2}$     &23.1$^{-1.9}_{+2.8}$      &121.6$^{+30.4}_{-25.5}$        &78.4$^{+16.9}_{-14.9}$        &39.1$^{+4.3}_{-3.7}$        &47.5$^{+5.1}_{-4.1}$  &   \\

\hline\hline
\multirow{2}{*}{Decay mode}       &\multirow{2}{*}{$M_f$}  &{$\underline{{|\Xi'_b~^2D_{\lambda\lambda} \frac{3}{2}^+ \rangle}(6459)}$}
               &{$\underline{{|\Xi'_b~^2D_{\lambda\lambda} \frac{5}{2}^+ \rangle}(6432)}$}
               &{$\underline{{|\Xi'_b~^4D_{\lambda\lambda}\frac{1}{2}^+ \rangle}(6447)}$}
               &{$\underline{{|\Xi'_b~^4D_{\lambda\lambda}\frac{3}{2}^+ \rangle}(6431)}$}
               &{$\underline{{|\Xi'_b~^4D_{\lambda\lambda}\frac{5}{2}^+ \rangle}(6420)}$}
               &{$\underline{{|\Xi'_b~^4D_{\lambda\lambda}\frac{7}{2}^+ \rangle}(6414)}$}\\
&   & $\Gamma_i$ (MeV)    & $\Gamma_i$ (MeV)    & $\Gamma_i$ (MeV)    & $\Gamma_i$ (MeV)     & $\Gamma_i$ (MeV)    & $\Gamma_i$ (MeV)  \\ \hline
$ \Xi_{b} \pi$ &5795           &2.02$^{+3.07}_{-1.72}$     &9.41$^{-1.51}_{+1.44}$     &4.41$^{+6.04}_{-3.64}$       &2.43$^{+2.93}_{-1.89}$       &2.42$^{-0.40}_{+0.40}$          &10.3$^{-1.73}_{+1.81}$       \\
$ \Xi'_{b} \pi$ &5935          &4.18$^{+2.37}_{-2.13}$     &2.82$^{-0.64}_{+0.81}$     &2.06$^{+1.10}_{-0.99}$       &1.01$^{+0.48}_{-0.46}$       &0.17$^{-0.04}_{+0.05}$          &0.71$^{-0.17}_{+0.22}$   \\
$\Xi'^*_{b} \pi$  &5955        &1.60$^{-0.21}_{+0.29}$     &1.59$^{+0.39}_{-0.35}$     &1.00$^{+0.47}_{-0.45}$       &4.03$^{+1.09}_{-0.99}$       &6.16$^{+0.46}_{-0.29}$          &0.79$^{+0.08}_{-0.03}$  \\
$ \Lambda_{b} K$ &5620         &1.04$^{+1.67}_{-0.90}$     &2.45$^{-0.89}_{+0.83}$     &2.39$^{+3.36}_{-1.98}$       &1.40$^{+1.47}_{-1.19}$       &1.29$^{-0.21}_{+0.20}$          &5.45$^{-0.90}_{+0.90}$   \\
$\Sigma_{b} K$   &5811         &7.80$^{+2.16}_{-2.31}$     &0.86$^{-0.22}_{+0.33}$     &4.03$^{+0.98}_{-1.07}$       &1.83$^{+0.37}_{-0.42}$       &0.04                            &0.16$^{-0.04}_{+0.06}$     \\
$\Sigma^{\ast}_{b} K$&5835     &0.85$^{-0.04}_{+0.08}$     &1.95$^{+0.28}_{-0.32}$     &1.77$^{+0.34}_{-0.39}$       &5.11$^{+0.74}_{-0.85}$       &4.77$^{+0.41}_{-0.45}$          &0.68$^{+0.08}_{-0.09}$    \\
$| \Lambda_{b}~^2P_{\lambda}\frac{1}{2}^- \rangle K$ &5912  &0.77$^{+0.11}_{-0.10}$    &0.06                        &1.54$^{-0.10}_{+0.14}$      &3.18$^{+0.81}_{-0.74}$        &0.04               &3.65$^{+0.79}_{-0.71}$       &              &        \\
$| \Lambda_{b}~^2P_{\lambda}\frac{3}{2}^- \rangle K$ &5920  &14.6$^{+3.07}_{-2.8}$     &0.17$^{+0.10}_{-0.07}$      &30.2$^{+7.48}_{-6.65}$      &5.87$^{+1.25}_{-1.13}$        &0.06                   &$\cdot\cdot\cdot$             &        \\
$| \Xi_{b}~^2P_{\lambda}\frac{1}{2}^- \rangle \pi$ &6120    &1.90$^{+0.14}_{-0.11}$    &1.35$^{+0.12}_{-0.09}$      &13.2$^{-1.19}_{+1.28}$      &7.36$^{+1.22}_{-1.13}$        &2.15$^{-0.20}_{+0.25}$        &11.0$^{+3.0}_{-2.6}$        \\
$|\Xi_{b}~^2P_{\lambda}\frac{3}{2}^- \rangle \pi$   &6130   &54.4$^{+8.22}_{-6.92}$    &2.80$^{-0.66}_{+0.93}$      &39.3$^{+12.0}_{-10.5}$      &10.5$^{+1.47}_{-1.29}$        &0.56$^{+0.03}_{<-0.01}$        &6.79$^{+0.07}_{<-0.01}$    \\
$| \Xi'_{b}~^2P_{\lambda}\frac{1}{2}^- \rangle \pi$  &6233  &0.47$^{+0.06}_{-0.05}$    &0.10$^{-0.01}_{+0.02}$       &0.28$^{-0.03}_{+0.04}$      &0.43$^{+0.11}_{-0.11}$        &0.01              &0.65$^{+0.14}_{-0.14}$    \\
$| \Xi'_{b}~^2P_{\lambda}\frac{3}{2}^- \rangle \pi$  &6234  &9.60$^{+1.87}_{-1.66}$    &0.14$^{+0.04}_{-0.02}$      &5.33$^{+1.31}_{-1.18}$      &0.72$^{+0.10}_{-0.09}$        &0.001              &0.18$^{+0.01}_{-0.012}$     \\
$| \Xi'_{b}~^4P_{\lambda}\frac{1}{2}^- \rangle \pi$  &6227  &0.03                      &0.43$^{+0.11}_{-0.10}$      &0.13$^{-0.02}_{+0.02}$      &5.80$^{+0.94}_{-0.87}$        &3.35$^{+0.61}_{-0.56}$        &$<$0.01        \\
$| \Xi'_{b}~^4P_{\lambda}\frac{3}{2}^- \rangle \pi$ &6224   &0.89$^{+0.08}_{-0.05}$    &0.21$^{+0.03}_{-0.03}$      &0.81$^{+0.03}_{-0.01}$      &3.16$^{+0.49}_{-0.43}$        &1.80$^{+0.59}_{-0.52}$        &0.12$^{<+0.01}_{<-0.01}$     \\
$| \Xi'_{b}~^4P_{\lambda}\frac{5}{2}^- \rangle \pi$ &6226   &0.72$^{+0.03}_{<-0.01}$    &0.96$^{+0.14}_{-0.13}$      &2.21$^{+0.65}_{-0.58}$      &4.80$^{+1.23}_{-1.12}$        &5.26$^{+0.89}_{-0.76}$        &0.20$^{+0.01}_{<-0.01}$     \\
Sum                 &   &100.8$^{+22.6}_{-18.4}$    &25.3$^{-2.7}_{+3.3}$     &108.7$^{+32.4}_{-26.0}$      &57.6$^{+14.7}_{-12.7}$            &28.1$^{+2.1}_{-1.7}$             &40.7$^{+1.3}_{-4.6}$     \\
\hline\hline
\end{tabular}
\end{center}
\end{table*}

\begin{table*}[htp]
\begin{center}
\caption{\label{rXipD} Partial widths of radiative decays for the $\lambda$-mode $D$-wave $\Xi'_c$ and  $\Xi'_b$
baryons, the masses (MeV) of which are taken from the quark model predictions of Ref.~\cite{Ebert:2011kk}. $M_f$ stands for the masses of $P$-wave
heavy baryons (MeV) in the final states, which are adopted from the RPP~\cite{Olive:2016xmw} and Ref.~\cite{Ebert:2011kk}. The superscript (subscript) stands for the uncertainty of a prediction with a $+ 10\%$ ($-10\%$) uncertainty of the oscillator parameter $\alpha_{\rho}$.}
\begin{tabular}{cc|cccccccccccccccccccccccccccccccc|}\hline\hline
\multirow{2}{*}{Decay mode}       &\multirow{2}{*}{$M_f$}    &$\underline{|\Xi^\prime_c~^2D_{\lambda\lambda} \frac{3}{2}^+ \rangle(3167)}$    &$\underline{|\Xi^\prime_c~^2D_{\lambda\lambda} \frac{5}{2}^+ \rangle(3166)}$   &$\underline{|\Xi^\prime_c~^4D_{\lambda\lambda}\frac{1}{2}^+ \rangle(3163)}$    &$\underline{|\Xi^\prime_c~^4D_{\lambda\lambda}\frac{3}{2}^+ \rangle(3160)}$    &$\underline{|\Xi^\prime_c~^4D_{\lambda\lambda}\frac{5}{2}^+ \rangle(3153)}$    &$\underline{|\Xi^\prime_c~^4D_{\lambda\lambda}\frac{7}{2}^+ \rangle(3147)}$\\
  &  &$\Gamma_i$ (keV)     &$\Gamma_i$ (keV)     &$\Gamma_i$ (keV)   &$\Gamma_i$ (keV)    &$\Gamma_i$ (keV)    &$\Gamma_i$ (keV)\\ \hline
$ |\Xi_c^{+}~ ^2P_{\lambda}\frac{1}{2}^- \rangle\gamma$   &2792       &7.91$^{-0.58}_{+0.45}$        &12.4$^{-0.7}_{+0.4}$      &9.33$^{-1.03}_{+1.09}$           &34.2$^{-3.5}_{+3.5}$            &12.4$^{-1.4}_{+1.4}$           &0.05              \\
$ |\Xi_c^{0}~ ^2P_{\lambda}\frac{1}{2}^- \rangle\gamma$   &           &0.0          &0.0          &0.00           &0.00         &0.00            &0.0                \\
$ |\Xi_c^{+}~ ^2P_{\lambda}\frac{3}{2}^- \rangle\gamma$   &2815       &24.4$^{-2.9}_{+3.2}$        &17.5$^{-1.9}_{+2.0}$         &7.49$^{-0.96}_{+1.12}$           &2.41$^{-0.36}_{+0.46}$          &22.0$^{-2.4}_{+2.4}$           &19.5$^{-1.9}_{+1.9}$            \\
$ |\Xi_c^{0}~ ^2P_{\lambda}\frac{3}{2}^- \rangle\gamma$   &           &0.0          &0.0          &0.00           &0.00         &0.00            &0.0               \\

$|\Xi'^{+}_c~ ^2P_{\lambda}\frac{1}{2}^- \rangle\gamma$   &2936       &0.02                       &0.32$^{-0.04}_{+0.05}$       &0.49$^{-0.08}_{+0.10}$    &0.47$^{-0.08}_{+0.10}$     &0.16$^{-0.03}_{+0.03}$            &$<0.01$                        \\
$|\Xi'^{0}_c~ ^2P_{\lambda}\frac{1}{2}^- \rangle\gamma$   &           &64.2$^{-10.3}_{+13.6}$     &6.38$^{-1.07}_{+1.41}$       &0.03                      &0.30$^{-0.04}_{+0.05}$     &0.10$^{-0.01}_{+0.02}$            &$<0.01$                 \\

$|\Xi'^{+}_c~ ^2P_{\lambda}\frac{3}{2}^- \rangle\gamma$   &2935       &0.45$^{-0.05}_{+0.06}$     &1.98$^{-0.35}_{+0.48}$       &0.13$^{-0.02}_{+0.03}$    &0.04                       &0.40$^{-0.06}_{+0.08}$            &0.35$^{-0.05}_{+0.07}$                      \\
$|\Xi'^{0}_c~ ^2P_{\lambda}\frac{3}{2}^- \rangle\gamma$   &           &29.2$^{-4.8}_{+6.4}$       &50.1$^{-8.5}_{+11.2}$        &0.08                      &0.03                       &0.25$^{-0.03}_{+0.04}$            &0.22$^{-0.03}_{+0.03}$                 \\

$|\Xi'^{+}_c~ ^4P_{\lambda}\frac{1}{2}^- \rangle\gamma$   & 2854      &1.36$^{-0.16}_{+0.18}$     &0.08                         &$<0.01$                   &0.54$^{-0.10}_{+0.16}$                  &0.05        &$<0.01$                       \\
$|\Xi'^{0}_c~ ^4P_{\lambda}\frac{1}{2}^- \rangle\gamma$   &           &0.69$^{-0.04}_{+0.02}$     &0.04                         &184.92$^{-27.2}_{+33.9}$  &74.7$^{-11.3}_{+14.2}$     &13.1$^{-2.0}_{+2.6}$           &2.71$^{-0.57}_{+0.85}$                     \\

$|\Xi'^{+}_c~ ^4P_{\lambda}\frac{3}{2}^- \rangle\gamma$   & 2912      &0.90$^{-0.13}_{+0.16}$     &0.40$^{-0.06}_{+0.06}$       &0.07                       &0.04                       &0.73$^{-0.13}_{+0.18}$           &0.02                     \\
$|\Xi'^{0}_c~ ^4P_{\lambda}\frac{3}{2}^- \rangle\gamma$   &           &0.54$^{-0.16}_{+0.08}$     &0.23$^{-0.03}_{+0.02}$       &42.00$^{-6.7}_{+8.7}$     &66.24$^{-10.5}_{+13.7}$     &43.1$^{-7.1}_{+9.3}$           &7.26$^{-1.19}_{+1.57}$                    \\

$|\Xi'^{+}_c~ ^4P_{\lambda}\frac{5}{2}^- \rangle\gamma$   & 2929      &0.19$^{-0.03}_{+0.03}$     &0.98$^{-0.15}_{+0.19}$       &0.02                       &0.04                                    &0.05                           &1.25$^{-0.22}_{+0.29}$                      \\
$|\Xi'^{0}_c~ ^4P_{\lambda}\frac{5}{2}^- \rangle\gamma$   &           &0.12$^{-0.02}_{+0.02}$     &0.61$^{-0.08}_{+0.10}$       &5.15$^{-0.97}_{+1.36}$     &14.00$^{-2.4}_{+3.2}$        &31.0$^{-5.1}_{+6.7}$           &33.7$^{-5.7}_{+7.7}$                    \\

\hline\hline
\multirow{2}{*}{Decay mode}       &\multirow{2}{*}{$M_f$}     &$\underline{|\Xi^\prime_b~^2D_{\lambda\lambda} \frac{3}{2}^+ \rangle(6459)}$    &$\underline{|\Xi^\prime_b~^2D_{\lambda\lambda} \frac{5}{2}^+ \rangle(6432)}$   &$\underline{|\Xi^\prime_b~^4D_{\lambda\lambda}\frac{1}{2}^+ \rangle(6447)}$    &$\underline{|\Xi^\prime_b~^4D_{\lambda\lambda}\frac{3}{2}^+ \rangle(6431)}$    &$\underline{|\Xi^\prime_b~^4D_{\lambda\lambda}\frac{5}{2}^+ \rangle(6420)}$    &$\underline{|\Xi^\prime_b~^4D_{\lambda\lambda}\frac{7}{2}^+ \rangle(6414)}$\\
   &  &$\Gamma_i$ (keV)     &$\Gamma_i$ (keV)     &$\Gamma_i$ (keV)   &$\Gamma_i$ (keV)    &$\Gamma_i$ (keV)    &$\Gamma_i$ (keV)\\ \hline
$ |\Xi_b^{0}~ ^2P_{\lambda}\frac{1}{2}^- \rangle\gamma$  &6120      &5.16$^{-0.43}_{+0.38}$       &10.00$^{-0.55}_{+0.25}$       &9.66$^{-1.02}_{+1.04}$      &29.7$^{-3.0}_{+3.0}$              &10.0$^{-1.1}_{+1.2}$             &0.05             \\
$ |\Xi_b^{-}~ ^2P_{\lambda}\frac{1}{2}^- \rangle\gamma$  &          &0.0         &0.0         &0.00       &0.0                 &0.0               &0.0                  \\
$ |\Xi_b^{0}~ ^2P_{\lambda}\frac{3}{2}^- \rangle\gamma$  &6130      &18.2$^{-2.2}_{+2.6}$       &16.7$^{-1.7}_{+1.8}$       &9.33$^{-1.15}_{+1.30}$      &2.60$^{-0.40}_{+0.51}$                &20.5$^{-2.1}_{+2.1}$             &17.7$^{-1.6}_{+1.6}$                      \\
$ |\Xi_b^{-}~ ^2P_{\lambda}\frac{3}{2}^- \rangle\gamma$  &          &0.0         &0.0         &0.00       &0.0                 &0.0               &0.0                           \\

$|\Xi'^{0}_b~ ^2P_{\lambda}\frac{1}{2}^- \rangle\gamma$  &6233      &14.9$^{-2.4}_{+2.9}$       &1.21$^{-0.20}_{+0.27}$      &0.07                         &0.16$^{-0.03}_{+0.02}$                  &0.05                                &$<0.01$                    \\
$|\Xi'^{-}_b~ ^2P_{\lambda}\frac{1}{2}^- \rangle\gamma$  &          &37.5$^{-5.6}_{+7.1}$       &3.72$^{-0.61}_{+0.79}$      &0.52$^{-0.08}_{+0.09}$       &0.68$^{-0.10}_{+0.12}$                &0.20$^{-0.03}_{+0.04}$              &$<0.01$                  \\

$|\Xi'^{0}_b~ ^2P_{\lambda}\frac{3}{2}^- \rangle\gamma$  &6234      &7.73$^{-1.24}_{+1.62}$     &6.88$^{-1.16}_{+1.54}$      &0.06                         &0.01                              &0.11$^{-0.02}_{+0.02}$              &0.09                     \\
$|\Xi'^{-}_b~ ^2P_{\lambda}\frac{3}{2}^- \rangle\gamma$  &          &22.4$^{-3.6}_{+4.5}$       &14.3$^{-2.4}_{+3.1}$        &0.25$^{-0.04}_{+0.05}$       &0.05                              &0.48$^{-0.06}_{+0.09}$              &0.41$^{-0.06}_{+0.07}$                    \\

$|\Xi'^{0}_b~ ^4P_{\lambda}\frac{1}{2}^- \rangle\gamma$  &6227      &0.21$^{-0.03}_{+0.02}$     &$<0.01$                     &16.7$^{-2.6}_{+3.2}$         &5.61$^{-0.89}_{+1.15}$                &0.81$^{-0.13}_{+0.17}$              &0.17$^{-0.04}_{+0.05}$                     \\
$|\Xi'^{-}_b~ ^4P_{\lambda}\frac{1}{2}^- \rangle\gamma$  &          &0.92$^{-0.11}_{+0.10}$     &0.03                        &43.9$^{-6.5}_{+8.1}$         &14.3$^{-2.2}_{+2.8}$                &2.32$^{-0.38}_{+0.48}$             &0.43$^{-0.10}_{+0.13}$                  \\

$|\Xi'^{0}_b~ ^4P_{\lambda}\frac{3}{2}^- \rangle\gamma$  &6224      &0.35$^{-0.05}_{+0.05}$     &0.09                        &7.36$^{-1.15}_{+1.49}$       &9.32$^{-1.48}_{+1.90}$                &5.58$^{-0.91}_{+1.21}$              &0.92$^{-0.15}_{+0.20}$                     \\
$|\Xi'^{-}_b~ ^4P_{\lambda}\frac{3}{2}^- \rangle\gamma$  &          &1.54$^{-0.20}_{+0.23}$     &0.39$^{-0.05}_{+0.06}$      &20.8$^{-3.2}_{+4.0}$         &24.4$^{-3.8}_{+4.8}$                &12.9$^{-2.1}_{+2.7}$            &2.56$^{-0.41}_{+0.55}$                      \\

$|\Xi'^{0}_b~ ^4P_{\lambda}\frac{5}{2}^- \rangle\gamma$  &6226      &0.10$^{-0.02}_{+0.02}$     &0.30$^{-0.05}_{+0.05}$      &1.14$^{-0.22}_{+0.32}$       &2.44$^{-0.42}_{+0.55}$                &4.70$^{-0.77}_{+1.01}$              &5.36$^{-0.91}_{+1.21}$                    \\
$|\Xi'^{-}_b~ ^4P_{\lambda}\frac{5}{2}^- \rangle\gamma$  &          &0.43$^{-0.06}_{+0.08}$     &1.31$^{-0.19}_{+0.24}$      &3.06$^{-0.61}_{+0.87}$       &6.76$^{-1.13}_{+1.50}$                &12.2$^{-2.0}_{+2.5}$              &10.0$^{-1.7}_{+2.3}$                \\
\hline\hline
\end{tabular}
\end{center}
\end{table*}

\subsection{$\Xi'_c$}

In the $\Xi'_c$ family, according to the quark model classification,
there are six $\lambda$-mode $1D$-wave excitations: $|\Xi'_c~^4D_{\lambda\lambda} \frac{1}{2}^+ \rangle$,
$|\Xi'_c~^4D_{\lambda\lambda} \frac{3}{2}^+ \rangle$, $|\Xi'_c~^2D_{\lambda\lambda} \frac{3}{2}^+ \rangle$,
$|\Xi'_c~^2D_{\lambda\lambda} \frac{5}{2}^+ \rangle$, $|\Xi'_c~^4D_{\lambda\lambda} \frac{5}{2}^+ \rangle$,
and $|\Xi'_c~^4D_{\lambda\lambda} \frac{7}{2}^+ \rangle$.
However, no $1D$-wave states have been established. The typical masses of the $\lambda$-mode
$1D$-wave $\Xi'_c$ excitations are $\sim 3.14$ within various quark model predictions (see Table~\ref{sp2}).

In Ref.~\cite{Liu:2012sj}, the strong decay properties of the $D$-wave
excited $\Xi'_c$ states were studied in their possible mass ranges.
However, we did not give correct predictions of the partial widths
of the $D$-wave excited $\Xi'_c$ states decaying into the
$P$-wave charmed baryons. In this work, we update our predictions,
which have been shown in Fig.~\ref{Xipc}.  To be more specific, taking the masses of the $1D$-wave states predicted within
the relativistic quark-diquark picture~\cite{Ebert:2011kk}, we give our predictions
in Table~\ref{Xicbp}.

\subsubsection{$J^P=1/2^+$ state}

If the mass of the $J^P=1/2^+$ state $|\Xi'_c~^4D_{\lambda\lambda} \frac{1}{2}^+ \rangle$ is less than 3.12 GeV,
this state should be a rather narrow state with a width of $\sim 10$ MeV; its main decay modes are
$\Xi_c\pi$, $\Lambda_cK$, and $\Sigma_cK$.
However, if the mass of $|\Xi'_c~^4D_{\lambda\lambda} \frac{1}{2}^+ \rangle$ is taken as the prediction 3163 MeV
in Ref.~\cite{Ebert:2011kk}, the decay channels into the $P$-wave charmed baryon final states should open, and the
decay channels $\Lambda_c(2625)K$ and $\Xi_c(2815)\pi$ will play dominant roles.
In this case, the $J^P=1/2^+$ state $|\Xi'_c~^4D_{\lambda\lambda} \frac{1}{2}^+ \rangle$ might be a broad state,
and the sum of the partial widths for the pionic and kaonic decays
can reach up to $\Gamma_{\mathrm{Sum}}\sim 120$ MeV (see Table~\ref{Xicbp}), while
the ratios between the partial widths for the $\Lambda_c(2625)K$ and $\Xi_c(2815)\pi$ channels
and $\Gamma_{\mathrm{Sum}}$ are predicted to be
\begin{eqnarray}
\frac{\Gamma[\Lambda_c(2625)K]}{\Gamma_{\mathrm{Sum}}}\simeq 0.33,\ \ \frac{\Gamma[\Xi_c(2815)\pi]}{\Gamma_{\mathrm{Sum}}}\simeq 0.33.
\end{eqnarray}
The decay channels $\Lambda_c(2625)K$ and $\Xi_c(2815)\pi$ may be ideal channels
for our search for $|\Xi'_c~^4D_{\lambda\lambda} \frac{1}{2}^+ \rangle$ in future experiments.

We also estimate its radiative decays into the $1P$-wave bottom baryon states. Our results are listed in Table~\ref{rXipD}.
It is found that $|\Xi_c'^{0}~^4D_{\lambda\lambda} \frac{1}{2}^+ \rangle$ might have
relatively large decay rates into $|\Xi_c'^{0}~ ^4P_{\lambda}\frac{1}{2}^- \rangle\gamma$, and the partial radiative decay width
is estimated to be $\mathcal{O}(100)$ keV. The branching fractions for this main radiative
decay process may reach up to $\mathcal{O}(10^{-3})$.

\subsubsection{$J^P=3/2^+$ states}

For $|\Xi'_c~^2D_{\lambda\lambda} \frac{3}{2}^+ \rangle$, if its mass is less than 3.12 GeV,
it should be a rather narrow state with a width of $\sim 10$ MeV, and its main decay modes are
$\Xi_c\pi$, $\Xi'_c\pi$, and $\Sigma_cK$. However, if the mass of $|\Xi'_c~^2D_{\lambda\lambda} \frac{3}{2}^+ \rangle$ is taken as the prediction 3167 MeV in Ref.~\cite{Ebert:2011kk}, the decay channels into the $P$-wave charmed baryon final states should open, and the
decay channels $\Lambda_c(2625)K$ and $\Xi_c(2815)\pi$ will play dominant roles.
In this case, the $J^P=3/2^+$ state $|\Xi'_c~^2D_{\lambda\lambda} \frac{3}{2}^+ \rangle$ might be a broad state,
and the sum of the partial widths for the pionic and kaonic decays
can reach up to $\Gamma_{\mathrm{Sum}}\sim 90$ MeV (see Table~\ref{Xicbp}), while
the ratios between the partial widths for the $\Lambda_c(2625)K$ and $\Xi_c(2815)\pi$ channels
and $\Gamma_{\mathrm{Sum}}$ are predicted to be
\begin{eqnarray}
\frac{\Gamma[\Lambda_c(2625)K]}{\Gamma_{\mathrm{Sum}}}\simeq 12\%,\ \ \frac{\Gamma[\Xi_c(2815)\pi]}{\Gamma_{\mathrm{Sum}}}\simeq 54\%.
\end{eqnarray}
The $\Lambda_c(2625)K$ and $\Xi_c(2815)\pi$ decay channels may be ideal channels for our search for $|\Xi'_c~^2D_{\lambda\lambda} \frac{3}{2}^+ \rangle$ in future experiments.

For the other $J^P=3/2^+$ state $|\Xi'_c~^4D_{\lambda\lambda} \frac{3}{2}^+ \rangle$, if its mass is less than 3.12 GeV,
it should be a rather narrow state with a width of $\sim 10$ MeV as well, and its main decay modes are
$\Xi_c\pi$, $\Xi_c'^*\pi$, and $\Sigma_c^*K$. However, if the mass of $|\Xi'_c~^4D_{\lambda\lambda} \frac{3}{2}^+ \rangle$ is taken as the prediction 3160 MeV in Ref.~\cite{Ebert:2011kk}, the decay channels into the $P$-wave charmed baryon final states should open, and the
decay channels $\Lambda_c(2625)K$, $\Xi_c(2790)\pi$, $\Xi_c(2815)\pi$, and $| \Xi'_{c}~^4P_{\lambda}\frac{1}{2}^- \rangle\pi$ will dominate the decays of $|\Xi'_c~^4D_{\lambda\lambda} \frac{3}{2}^+ \rangle$.
In this case, the sum of the partial widths for the pionic and kaonic decays
can reach up to $\Gamma_{\mathrm{Sum}}\sim 80$ MeV (see Table~\ref{Xicbp}), while
the ratios between the partial widths for the $\Lambda_c(2625)K$, $\Xi_c(2790)\pi$, and $\Xi_c(2815)\pi$ channels
and $\Gamma_{\mathrm{Sum}}$ are predicted to be
\begin{eqnarray}
\frac{\Gamma[\Lambda_c(2625)K,\Xi_c(2790)\pi,\Xi_c(2815)\pi]}{\Gamma_{\mathrm{Sum}}}\simeq 10\%,12\%,16\%.
\end{eqnarray}
The $\Lambda_c(2625)K$, $\Xi_c(2790)\pi$, and $\Xi_c(2815)\pi$ decay channels may be ideal channels for a search for $|\Xi'_c~^4D_{\lambda\lambda} \frac{3}{2}^+ \rangle$ in future experiments.

The radiative decays of these $J^P=3/2^+$ states into the $1P$-wave charmed baryon states
are also estimated. Our results are listed in Table~\ref{rXipD}.
It is found that for the $|\Xi'_c~^2D_{\lambda\lambda} \frac{3}{2}^+ \rangle$ state the main radiative decay processes are $|\Xi_c'^{+}~^2D_{\lambda\lambda} \frac{3}{2}^+ \rangle\to \Xi_c^+(2815)\gamma$,
$|\Xi_c'^{0}~^2D_{\lambda\lambda} \frac{3}{2}^+ \rangle\to |\Xi'^{0}_c~ ^2P_{\lambda}\frac{1}{2}^- \rangle\gamma,|\Xi'^{0}_c~ ^2P_{\lambda}\frac{3}{2}^- \rangle\gamma$, while for the $|\Xi'_c~^4D_{\lambda\lambda} \frac{3}{2}^+ \rangle$ state, the main radiative decay processes are $|\Xi_c'^{+}~^4D_{\lambda\lambda} \frac{3}{2}^+ \rangle\to \Xi_c^+(2790)\gamma$, $|\Xi_c'^{0}~^4D_{\lambda\lambda} \frac{3}{2}^+ \rangle\to |\Xi'^{0}_c~ ^4P_{\lambda}\frac{1}{2}^- \rangle\gamma,|\Xi'^{0}_c~ ^4P_{\lambda}\frac{3}{2}^- \rangle\gamma$. Their partial radiative decay widths are estimated to be $\mathcal{O}(10)$ keV, while the branching fractions may reach up to $\mathcal{O}(10^{-4})-\mathcal{O}(10^{-3})$.

\subsubsection{$J^P=5/2^+$ states}

For $|\Xi'_c~^2D_{\lambda\lambda} \frac{5}{2}^+ \rangle$, if its mass is less than 3.12 GeV,
it should be a rather narrow state with a width of $\sim 10$ MeV, and its main decay modes are
$\Xi_c\pi$, $\Xi'_c\pi$. However, if the mass of $|\Xi'_c~^2D_{\lambda\lambda} \frac{3}{2}^+ \rangle$ is taken as the prediction 3166 MeV in Ref.~\cite{Ebert:2011kk}, the decay channels into the $P$-wave charmed baryon final states should open, and the $\Xi_c(2815)\pi$ decay mode together with $\Xi_c\pi$ and $\Xi'_c\pi$ dominates its decays.
In this case, the width of $|\Xi'_c~^2D_{\lambda\lambda} \frac{5}{2}^+ \rangle$ might be still fairly narrow,
and the sum of the partial widths for the pionic and kaonic decays
is $\Gamma_{\mathrm{Sum}}\sim 20$ MeV (see Table~\ref{Xicbp}), while
the ratios between the partial widths for the $\Xi_c\pi$, $\Xi'_c\pi$ and $\Xi_c(2815)\pi$ channels
and $\Gamma_{\mathrm{Sum}}$ are predicted to be
\begin{eqnarray}
\frac{\Gamma[\Xi_c\pi,\Xi'_c\pi,\Xi_c(2815)\pi]}{\Gamma_{\mathrm{Sum}}}\simeq 22\%,12\%,15\%.
\end{eqnarray}
The $\Xi_c\pi$, $\Xi'_c\pi$, and $\Xi_c(2815)\pi$ decay modes may be ideal modes for our search for $|\Xi'_c~^2D_{\lambda\lambda} \frac{5}{2}^+ \rangle$ in future experiments.

For the other $J^P=5/2^+$ state $|\Xi'_c~^4D_{\lambda\lambda} \frac{5}{2}^+ \rangle$, if its mass is less than 3.12 GeV,
it should be a rather narrow states with a width of $\sim 10$ MeV as well, its main decay modes are
$\Xi_c'^*\pi$ and $\Sigma_c^*K$. However, if the mass of $|\Xi'_c~^4D_{\lambda\lambda} \frac{5}{2}^+ \rangle$ is taken as the prediction 3153 MeV in Ref.~\cite{Ebert:2011kk}, the decay channels into the $P$-wave charmed baryon final states should open, and the $\Xi_c(2790)\pi$, $| \Xi'_{c}~^4P_{\lambda}\frac{1}{2}^- \rangle\pi$, and $| \Xi'_{c}~^4P_{\lambda}\frac{5}{2}^- \rangle\pi$ decay modes together with $\Xi_c'^*\pi$ and $\Sigma_c^*K$ will dominate the decays of $|\Xi'_c~^4D_{\lambda\lambda} \frac{5}{2}^+ \rangle$.
In this case, the sum of the partial widths for the pionic and kaonic decays
can reach up to $\Gamma_{\mathrm{Sum}}\sim 40$ MeV (see Table~\ref{Xicbp}), while
the ratios between the partial widths for the $\Xi_c'^*\pi$, $\Sigma_c^*K$, and $\Xi_c(2790)\pi$ channels
and $\Gamma_{\mathrm{Sum}}$ are predicted to be
\begin{eqnarray}
\frac{\Gamma[\Xi_c'^*\pi,\Sigma_c^*K,\Xi_c(2790)\pi]}{\Gamma_{\mathrm{Sum}}}\simeq 11\%,11\%,9\%.
\end{eqnarray}
The decay modes $\Xi_c'^*\pi$, $\Sigma_c^*K$, and $\Xi_c(2790)\pi$ may be ideal modes for our search for $|\Xi'_c~^4D_{\lambda\lambda} \frac{5}{2}^+ \rangle$ in future experiments.

The radiative decays of these $J^P=5/2^+$ states into the $1P$-wave charmed baryon states
are also estimated. Our results are listed in Table~\ref{rXipD}.
It is found that for the $|\Xi'_c~^2D_{\lambda\lambda} \frac{5}{2}^+ \rangle$ state the main radiative decay processes are $|\Xi_c'^{+}~^2D_{\lambda\lambda} \frac{5}{2}^+ \rangle\to \Xi_c^+(2790,2815)\gamma$,
$|\Xi_c'^{0}~^2D_{\lambda\lambda} \frac{5}{2}^+ \rangle\to |\Xi'^{0}_c~ ^2P_{\lambda}\frac{3}{2}^- \rangle\gamma$, while for the $|\Xi'_c~^4D_{\lambda\lambda} \frac{5}{2}^+ \rangle$ state, the main radiative decay processes are $|\Xi_c'^{+}~^4D_{\lambda\lambda} \frac{5}{2}^+ \rangle\to \Xi_c^+(2790,2815)\gamma$, $|\Xi_c'^{0}~^4D_{\lambda\lambda} \frac{3}{2}^+ \rangle\to |\Xi'^{0}_c~ ^4P_{\lambda}\frac{1}{2}^- \rangle\gamma,|\Xi'^{0}_c~ ^4P_{\lambda}\frac{3}{2}^- \rangle\gamma,|\Xi'^{0}_c~ ^4P_{\lambda}\frac{5}{2}^- \rangle\gamma$. Their partial radiative decay widths are estimated to be $\mathcal{O}(10)$ keV, while the branching fractions may reach up to $\mathcal{O}(10^{-4})-\mathcal{O}(10^{-3})$.

\subsubsection{$J^P=7/2^+$ state}

If the mass of the $J^P=7/2^+$ state $|\Xi'_c~^4D_{\lambda\lambda} \frac{7}{2}^+ \rangle$ is less than 3.12 GeV, it should be a rather narrow state with a width of a few MeV, and its main decay modes are $\Xi_c\pi$ and $\Lambda_cK$.
However, if the mass of $|\Xi'_c~^4D_{\lambda\lambda} \frac{7}{2}^+ \rangle$ is taken to be 3147 MeV as predicted in Ref.~\cite{Ebert:2011kk}, the decay channels into the $P$-wave charmed baryon final states, $\Lambda_c(2595) K$, $\Xi_c(2790)\pi$, and $\Xi_c(2815)\pi$,
become dominant. In this case, the sum of the partial widths for the pionic and kaonic decays
can reach up to $\Gamma_{\mathrm{Sum}}\sim 50$ MeV (see Table~\ref{Xicbp}), while
the ratios between the partial widths for the main channels and $\Gamma_{\mathrm{Sum}}$ are predicted to be
\begin{eqnarray}
\frac{\Gamma[\Xi_c\pi,\Lambda_c(2595) K,\Xi_c(2790)\pi,\Xi_c(2815)\pi]}{\Gamma_{\mathrm{Sum}}}\nonumber\\
\ \ \ \ \ \simeq 12\%,20\%,30\%,19\%.
\end{eqnarray}
The $\Xi_c\pi$, $\Lambda_cK$, $\Lambda_c(2595) K$, $\Xi_c(2790)\pi$, and $\Xi_c(2815)\pi$ decay modes may be ideal modes for our search for $|\Xi'_c~^4D_{\lambda\lambda} \frac{7}{2}^+ \rangle$ in future experiments.

The radiative decays of $|\Xi'_c~^4D_{\lambda\lambda} \frac{7}{2}^+ \rangle$ into the $1P$-wave charmed baryon states
are also estimated. Our results are listed in Table~\ref{rXipD}.
It is found that the main radiative decay processes are $|\Xi_c'^{+}~^2D_{\lambda\lambda} \frac{7}{2}^+ \rangle\to \Xi_c^+(2815)\gamma$,
$|\Xi_c'^{0}~^2D_{\lambda\lambda} \frac{7}{2}^+ \rangle\to |\Xi'^{0}_c~ ^2P_{\lambda}\frac{5}{2}^- \rangle\gamma$. Their partial radiative decay widths are estimated to be $\mathcal{O}(10)$ keV, while the branching fractions may be $\mathcal{O}(10^{-4})$.

\subsection{$\Xi'_b$}

In the $\Xi'_b$ family, there are six $\lambda$-mode $1D$-wave excitations: $|\Xi'_b~^4D_{\lambda\lambda} \frac{1}{2}^+ \rangle$,
$|\Xi'_b~^4D_{\lambda\lambda} \frac{3}{2}^+ \rangle$, $|\Xi'_b~^2D_{\lambda\lambda} \frac{3}{2}^+ \rangle$,
$|\Xi'_b~^2D_{\lambda\lambda} \frac{5}{2}^+ \rangle$, $|\Xi'_b~^4D_{\lambda\lambda} \frac{5}{2}^+ \rangle$
and $|\Xi'_b~^4D_{\lambda\lambda} \frac{7}{2}^+ \rangle$.
However, no $1D$-wave states have been established. The typical masses of the $\lambda$-mode
$1D$-wave $\Xi'_b$ excitations are $\sim 6.44$ GeV within various quark model predictions (see Table~\ref{sp2}).
In the possible mass ranges, we study their
strong decay transitions by emitting one light pseudoscalar meson within the ChQM.
Our results are shown in Fig.~\ref{Xipb}.
To be more specific, taking the masses of the $1D$-wave states
obtained in the relativistic quark-diquark picture~\cite{Ebert:2011kk}, we give the predicted
widths in Table~\ref{Xicbp}.

\subsubsection{$J^P=1/2^+$ state}

The $J^P=1/2^+$ state $|\Xi'_b~^4D_{\lambda\lambda} \frac{1}{2}^+ \rangle$ might be a broad
state with a width of $\mathcal{O}(100)$ MeV, which obviously depends on its mass. This state might mainly decay into the $P$-wave bottom baryons
via pionic decay modes $|\Xi_{b}~^2P_{\lambda}\frac{1}{2}^- \rangle \pi$ and $|\Xi_{b}~^2P_{\lambda}\frac{3}{2}^- \rangle \pi$ and kaonic decay mode $\Lambda_b(5920)K$. If its mass is taken as the prediction 6447 MeV in Ref.~\cite{Ebert:2011kk}, the sum of the partial widths for the pionic and kaonic decays can reach up to $\Gamma_{\mathrm{Sum}}\sim 110$ MeV (see Table~\ref{scb}), while
the ratio between the partial width for the $\Lambda_b(5920)K$ channel
and $\Gamma_{\mathrm{Sum}}$ is predicted to be
\begin{eqnarray}
\frac{\Gamma[\Lambda_b(5920)K]}{\Gamma_{\mathrm{Sum}}}\simeq 27\%.
\end{eqnarray}
The $\Lambda_b(5920)K$ decay channel may be an ideal channel for our search for $|\Xi'_b~^4D_{\lambda\lambda} \frac{1}{2}^+ \rangle$ in future experiments.

We also estimate its radiative decays into the $1P$-wave bottom baryon states. Our results are listed in Table~\ref{rXipD}.
It is found that $|\Xi_b'^{-}~^4D_{\lambda\lambda} \frac{1}{2}^+ \rangle$ might have
relatively large decay rates into $|\Xi_b'^{-}~ ^4P_{\lambda}\frac{1}{2}^- \rangle\gamma$
and $|\Xi_b'^{-}~ ^4P_{\lambda}\frac{3}{2}^- \rangle\gamma$, and the partial radiative decay width
is estimated to be $\mathcal{O}(10)$ keV. The branching fractions for these main radiative
decay processes may be $\mathcal{O}(10^{-4})$.

\subsubsection{$J^P=3/2^+$ states}

The $J^P=3/2^+$ state $|\Xi'_b~^2D_{\lambda\lambda} \frac{3}{2}^+ \rangle$ has a width of $O(10)-O(100)$ MeV, which significantly depends on the mass. It might mainly decay into the $P$-wave bottom baryons via the $|\Xi_{b}~^2P_{\lambda}\frac{3}{2}^- \rangle \pi$ and $\Lambda_b(5920)K$ decay modes. The decay rate into $\Sigma_bK$ is also sizeable. If its mass is taken as the prediction in Ref.~\cite{Ebert:2011kk},  the sum of the partial widths for the pionic and kaonic decays can reach up to $\Gamma_{\mathrm{Sum}}\sim 100$ MeV (see Table~\ref{Xicbp}), while
the ratios between the partial widths for the $\Sigma_bK$ and $\Lambda_b(5920)K$ channels
and $\Gamma_{\mathrm{Sum}}$ are predicted to be
\begin{eqnarray}
\frac{\Gamma[\Sigma_bK]}{\Gamma_{\mathrm{Sum}}}\simeq 7\%,\ \ \frac{\Gamma[\Lambda_b(5920)K]}{\Gamma_{\mathrm{Sum}}}\simeq 14\%.
\end{eqnarray}
The $\Sigma_bK$ and $\Lambda_b(5920)K$ decay modes may be ideal modes for our search for $|\Xi'_b~^2D_{\lambda\lambda} \frac{3}{2}^+ \rangle$ in future experiments.

For the other $J^P=3/2^+$ state $|\Xi'_b~^4D_{\lambda\lambda} \frac{3}{2}^+ \rangle$,
one finds that it has large decay rates into the $P$-wave states through the decay modes $\Lambda_b(5912)K$, $\Lambda_b(5920)K$, $|\Xi_{b}~^2P_{\lambda}\frac{1}{2}^- \rangle \pi$, $|\Xi_{b}~^2P_{\lambda}\frac{3}{2}^- \rangle \pi$, $|\Xi_{b}'~^4P_{\lambda}\frac{1}{2}^- \rangle \pi$, and $|\Xi_{b}'~^4P_{\lambda}\frac{5}{2}^- \rangle \pi$. Furthermore, the decay rates into $\Xi_b'^*\pi$ and $\Sigma_b^*K$ are sizable as well.
Its width should be about a factor of 2 narrower than that of $|\Xi'_b~^2D_{\lambda\lambda} \frac{3}{2}^+ \rangle$. If its mass is taken as the predictions in Ref.~\cite{Ebert:2011kk}, the sum of the partial widths of the poinic and kaonic decays can reach up to $\Gamma_{\mathrm{Sum}}\sim 60$ MeV (see Table~\ref{scb}), while the ratios between the partial widths for the $\Xi_b'^*\pi$, $\Sigma_b^*K$, $\Lambda_b(5912)K$, and $\Lambda_b(5920)K$ decay modes and $\Gamma_{\mathrm{Sum}}$ are predicted to be
\begin{eqnarray}
 \frac{\Gamma[\Xi_b'^*\pi,\Sigma_b^*K,\Lambda_b(5912)K,\Lambda_b(5920)K]}{\Gamma_{\mathrm{Sum}}}\nonumber\\
 \simeq 7\%,8\%,5\%,10\%.
\end{eqnarray}
The $\Xi_b'^*\pi$, $\Sigma_b^*K$, $\Lambda_b(5912)K$, and $\Lambda_b(5920)K$ decay modes may be ideal modes for our search for $|\Xi'_b~^4D_{\lambda\lambda} \frac{3}{2}^+ \rangle$ in future experiments.

We also estimate the radiative decays of these $J^P=3/2^+$ states into the $1P$-wave bottom baryon states. Our results are listed in Table~\ref{rXipD}. It is found that for $|\Xi_b'~^2D_{\lambda\lambda} \frac{3}{2}^+ \rangle$ the main radiative processes are $|\Xi_b'^{0}~^2D_{\lambda\lambda} \frac{3}{2}^+ \rangle\to |\Xi_b^{0}~ ^2P_{\lambda}\frac{3}{2}^- \rangle\gamma$ and $|\Xi_b'^{-}~^2D_{\lambda\lambda} \frac{3}{2}^+ \rangle\to |\Xi_b'^{-}~ ^2P_{\lambda}\frac{3}{2}^- \rangle\gamma,|\Xi_b'^{-}~ ^2P_{\lambda}\frac{3}{2}^- \rangle\gamma$, while for $|\Xi_b'~^4D_{\lambda\lambda} \frac{3}{2}^+ \rangle$, the main radiative processes are $|\Xi_b'^{0}~^4D_{\lambda\lambda} \frac{3}{2}^+ \rangle\to |\Xi_b^{0}~ ^2P_{\lambda}\frac{1}{2}^- \rangle\gamma$ and $|\Xi_b'^{-}~^4D_{\lambda\lambda} \frac{3}{2}^+ \rangle\to |\Xi_b'^{-}~ ^4P_{\lambda}\frac{1}{2}^- \rangle\gamma,|\Xi_b'^{-}~ ^4P_{\lambda}\frac{3}{2}^- \rangle\gamma$. Their partial radiative decay widths
are estimated to be $\mathcal{O}(10)$ keV, while the branching fractions may be $\mathcal{O}(10^{-4})$.

\subsubsection{$J^P=5/2^+$ states}

The $J^P=5/2^+$ state $|\Xi'_b~^2D_{\lambda\lambda} \frac{5}{2}^+ \rangle$ might be a narrow state with a width of a few tens of MeV.
Its dominant decay mode is $\Xi_b\pi$. If the mass of $|\Xi'_b~^2D_{\lambda\lambda} \frac{5}{2}^+ \rangle$ is taken as the prediction 6432 MeV in Ref.~\cite{Ebert:2011kk},
the sum of the partial widths for the pionic and kaonic decays
is $\Gamma_{\mathrm{Sum}}\sim 25$ MeV (see Table~\ref{Xicbp}), while
the ratio between the partial width for the $\Xi_b\pi$ channel
and $\Gamma_{\mathrm{Sum}}$ is predicted to be
\begin{eqnarray}
\frac{\Gamma[\Xi_b\pi]}{\Gamma_{\mathrm{Sum}}}\simeq 37\%.
\end{eqnarray}
To look for this state, the $\Xi_b\pi$ decay mode is worth observing in future experiments.

The other $J^P=5/2^+$ state $|\Xi'_b~^4D_{\lambda\lambda} \frac{5}{2}^+ \rangle$ might be also a narrow state with a width of a few tens of MeV.
It has large decay rates into $\Xi_b'^*\pi$ and $\Sigma_b^*K$.
If its mass is taken as the predictions in Ref.~\cite{Ebert:2011kk}, the sum of the partial widths of the poinic and kaonic decays is $\Gamma_{\mathrm{Sum}}\sim 30$ MeV (see Table~\ref{scb}), while the ratios between the partial widths for the $\Xi_b'^*\pi$ and $\Sigma_b^*K$ decay modes and $\Gamma_{\mathrm{Sum}}$ are predicted to be
\begin{eqnarray}
 \frac{\Gamma[\Xi_b'^*\pi]}{\Gamma_{\mathrm{Sum}}}
 \simeq 20\%, \ \ \frac{\Gamma[\Sigma_b^*K]}{\Gamma_{\mathrm{Sum}}}
 \simeq 16\%.
\end{eqnarray}
To look for this state, the $\Xi_b'^*\pi$ and $\Sigma_b^*K$ decay modes are worth observing in future experiments.

We also estimate the radiative decays of these $J^P=5/2^+$ states into the $1P$-wave bottom baryon states. Our results are listed in Table~\ref{rXipD}. It is found that for $|\Xi_b'~^2D_{\lambda\lambda} \frac{5}{2}^+ \rangle$ the main radiative processes are $|\Xi_b'^{0}~^2D_{\lambda\lambda} \frac{5}{2}^+ \rangle\to |\Xi_b^{0}~ ^2P_{\lambda}\frac{3}{2}^- \rangle\gamma$ and $|\Xi_b'^{-}~^2D_{\lambda\lambda} \frac{5}{2}^+ \rangle\to |\Xi_b'^{-}~ ^2P_{\lambda}\frac{3}{2}^- \rangle\gamma$, while for $|\Xi_b'~^4D_{\lambda\lambda} \frac{5}{2}^+ \rangle$, the main radiative processes are $|\Xi_b'^{0}~^4D_{\lambda\lambda} \frac{5}{2}^+ \rangle\to |\Xi_b^{0}~ ^2P_{\lambda}\frac{3}{2}^- \rangle\gamma$ and $|\Xi_b'^{-}~^4D_{\lambda\lambda} \frac{5}{2}^+ \rangle\to |\Xi_b'^{-}~ ^4P_{\lambda}\frac{3}{2}^- \rangle\gamma,|\Xi_b'^{-}~ ^4P_{\lambda}\frac{5}{2}^- \rangle\gamma$. Their partial radiative decay widths
are estimated to be $\sim 10-20$ keV, while the branching fractions is no more than $\mathcal{O}(10^{-4})$.

\subsubsection{$J^P=7/2^+$ state}

The $J^P=7/2^+$ state $|\Xi'_b~^4D_{\lambda\lambda} \frac{7}{2}^+ \rangle$ might have a relatively narrow width of $\mathcal{O}(10)$ MeV.
This state mainly decays into $\Xi_b\pi$, $\Lambda_bK$, $\Lambda_b(5912) K$, $|\Xi_{b}~^2P_{\lambda}\frac{1}{2}^- \rangle \pi$, and $|\Xi_{b}~^2P_{\lambda}\frac{3}{2}^- \rangle \pi$ channels. If one adopts the predicted mass in Ref.~\cite{Ebert:2011kk}, the sum of the partial widths of the pionic and kaonic decays
is estimated to be $\Gamma_{\mathrm{Sum}}\sim 40$ MeV (see Table~\ref{scb}), and the ratios between the partial widths for the $\Xi_b\pi$, $\Lambda_bK$, and $\Lambda_b(5912) K$ decay modes and $\Gamma_{\mathrm{Sum}}$ are predicted to be
\begin{eqnarray}
\frac{\Gamma[\Xi_b\pi,\Lambda_bK,\Lambda_b(5912) K]}{\Gamma_{\mathrm{Sum}}}\simeq 24\%,13\%,8\%.
\end{eqnarray}
The $\Xi_b\pi$, $\Lambda_bK$, and $\Lambda_b(5912) K$ decay modes may be ideal modes for our search for $|\Xi'_b~^4D_{\lambda\lambda} \frac{7}{2}^+ \rangle$ in future experiments.

We also estimate its radiative decays into the $1P$-wave bottom baryon states. Our results are listed in Table~\ref{rXipD}.
It is found that the decay rates for these radiative transitions are small. Their branching
fractions may be no more than $\mathcal{O}(10^{-4})$.

\begin{table*}[htp]
\begin{center}
\caption{\label{omigeD} Partial widths of strong and radiative decays for the $\lambda$-mode $D$-wave $\Omega_c$ and  $\Omega_b$
baryons, the masses (MeV) of which  are taken from the quark model predictions of Ref.~\cite{Ebert:2011kk}. $M_f$ stands for the masses of $P$-wave
heavy baryons (MeV) in the final states, which are adopted from the RPP~\cite{Olive:2016xmw} and Refs.~\cite{Ebert:2011kk,Aaij:2017nav}.
The units for the partial widths of radiative and strong decays are keV and MeV, respectively. The superscript (subscript) stands for the uncertainty of a prediction with a $+ 10\%$ ($-10\%$) uncertainty of the oscillator parameter $\alpha_{\rho}$.}
\scalebox{1.0}{
\begin{tabular}{cc|ccccccccccccccc}\hline\hline
 \multirow{2}{*}{Decay mode}       &\multirow{2}{*}{$M_f$}  &\multicolumn{2}{c}{$\underline{|\Omega_{c}~^2D_{\lambda\lambda} \frac{3}{2}^+\rangle(3282)}$}      &\multicolumn{2}{c}{$\underline{|\Omega_{c}~^2D_{\lambda\lambda} \frac{5}{2}^+\rangle(3286)}$}      &\multicolumn{2}{c}{$\underline{|\Omega_{c}~^4D_{\lambda\lambda}\frac{1}{2}^+\rangle (3287)}$}        &\multicolumn{2}{c}{$\underline{|\Omega_{c}~^4D_{\lambda\lambda} \frac{3}{2}^+\rangle(3298)}$}       &\multicolumn{2}{c}{$\underline{|\Omega_{c}~^4D_{\lambda\lambda} \frac{5}{2}^+\rangle(3297)}$}       &\multicolumn{2}{c}{$\underline{|\Omega_{c}~^4D_{\lambda\lambda} \frac{7}{2}^+\rangle(3283)}$}       \\
& &$\Gamma_i$   &$\mathcal{B}_i$ (\%)   &$\Gamma_i$   &$\mathcal{B}_i$ (\%)    &$\Gamma_i$   &$\mathcal{B}_i$ (\%)   &$\Gamma$   &$\mathcal{B}_i$ (\%)   &$\Gamma_i$   &$\mathcal{B}_i$ (\%)   &$\Gamma_i$   &$\mathcal{B}_i$ (\%)              \\
\hline
 $ \Xi_cK$                                           &2470 &7.97$^{+3.85}_{-3.61}$    &44.7      &5.16$^{-1.13}_{+1.42}$    &57.3    &15.9$^{+7.8}_{-7.2}$   &71.4        &7.90$^{+4.08}_{-3.73}$     &48.2      &1.65$^{-0.35}_{+0.44}$        &19.8     &6.43$^{-1.43}_{+1.77}$          &81.7      \\
 $ \Xi_c^{'}K$                                       &2575 &9.26$^{+2.34}_{-2.55}$    &51.9      &1.77$^{-0.45}_{+0.64}$    &19.7    &4.71$^{+1.22}_{-1.34}$    &20.8     &2.44$^{+0.68}_{-0.73}$     &14.8      &0.15$^{-0.04}_{+0.05}$        &1.79     &0.54$^{-0.14}_{+0.20}$          &6.73      \\
 $ \Xi'^*_{c}K$                                      &2645 &0.50$^{-0.01}_{+0.02}$    &2.8       &2.05$^{+0.28}_{-0.32}$    &22.8    &1.65$^{+0.25}_{-0.31}$    &7.4      &5.97$^{+0.90}_{-1.03}$     &36.4      &6.61$^{-0.60}_{+0.64}$        &79.2     &0.88$^{+0.11}_{-0.12}$          &11.2      \\
 $ |\Omega_c~1^2P_{\lambda} \frac{1}{2}^-\rangle \gamma $ &3000 &91.8$^{-14.5}_{+18.8}$  &0.51      &9.79$^{-1.60}_{+2.12}$  &0.11       &0.03                 &$<$0.01   &0.49$^{-0.05}_{+0.06}$   &$<$0.01   &0.19$^{-0.02}_{+0.02}$      &$<$0.01  &$\simeq0.0$ &$<$0.01   \\
 $ |\Omega_c~1^2P_{\lambda} \frac{3}{2}^-\rangle \gamma $ &3066 &18.4$^{-3.1}_{+4.0}$   &0.10      &39.7$^{-6.7}_{+8.9}$    &0.44       &0.04                  &$<$0.01    &0.02                &$<$0.01   &0.17$^{-0.02}_{+0.02}$      &$<$0.01  &0.13$^{-0.02}_{+0.02}$     &$<$0.01   \\
 $ |\Omega_c~1^4P_{\lambda} \frac{1}{2}^-\rangle \gamma $ &3050 &0.13$^{-0.02}_{+0.01}$  &$<$0.01   &0.008                   &$<$0.01    &70.3$^{-11.3}_{+14.8}$ &0.32     &34.7$^{-5.6}_{+7.3}$    &0.21      &5.56$^{-0.91}_{+1.20}$      &0.07     &1.07$^{-0.21}_{+0.30}$      &0.01      \\
 $ |\Omega_c~1^4P_{\lambda} \frac{3}{2}^-\rangle \gamma $ &3050 &0.2$^{-0.03}_{+0.03}$   &$<$0.01   &0.09                    &$<$0.01    &27.5$^{-4.5}_{+5.9}$   &0.12     &54.1$^{-8.8}_{+11.4}$   &0.33      &39.8$^{-6.6}_{+8.6}$     &0.47     &5.62$^{-0.94}_{+1.23}$      &0.07      \\
 $ |\Omega_c~1^4P_{\lambda} \frac{5}{2}^-\rangle \gamma $ &3090 &0.03                    &$<$0.01   &0.14$^{-0.02}_{+0.03}$  &$<$0.01    &2.44$^{-0.45}_{+0.62}$ &$<$0.10  &7.95$^{-1.36}_{+1.82}$   &0.05      &19.6$^{-3.3}_{+4.3}$ &0.24     &23.1$^{-3.9}_{+5.3}$     &0.29      \\
 Sum                                                    & &17.84$^{+6.16}_{-6.16}$   &          &9.03$^{-1.30}_{+1.73}$    &        &22.33$^{+9.25}_{-8.83}$   &           &16.21$^{+5.64}_{-5.47}$    &          &8.34$^{-0.98}_{+1.11}$       &         &7.82$^{-1.46}_{+1.85}$        &          \\
\hline\hline
\multirow{2}{*}{Decay mode}       &\multirow{2}{*}{$M_f$} &\multicolumn{2}{c}{$\underline{|\Omega_{b}~^2D_{\lambda\lambda} \frac{3}{2}^+\rangle(6530)}$}
&\multicolumn{2}{c}{$\underline{|\Omega_{b}~^2D_{\lambda\lambda} \frac{5}{2}^+\rangle(6520)}$}
&\multicolumn{2}{c}{$\underline{|\Omega_{b}~^4D_{\lambda\lambda} \frac{1}{2}^+\rangle(6540)}$}
&\multicolumn{2}{c}{$\underline{|\Omega_{b}~^4D_{\lambda\lambda} \frac{3}{2}^+\rangle(6549)}$}
&\multicolumn{2}{c}{$\underline{|\Omega_{b}~^4D_{\lambda\lambda} \frac{5}{2}^+\rangle(6529)}$}
&\multicolumn{2}{c}{$\underline{|\Omega_{b}~^4D_{\lambda\lambda} \frac{7}{2}^+\rangle(6517)}$}       \\
& &$\Gamma_i$   &$\mathcal{B}_i$ (\%)   &$\Gamma_i$   &$\mathcal{B}_i$ (\%)    &$\Gamma_i$   &$\mathcal{B}_i$ (\%)   &$\Gamma$   &$\mathcal{B}_i$ (\%)   &$\Gamma_i$   &$\mathcal{B}_i$ (\%)   &$\Gamma_i$   &$\mathcal{B}_i$ (\%)              \\
\hline
 $\Xi_bK$                    &  5795    &11.2$^{+5.5}_{-5.1}$      &55.5       &6.18$^{-1.36}_{+1.71}$     &75.83     &22.1$^{+11.7}_{-10.5}$   &76.98    &10.9$^{+6.1}_{-5.4}$  &53.48     &2.07$^{-0.49}_{+0.48}$     &28.09    &7.83$^{-1.69}_{+2.13}$        &90.84       \\
 $\Xi_b'K$                   &  5935    &8.43$^{+1.17}_{-1.38}$    &41.8       &0.30$^{-0.09}_{+0.13}$     &3.68      &4.69$^{+0.73}_{-0.86}$    &16.34    &2.55$^{+0.44}_{-0.51}$   &12.51         &0.03                       &0.41     &0.09        &1.04        \\
 $\Xi_b'^{*}K$               &  5955    &0.51$^{-0.09}_{-0.10}$    &2.52       &1.65$^{+0.15}_{-0.16}$     &20.25     &1.86$^{+0.23}_{-0.27}$    &6.48     &6.87$^{+1.85}_{-1.01}$   &33.71         &5.24$^{+0.37}_{-0.43}$     &71.10    &0.68$^{+0.06}_{-0.07}$        &7.89         \\
 $ |\Omega_b~1^2P_{\lambda} \frac{1}{2}^-\rangle \gamma $ &6301   &33.2$^{-5.1}_{+6.5}$       &0.16       &4.03$^{-0.66}_{+0.86}$   &0.05       &0.55$^{-0.08}_{+0.10}$    &$<$0.01  &1.20$^{-0.16}_{+0.19}$ &$<$0.01   &0.32$^{-0.05}_{+0.05}$   &$<$0.01   &$\simeq0.0$ &$<$0.01       \\
 $ |\Omega_b~1^2P_{\lambda} \frac{3}{2}^-\rangle \gamma $ & 6304  &17.9$^{-2.9}_{+3.7}$       &0.09       &16.8$^{-2.8}_{+3.6}$     &0.21       &0.26$^{-0.04}_{+0.05}$     &$<$0.01                &0.10$^{-0.02}_{+0.02}$ &$<$0.01   &0.73$^{-0.10}_{+0.12}$   &$<$0.01   &0.56$^{-0.07}_{+0.09}$      &$<$0.01       \\
 $ |\Omega_b~1^4P_{\lambda} \frac{1}{2}^-\rangle \gamma $ & 6312  &0.47$^{-0.06}_{+0.07}$     &$<$0.01    &0.024                    &$<$0.01    &40.4$^{-6.1}_{+7.5}$       &0.14                   &18.8$^{-2.9}_{+3.5}$    &0.09      &2.72$^{-0.43}_{+0.57}$   &0.04      &0.50$^{-0.11}_{+0.15}$      &$<$0.01       \\
 $ |\Omega_b~1^4P_{\lambda} \frac{3}{2}^-\rangle \gamma $ & 6311  &0.74$^{-0.10}_{+0.13}$     &$<$0.01    &0.27$^{-0.04}_{+0.04}$   &$<$0.01    &17.9$^{-2.8}_{+3.5}$       &0.06                   &31.0$^{-4.7}_{+5.9}$ &0.15      &15.4$^{-2.5}_{+3.2}$  &0.21      &2.70$^{-0.44}_{+0.58}$      &0.03          \\
 $ |\Omega_b~1^4P_{\lambda} \frac{5}{2}^-\rangle \gamma $ & 6311  &0.21$^{-0.03}_{+0.04}$     &$<$0.01    &0.91$^{-0.13}_{+0.17}$   &$<$0.01    &2.83$^{-0.56}_{+0.81}$      &0.01                  &8.98$^{-1.50}_{+1.98}$ &0.04      &14.6$^{-3.3}_{+2.0}$  &0.20      &12.1$^{-2.0}_{+2.6}$     &0.14      \\
 Sum                                &   &20.19$^{+6.57}_{-6.57}$      &           &8.15$^{-1.30}_{+1.68}$     &           &28.71$^{+12.65}_{-11.61}$  &         &20.38$^{+8.38}_{-6.91}$  &          &7.37$^{-0.11}_{+0.04}$     &          &8.62$^{-1.63}_{+2.06}$ &\\
\hline\hline
\end{tabular}}
\end{center}
\end{table*}

\subsection{$\Omega_c$}

In the $\Omega_c$ family, there are six $\lambda$-mode $1D$-wave excitations: $|\Omega_c~^4D_{\lambda\lambda} \frac{1}{2}^+ \rangle$,
$|\Omega_c~^4D_{\lambda\lambda} \frac{3}{2}^+ \rangle$, $|\Omega_c~^2D_{\lambda\lambda} \frac{3}{2}^+ \rangle$,
$|\Omega_c~^2D_{\lambda\lambda} \frac{5}{2}^+ \rangle$, $|\Omega_c~^4D_{\lambda\lambda} \frac{5}{2}^+ \rangle$,
and $|\Omega_c~^4D_{\lambda\lambda} \frac{7}{2}^+ \rangle$. However, no $D$-wave states have been established.
The typical masses of the $\lambda$-mode $1D$-wave $\Omega_c$ excitations
are $\sim 3.3$ within various quark model predictions (see Table~\ref{sp2}). In the possible mass ranges, we study their
strong decay transitions within the ChQM.
Our results are shown in Fig.~\ref{omcb}. To be more specific, taking the masses of the $1D$-wave states
obtained in the relativistic quark-diquark picture~\cite{Ebert:2011kk}, we give the predicted
widths in Table~\ref{omigeD}.

\subsubsection{$J^P=1/2^+$ state}

The $J^P=1/2^+$ state $|\Omega_c~^4D_{\lambda\lambda} \frac{1}{2}^+ \rangle$ may be a narrow
state with a width of $\Gamma_{\mathrm{total}}\sim 20$ MeV. Its decays are dominated by the $\Xi_c(2470)K$.
The decay rates into $\Xi_c'(2575)K$ and $\Xi_c'^*(2645)K$ are sizeable as well.
The branching fractions for the $\Xi_c(2470)K$, $\Xi_c'(2575)K$, and $\Xi_c'^*(2645)K$ modes are predicted to be
\begin{eqnarray}
\frac{\Gamma[\Xi_cK,\Xi_c'K,\Xi_c'^*K]}{\Gamma_{\mathrm{total}}}\simeq 71\%,21\%,7\%.
\end{eqnarray}
From Fig.~\ref{omcb}, it is found that the strong decay properties of $|\Omega_c~^4D_{\lambda\lambda} \frac{1}{2}^+ \rangle$
are less sensitive to its mass. The $\Xi_c(2470)K$ and $\Xi_c'(2575)K$ may be optimal channels for us to search for
this missing $J^P=1/2^+$ state $|\Omega_c~^4D_{\lambda\lambda} \frac{1}{2}^+ \rangle$.

We also estimate its radiative decays into the $1P$-wave charmed baryon states. Our results are listed in Table~\ref{omigeD}.
It is found that $|\Omega_c^{0}~^4D_{\lambda\lambda} \frac{1}{2}^+ \rangle$ might have
relatively large decay rates into $|\Omega_c^{0}~ ^4P_{\lambda}\frac{1}{2}^- \rangle\gamma$
and $|\Omega_c^{0}~ ^4P_{\lambda}\frac{3}{2}^- \rangle\gamma$, and the partial radiative decay width
is estimated to be $\mathcal{O}(10)$ keV. The branching fractions for these main radiative
decay processes may be $\mathcal{O}(10^{-3})$.

\subsubsection{$J^P=3/2^+$ states}

For the $J^P=3/2^+$ state $|\Omega_c~^2D_{\lambda\lambda} \frac{3}{2}^+ \rangle$,
the width is predicted to be $\Gamma_{\mathrm{total}}\sim 18$ MeV, which is less sensitive to
the phase space of strong decays (see Fig.~\ref{omcb}). It is found that $\Xi_c(2470)K$ together with
$\Xi_c'(2575)K$ governs the decays of $|\Omega_c~^2D_{\lambda\lambda} \frac{3}{2}^+ \rangle$.
Their branching fractions are predicted to be
\begin{eqnarray}
\frac{\Gamma[\Xi_cK,\Xi_c'K]}{\Gamma_{\mathrm{total}}}\simeq 45\%,52\%.
\end{eqnarray}
The $\Xi_c(2470)K$ and $\Xi_c'(2575)K$ may be ideal channels for us to search for
this missing $J^P=3/2^+$ state $|\Omega_c~^2D_{\lambda\lambda} \frac{3}{2}^+ \rangle$.

For the other $J^P=3/2^+$ state $|\Omega_c~^4D_{\lambda\lambda} \frac{3}{2}^+ \rangle$,
if its mass is below the threshold of $\Xi_c(2790)K$, the width is predicted to be
$\Gamma_{\mathrm{total}}\sim 16$ MeV, which is less sensitive to
the phase space of strong decays (see Fig.~\ref{omcb}). In this case, both $\Xi_c(2470)K$
and $\Xi_c'^*(2645)K$ are the dominant decay channels of $|\Omega_c~^2D_{\lambda\lambda} \frac{3}{2}^+ \rangle$,
while the decay rate into $\Xi_c'(2575)K$ is sizeable as well.
Their branching fractions are predicted to be
\begin{eqnarray}
\frac{\Gamma[\Xi_cK,\Xi_c'^*K,\Xi_c'K]}{\Gamma_{\mathrm{total}}}\simeq 48\%,36\%,15\%.
\end{eqnarray}
However, if the $\Xi_c(2790)K$ decay channel is open, this decay mode should be
the dominant decay mode. The decay width of $|\Omega_c~^4D_{\lambda\lambda} \frac{3}{2}^+ \rangle$
can reach up to $30-40$ MeV. The $\Xi_c(2470)K$ and $\Xi_c'^*(2645)K$ may be ideal channels for us to search for
this missing $J^P=3/2^+$ state $|\Omega_c~^4D_{\lambda\lambda} \frac{3}{2}^+ \rangle$.

We also estimate the radiative decays of these $J^P=3/2^+$ $\Omega_c$ states into the $1P$-wave charmed baryon states. Our results are listed in Table~\ref{omigeD}. It is found that $|\Omega_c^{0}~^2D_{\lambda\lambda} \frac{3}{2}^+ \rangle\to |\Omega_c^{0}~ ^2P_{\lambda}\frac{1}{2}^- \rangle\gamma$ and $|\Omega_c^{0}~^4D_{\lambda\lambda} \frac{3}{2}^+ \rangle\to |\Omega_c^{0}~ ^4P_{\lambda}\frac{1}{2}^- \rangle\gamma,|\Omega_c^{0}~ ^4P_{\lambda}\frac{3}{2}^- \rangle\gamma$ might have relatively large decay rates.
The partial radiative decay widths are estimated to be $\mathcal{O}(10)$ keV, while their branching
fractions may reach up to $\mathcal{O}(10^{-3})$.

\subsubsection{$J^P=5/2^+$ states}

For the $J^P=5/2^+$ state $|\Omega_c~^2D_{\lambda\lambda} \frac{5}{2}^+ \rangle$, if we take its mass as the prediction 3286 MeV in
Ref.~\cite{Ebert:2011kk}, the width is predicted to be $\Gamma_{\mathrm{total}}\sim 9$ MeV, which is slightly dependent on
the phase space of strong decays (see Fig.~\ref{omcb}). The decays of $|\Omega_c~^2D_{\lambda\lambda} \frac{3}{2}^+ \rangle$
are governed by the $\Xi_c(2470)K$ channel, while the decay rates into $\Xi_c'(2575)K$ and $\Xi_c'^*(2645)K$ are sizeable as well.
Their branching fractions are predicted to be
\begin{eqnarray}
\frac{\Gamma[\Xi_cK,\Xi_c'^*K,\Xi_c'K]}{\Gamma_{\mathrm{total}}}\simeq 57\%,23\%,20\%.
\end{eqnarray}
The $\Xi_c(2470)K$, $\Xi_c'(2575)K$, and $\Xi_c'^*(2645)K$ may be ideal channels for us to search for
this missing $J^P=5/2^+$ state $|\Omega_c~^2D_{\lambda\lambda} \frac{5}{2}^+ \rangle$.

For the other $J^P=5/2^+$ state $|\Omega_c~^4D_{\lambda\lambda} \frac{5}{2}^+ \rangle$,
if we take its mass as the prediction 3297 MeV in
Ref.~\cite{Ebert:2011kk}, the width is predicted to be
$\Gamma_{\mathrm{total}}\sim 8$ MeV, which is slightly dependent on
the phase space of strong decays (see Fig.~\ref{omcb}).
The $\Xi_c'^*(2645)K$ are the dominant decay channels of $|\Omega_c~^2D_{\lambda\lambda} \frac{3}{2}^+ \rangle$,
while the decay rate into $\Xi_c(2470)K$ is sizeable as well.
Their branching fractions are predicted to be
\begin{eqnarray}
\frac{\Gamma[\Xi_c'^*K,\Xi_cK]}{\Gamma_{\mathrm{total}}}\simeq 78\%,19\%.
\end{eqnarray}
The $\Xi_c(2470)K$ and $\Xi_c'^*(2645)K$ may be ideal channels for us to search for
this missing $J^P=5/2^+$ state $|\Omega_c~^4D_{\lambda\lambda} \frac{5}{2}^+ \rangle$.

We also estimate the radiative decays of the $J^P=5/2^+$ $\Omega_c$ states into the $1P$-wave charmed baryon states. Our results are listed in Table~\ref{omigeD}. It is found that $|\Omega_c^{0}~^2D_{\lambda\lambda} \frac{5}{2}^+ \rangle\to |\Omega_c^{0}~ ^2P_{\lambda}\frac{3}{2}^- \rangle\gamma$ and $|\Omega_c^{0}~^4D_{\lambda\lambda} \frac{5}{2}^+ \rangle\to |\Omega_c^{0}~ ^4P_{\lambda}\frac{3}{2}^- \rangle\gamma$ might have
relatively large decay rates. The partial radiative decay widths are estimated to be $\mathcal{O}(10)$ keV, while their branching
fractions may reach up to $\mathcal{O}(10^{-3})$.

\subsubsection{$J^P=7/2^+$ state}

The $J^P=7/2^+$ state $|\Omega_c~^4D_{\lambda\lambda} \frac{7}{2}^+ \rangle$
may be a narrow state with a width of a few MeV.
If its mass is taken to be 3283 MeV as predicted in
Ref.~\cite{Ebert:2011kk}, the width is predicted to be
$\Gamma_{\mathrm{total}}\sim 8$ MeV, which shows some sensitivities to
the phase space of strong decays (see Fig.~\ref{omcb}).
The decays of $|\Omega_c~^4D_{\lambda\lambda} \frac{7}{2}^+ \rangle$ are governed by
the $\Xi_c(2470)K$ channel. The branching fraction is predicted to be
\begin{eqnarray}
\frac{\Gamma[\Xi_cK]}{\Gamma_{\mathrm{total}}}\simeq 80\%.
\end{eqnarray}
The $\Xi_c(2470)K$ may be an ideal channel for our search for
this missing $J^P=7/2^+$ state $|\Omega_c~^4D_{\lambda\lambda} \frac{7}{2}^+ \rangle$.

We also estimate its radiative decays into the $1P$-wave charmed baryon states. Our results are listed in Table~\ref{omigeD}.
It is found that $|\Omega_c^{0}~^4D_{\lambda\lambda} \frac{7}{2}^+ \rangle$ might have
a relatively large decay rate into $|\Omega_c^{0}~ ^4P_{\lambda}\frac{5}{2}^- \rangle\gamma$,
and the partial radiative decay width is estimated to be $\mathcal{O}(10)$ keV. The branching fraction for this main radiative
decay process may be $\mathcal{O}(10^{-3})$.

\subsection{$\Omega_b$}

In the $\Omega_b$ family, according to the quark model classification,
there are six $\lambda$-mode $1D$-wave excitations: $|\Omega_b~^4D_{\lambda\lambda} \frac{1}{2}^+ \rangle$,
$|\Omega_b~^4D_{\lambda\lambda} \frac{3}{2}^+ \rangle$, $|\Omega_b~^2D_{\lambda\lambda} \frac{3}{2}^+ \rangle$,
$|\Omega_b~^2D_{\lambda\lambda} \frac{5}{2}^+ \rangle$, $|\Omega_b~^4D_{\lambda\lambda} \frac{5}{2}^+ \rangle$,
and $|\Omega_b~^4D_{\lambda\lambda} \frac{7}{2}^+ \rangle$. However, no $D$-wave states have been established.
The typical masses of the $\lambda$-mode $1D$-wave $\Omega_b$ excitations
are $\sim 6.6$ GeV within various quark model predictions (see Table~\ref{sp2}). In these possible mass ranges, we study their
strong decay transitions by emitting one light pseudoscalar meson within the ChQM.
Our results are shown in Fig.~\ref{omcb}. To be more specific, taking the masses of the $1D$-wave states
obtained in the relativistic quark-diquark picture~\cite{Ebert:2011kk}, we give the predicted
widths in Table~\ref{omigeD}. %The strong decay properties can be clearly seen from the figure and table.

\subsubsection{$J^P=1/2^+$ state}

The $J^P=1/2^+$ state $|\Omega_b~^4D_{\lambda\lambda} \frac{1}{2}^+ \rangle$ may be a narrow
state with a width of $\Gamma_{\mathrm{total}}\sim 30$ MeV. Its decays are dominated by the $\Xi_b(5795)K$.
The decay rates into $\Xi_b'(5935)K$ and $\Xi_b'^*(5955)K$ are sizeable as well.
The branching fractions for the $\Xi_b(5795)K$, $\Xi_b'(5935)K$, and $\Xi_b'^*(5955)K$ modes are predicted to be
\begin{eqnarray}
\frac{\Gamma[\Xi_bK,\Xi_b'K,\Xi_b'^*K]}{\Gamma_{\mathrm{total}}}\simeq 76\%,16\%,7\%.
\end{eqnarray}
From Fig.~\ref{omcb}, it is found that the strong decay properties of $|\Omega_b~^4D_{\lambda\lambda} \frac{1}{2}^+ \rangle$
are less sensitive to its mass.  $\Xi_b(5795)K$ and $\Xi_b'(5935)K$ may be ideal channels for us to search for
this missing $J^P=1/2^+$ state $|\Omega_b~^4D_{\lambda\lambda} \frac{1}{2}^+ \rangle$.

We also estimate its radiative decays into the $1P$-wave bottom baryon states. Our results are listed in Table~\ref{omigeD}.
It is found that $|\Omega_b^{-}~^4D_{\lambda\lambda} \frac{1}{2}^+ \rangle$ might have
a relatively large decay rate into $|\Omega_b^{-}~ ^4P_{\lambda}\frac{1}{2}^- \rangle\gamma$. The partial radiative decay width
of this process is estimated to be $\mathcal{O}(10)$ keV, while the branching fraction may be $\mathcal{O}(10^{-3})$.

\subsubsection{$J^P=3/2^+$ states}

For the $J^P=3/2^+$ state $|\Omega_b~^2D_{\lambda\lambda} \frac{3}{2}^+ \rangle$,
the width is predicted to be $\Gamma_{\mathrm{total}}\sim 20\pm 5$ MeV, which is slightly dependent on
the phase space of strong decays (see Fig.~\ref{omcb}). It is found that $\Xi_b(5795)K$ together with
$\Xi_b'(5935)K$ governs the decays of $|\Omega_b~^2D_{\lambda\lambda} \frac{3}{2}^+ \rangle$.
If its mass is taken to be 6530 MeV as predicted in Ref.~\cite{Ebert:2011kk},
the branching fractions for the $\Xi_b(5795)K$ and $\Xi_b'(5935)K$ modes are predicted to be
\begin{eqnarray}
\frac{\Gamma[\Xi_bK,\Xi_b'K]}{\Gamma_{\mathrm{total}}}\simeq 55\%,42\%,
\end{eqnarray}
The $\Xi_b(5795)K$ and $\Xi_b'(5935)K$ may be ideal channels for us to search for
this missing $J^P=3/2^+$ state $|\Omega_b~^2D_{\lambda\lambda} \frac{3}{2}^+ \rangle$.

For the other $J^P=3/2^+$ state $|\Omega_b~^4D_{\lambda\lambda} \frac{3}{2}^+ \rangle$,
if its mass is taken to be 6549 MeV as predicted in Ref.~\cite{Ebert:2011kk},
the width is predicted to be $\Gamma_{\mathrm{total}}\sim 20$ MeV.
The decays are dominated by the $\Xi_b(5795)K$ and $\Xi_b'^*(5955)K$ channels,
while the decay rate into $\Xi_b'(5935)K$ is sizeable as well.
Their branching fractions are predicted to be
\begin{eqnarray}
\frac{\Gamma[\Xi_bK,\Xi_b'^*K,\Xi_b'K]}{\Gamma_{\mathrm{total}}}\simeq 53\%,34\%,12\%,
\end{eqnarray}
which is sensitive to the phase space of strong decays (see Fig.~\ref{omcb}).
 $\Xi_b(5795)K$ and $\Xi_b'^*(5955)K$ may be ideal channels for our search for
this missing $J^P=3/2^+$ state $|\Omega_b~^4D_{\lambda\lambda} \frac{3}{2}^+ \rangle$.

We also estimate the radiative decays of these $J^P=3/2^+$ $\Omega_b$ states into the $1P$-wave bottom baryon states. Our results are listed in Table~\ref{omigeD}. It is found that $|\Omega_b^{-}~^2D_{\lambda\lambda} \frac{3}{2}^+ \rangle\to |\Omega_b^{-}~ ^2P_{\lambda}\frac{1}{2}^- \rangle\gamma$ and $|\Omega_b^{-}~^4D_{\lambda\lambda} \frac{3}{2}^+ \rangle\to |\Omega_b^{-}~ ^4P_{\lambda}\frac{3}{2}^- \rangle\gamma$ might have relatively large decay rates. The partial radiative decay widths are estimated to be $\mathcal{O}(10)$ keV, while their branching
fractions may reach up to $\mathcal{O}(10^{-3})$.

\subsubsection{$J^P=5/2^+$ states}

For the $J^P=5/2^+$ state $|\Omega_b~^2D_{\lambda\lambda} \frac{5}{2}^+ \rangle$, if we take its mass as the prediction 6520 MeV in
Ref.~\cite{Ebert:2011kk}, the width is predicted to be $\Gamma_{\mathrm{total}}\sim 8$ MeV. The decays of $|\Omega_b~^2D_{\lambda\lambda} \frac{3}{2}^+ \rangle$ are governed by the $\Xi_b(5795)K$ channel, while the decay rate into $\Xi_b'^*(5955)K$ are sizeable as well.
Their branching fractions is predicted to be
\begin{eqnarray}
\frac{\Gamma[\Xi_bK,\Xi_b'^*K]}{\Gamma_{\mathrm{total}}}\simeq 76\%,20\%.
\end{eqnarray}
It should be pointed out that the decay properties of $|\Omega_b~^2D_{\lambda\lambda} \frac{5}{2}^+ \rangle$
show some sensitivities to  its mass (see Fig.~\ref{omcb}).
 $\Xi_b(5795)K$ and $\Xi_b'^*(5955)K$ may be ideal channels for us to search for
this missing $J^P=5/2^+$ state $|\Omega_b~^2D_{\lambda\lambda} \frac{5}{2}^+ \rangle$.

For the other $J^P=5/2^+$ state $|\Omega_b~^4D_{\lambda\lambda} \frac{5}{2}^+ \rangle$,
if we take its mass as the prediction 6549 MeV in
Ref.~\cite{Ebert:2011kk}, the width is predicted to be
$\Gamma_{\mathrm{total}}\sim 8$ MeV.
$\Xi_b'^*(5955)K$ is the dominant decay channel of $|\Omega_b~^4D_{\lambda\lambda} \frac{5}{2}^+ \rangle$,
while the decay rate into $\Xi_b(5790)K$ is sizeable as well.
Their branching fractions are predicted to be
\begin{eqnarray}
\frac{\Gamma[\Xi_b'^*K,\Xi_bK]}{\Gamma_{\mathrm{total}}}\simeq 74\%,25\%.
\end{eqnarray}
The decay properties of $|\Omega_b~^4D_{\lambda\lambda} \frac{5}{2}^+ \rangle$
show some uncertainties with its mass changes (see Fig.~\ref{omcb}).
$\Xi_b(5790)K$ and $\Xi_b'^*(5955)K$ may be ideal channels for us to search for
this missing $J^P=5/2^+$ state $|\Omega_b~^4D_{\lambda\lambda} \frac{5}{2}^+ \rangle$.

We also estimate the radiative decays of the $J^P=5/2^+$ $\Omega_b$ states into the $1P$-wave bottom baryon states. Our results are listed in Table~\ref{omigeD}. It is found that $|\Omega_b^{-}~^2D_{\lambda\lambda} \frac{5}{2}^+ \rangle\to |\Omega_b^{-}~ ^2P_{\lambda}\frac{3}{2}^- \rangle\gamma$ and $|\Omega_b^{-}~^4D_{\lambda\lambda} \frac{5}{2}^+ \rangle\to |\Omega_b^{-}~ ^4P_{\lambda}\frac{3}{2}^- \rangle\gamma,|\Omega_b^{-}~ ^4P_{\lambda}\frac{5}{2}^- \rangle\gamma$ might have relatively large decay rates. The partial radiative decay widths are estimated to be $\mathcal{O}(10)$ keV, while their branching fractions may reach up to $\mathcal{O}(10^{-3})$.

\subsubsection{$J^P=7/2^+$ state}

The $J^P=7/2^+$ state $|\Omega_b~^4D_{\lambda\lambda} \frac{7}{2}^+ \rangle$
may be a narrow state with a width of a few MeV or a few tens of MeV.
If its mass is taken to be 6517 MeV as predicted in
Ref.~\cite{Ebert:2011kk}, the width is predicted to be
$\Gamma_{\mathrm{total}}\sim 8$ MeV, which shows some sensitivities to
the phase space of strong decays (see Fig.~\ref{omcb}).
The decays of $|\Omega_b~^4D_{\lambda\lambda} \frac{7}{2}^+ \rangle$ are governed by
the $\Xi_b(5970)K$ channel. The branching fraction is predicted to be
\begin{eqnarray}
\frac{\Gamma[\Xi_bK]}{\Gamma_{\mathrm{total}}}\simeq 89\%.
\end{eqnarray}
$\Xi_b(5970)K$ may be an ideal channel for us to search for
this missing $J^P=7/2^+$ state $|\Omega_b~^4D_{\lambda\lambda} \frac{7}{2}^+ \rangle$.

We also estimate its radiative decays into the $1P$-wave bottom baryon states. Our results are listed in Table~\ref{omigeD}.
It is found that $|\Omega_b^{-}~^4D_{\lambda\lambda} \frac{7}{2}^+ \rangle$ might have
a relatively large decay rate into $|\Omega_b^{-}~ ^4P_{\lambda}\frac{5}{2}^- \rangle\gamma$,
and the partial radiative decay width is estimated to be $\mathcal{O}(10)$ keV. The branching fraction for this main radiative
decay process may be $\mathcal{O}(10^{-3})$.

\section{Summary}\label{Summa}

In this work, we carry out a systematic study of the strong decays
with emission of one light pseudoscalar meson and the radiative decays
with emission one photon of the low-lying
$D$-wave singly heavy baryons in a constituent quark model.
Our results may provide helpful references to establish
these missing $D$-wave singly heavy baryons in future experiments.
Several key results from this study can be learned as follows.

The $\lambda$-mode $1D$-wave $J^P=3/2^+$ and $J^P=5/2^+$ excitations of
$\bar{\mathbf{3}}_F$ (i.e., $|^2D_{\lambda\lambda} \frac{3}{2}^+ \rangle$ and $|^2D_{\lambda\lambda} \frac{5}{2}^+ \rangle$) in the $\Lambda_{c}$ and $\Xi_{c}$ families might have been observed in experiments. Both $\Lambda_c(2860)$ and $\Xi_c(3050)$ seem to favor the $J^P=3/2^+$ excitation
$|^2D_{\lambda\lambda} \frac{3}{2}^+ \rangle$ of $\bar{\mathbf{3}}_F$, while $\Lambda_c(2880)$ and $\Xi_c(3080)$ seem to favor assigning
the $J^P=5/2^+$ excitation $|^2D_{\lambda\lambda} \frac{5}{2}^+ \rangle$. The nature of $\Xi_c(3050)$ and $\Xi_c(3080)$
may be tested by the radiative transitions $\Xi_c(3055)^0\to \Xi_c(2790)^0  \gamma$ and $\Xi_c(3080)^0 \to \Xi_c(2815)^0 \gamma$, respectively.

The missing $\lambda$-mode $1D$-wave $J^P=3/2^+$ and $J^P=5/2^+$ excitations of $\bar{\mathbf{3}}_F$ in the $\Lambda_{b}$ and $\Xi_{b}$ families
have a large potential to be found in forthcoming experiments. They might be narrow states with a width of $\sim 10$ MeV. In the $\Lambda_{b}$ family, the $J^P=3/2^+$ state might be established in the $\Sigma_b\pi$ and $\Lambda_b(5912) \gamma$ final states, while the $J^P=5/2^+$ state might be established in the $\Sigma_b^*\pi$ and $\Lambda_b(5920) \gamma$ final states.
In the $\Xi_{b}$ family, the $J^P=3/2^+$ state might be established in the $\Xi_b'\pi$, $\Xi_b'^*\pi$ and $\Sigma_b K$ final states, while the $J^P=5/2^+$ state might be established in the $\Sigma_b^* K$ and $\Xi_b'^*\pi$ final states.

The $\lambda$-mode $1D$-wave excitations of $\mathbf{6}_F$ in the $\Omega_{c}$ and $\Omega_b$ families have a large potential to be found in forthcoming experiments as well. They are fairly narrow states with a width of a few MeV or a few tens of MeV.
The kaonic decay channels $\Xi_c(2470)K$, $\Xi_c'(2575)K$, and $\Xi_c'^*(2645)K$ may be ideal channels for us to search for
these missing $1D$-wave excited $\Omega_c$ states, while the kaonic decay channels $\Xi_b K$, $\Xi_b' K$, and $\Xi_b'^* K$
may be ideal channels for us to search for these missing $1D$-wave excited $\Omega_b$ states.

The $\lambda$-mode $1D$-wave excitations in the $\Sigma_{c(b)}$ and $\Xi_{c(b)}'$ families appear to have relatively broad widths.
The sum of the partial widths with emission of a one-pion meson and one-kaon meson
is about $50\sim 200$ MeV.  These $1D$-wave states might have large decay rates into the $1P$-wave
heavy baryon states via the pionic and/or kaonic decays. The $\Lambda_c(2595) \pi$ and $\Lambda_c(2625) \pi$ channels may be ideal channels for looking for the missing $1D$-wave excitations in the $\Sigma_{c}$ family. The $\Xi_c(2790) \pi$, $\Xi_c(2815) \pi$,
$\Lambda_c(2595) K$, and $\Lambda_c(2625) K$ decay channels may be ideal channels for looking for the missing $1D$-wave excitations in the $\Xi_{c}'$ family.  The $\Lambda_b(5912)\pi$, and $\Lambda_b(5920)\pi$ decay channels may be ideal channels for looking for the missing $D$-wave excitations in the $\Sigma_b$ family.

Finally, it should be pointed out that some of our predictions bear a fairly large uncertainty from the nonrelativistic harmonic
oscillator wave functions adopted in the calculations. Considering a $10\%$ uncertainty of the oscillator parameter,
one finds that the uncertainty of our predictions can reach up to $\sim 30\%$. In some senses, our results are only a semiquantitative estimation based on the SU(3) symmetry. Fortunately, it is found that most of the featured results of the singly heavy baryons predicted in the present work and previous work~\cite{Wang:2017kfr} are consistent with other model approaches and observations, which indicates that the constituent quark model can still serve as a useful tool for investigating the heavy baryon mesonic decays and radiative transitions.

\section*{  Acknowledgments }

We thank Qiang Zhao for helpful discussions. This work is supported by the National
Natural Science Foundation of China under Grants No. 11775078.

%\appendix
%%%%%%%%%%%%%%%%%%%%%%%%%%%%%%%%%%%%%%%%%%%%%%%%%%%%%%%%%%%%%%%%%%555

\bibliographystyle{unsrt}

\end{document}